\documentclass[10pt,letterpaper]{article}
\usepackage{causality-preprint}
\usepackage{amsmath}
\usepackage{booktabs}
\usepackage{array}
\usepackage{mathtools}
\usepackage{microtype}
\usepackage{xcolor}
\usepackage{tikz}

\usepackage{xspace}

\newcommand*\ie{\text{i.e.,}\xspace}

\newcommand{\PiP}{\Pi}
\newcommand{\Dx}{D^{\mathsf{x}}}
\newcommand{\Univ}{\mathcal{U}}
\newcommand{\EDB}{\Sigma_{\mathsf{E}}}
\newcommand{\IDB}{\Sigma_{\mathsf{I}}}
\newcommand{\WFM}{\mathfrak{M}}
\newcommand{\Const}{\mathsf{const}}
\newcommand{\Ground}{\mathsf{ground}}
\newcommand{\Val}{\mathsf{Val}}
\newcommand{\Bthree}{\mathbb{B}_{3}}
\newcommand{\True}{\top}
\newcommand{\Undef}{\mathpalette\UndefSymbol{\mathsf{u}}}
\newcommand{\UndefSymbol}[2]{%
  \begingroup
  \setbox0=\hbox{$\mathsurround=0pt#1\epsilon$}%
  \setbox2=\hbox to\wd0{\hss$\mathsurround=0pt#1#2$\hss}%
  \ht2=\ht0 \dp2=\dp0 \box2
  \endgroup}
\newcommand{\False}{\bot}
\newcommand{\Nat}{\mathbb{N}}
\newcommand{\Inf}{\infty}

\newcommand{\Ans}{\mathsf{ans}}
\newcommand{\Resp}{\mathsf{Resp}}
\newcommand{\Kappa}{\kappa}
\newcommand{\DetRad}{\delta}
\newcommand{\DirRad}{\delta}
\newcommand{\Dist}{\mathsf{d}}
\newcommand{\toggle}{\mathbin{\triangle}}
\newcommand{\Cause}{\mathsf{Cause}}
\newcommand{\DetRadius}{\mathsf{DetRadius}}
\newcommand{\NF}{\mathsf{NF}}

\newcommand{\Win}{\mathsf{Win}}
\newcommand{\Move}{\mathsf{Move}}
\newcommand{\Succ}{\mathsf{Succ}}
\newcommand{\Rank}{\mathcal{R}}
\newcommand{\Draw}{\mathcal{D}}
\newcommand{\Req}{\mathcal{Q}}
\newcommand{\Certs}{\mathfrak{C}}
\newcommand{\ReqOf}{\mathsf{term}}
\newcommand{\Reqs}{\mathfrak{Q}}
\newcommand{\RawReqs}{\widetilde{\mathfrak{Q}}}
\newcommand{\Pairs}{\mathfrak{P}}
\newcommand{\Flip}{\mathsf{flip}}
\newcommand{\ReqJoin}{\mathbin{\wedge}}
\newcommand{\Cost}{\mathsf{cost}}
\newcommand{\Pot}{\overline{M}}
\newcommand{\Mx}{M^{\mathsf{x}}}
\newcommand{\Mu}{M^{\mathsf{u}}}
\newcommand{\Parent}{\mathsf{par}}

\newcommand{\Wset}{W}
\newcommand{\Lset}{L}
\newcommand{\Dset}{D}
\newcommand{\Wp}{W^{+}}
\newcommand{\Wm}{W^{-}}
\newcommand{\rhoF}{\rho}

\newcommand{\Hyp}{\mathcal{H}}
\newcommand{\Idx}{J}
\newcommand{\Cov}{\mathsf{cov}}
\newcommand{\Trans}{\tau_{\mathsf{tr}}}
\newcommand{\VC}{\mathsf{vc}}

\newcommand{\Approve}{\mathsf{approve}}
\newcommand{\Eligible}{\mathsf{eligible}}
\newcommand{\Blocked}{\mathsf{blocked}}
\newcommand{\Violation}{\mathsf{violation}}
\newcommand{\HighRisk}{\mathsf{highRisk}}
\newcommand{\Cleared}{\mathsf{cleared}}
\newcommand{\Audit}{\mathsf{audit}}
\newcommand{\Clearance}{\mathsf{clearance}}

\newcommand{\CountMin}{\mathsf{CountMin}}
\newcommand{\FPNPlog}{\mathsf{FP}^{\mathsf{NP}[\log]}}
\newcommand{\SharpP}{\#\mathsf{P}}
\newcommand{\herbrand}{\mathfrak{B}}

\newif\ifappendixproofs
\newcommand{\registerfullproof}[1]{%
  \csgdef{prooflabel@\getrefnumber{#1}}{#1}}
\makeatletter
\newcommand{\proofresultref}[1]{%
  \protected@write\@auxout{}{\string\registerfullproof{#1}}%
  \hypertarget{fullproof.#1}{}\originalproofref{#1}}
\makeatother
\newcommand{\linkedproofheading}[1]{%
  \begingroup
    \let\originalproofref\ref
    \let\ref\proofresultref
    #1%
  \endgroup}
\newcommand{\resultheading}[2]{%
  \ifcsdef{prooflabel@\thetheorem}
    {\hyperlink{fullproof.\csuse{prooflabel@\thetheorem}}{\thmname{#1}\thmnumber{ #2}}}
    {\thmname{#1}\thmnumber{ #2}}}
\NewCommandCopy\articleproof\proof
\NewCommandCopy\endarticleproof\endproof
\renewenvironment{proof}[1][\proofname]{%
  \ifappendixproofs
    \articleproof[\linkedproofheading{#1}]%
  \else\articleproof[#1]\fi
}{\endarticleproof}

\newtheoremstyle{result}{7pt plus 1pt}{6pt plus 1pt}{\itshape}{}{\bfseries}{.}{.5em}{%
  \resultheading{#1}{#2}\thmnote{ \normalfont(#3)}}
\newtheoremstyle{definitiontext}{7pt plus 1pt}{6pt plus 1pt}{\normalfont}{}{\bfseries}{.}{.5em}{%
  \resultheading{#1}{#2}\thmnote{ \normalfont(#3)}}
\theoremstyle{result}
\newtheorem{theorem}{Theorem}
\newtheorem{lemma}[theorem]{Lemma}
\newtheorem{corollary}[theorem]{Corollary}
\newtheorem{proposition}[theorem]{Proposition}
\theoremstyle{definitiontext}
\newtheorem{definition}[theorem]{Definition}
\newtheorem{example}[theorem]{Example}
\newcommand{\keepresultheading}[1]{\BeforeBeginEnvironment{#1}{\Needspace{3\baselineskip}}}
\forcsvlist{\keepresultheading}{theorem,lemma,corollary,proposition,definition,example}

\tikzset{every picture/.append style={scale=1.10,transform shape}}
\title{Causal Resolution of Undefined Answers in Datalog with Negation}
\usepackage{orcidlink} 
\author{\texorpdfstring{%
  \mbox{\normalfont\sffamily\large
    \MakeUppercase{Ratan Bahadur Thapa}%
    \textsuperscript{\orcidlink{0009-0000-2368-5928}},
    University of Stuttgart, Germany}%
}{Ratan Bahadur Thapa}}
\begin{document}
\maketitle
\vspace{.25\baselineskip}
\begingroup
\noindent\textbf{Abstract.}
Unstratified negation may leave the answer to a Boolean Datalog query
undefined under the well-founded semantics. We study database causality in
this setting by asking which fact insertions and deletions realize a
prescribed transition and which other changes preserve the observed query answer while making a designated fact pivotal. Such a set of changes to other facts is a contingency for the designated fact and the prescribed transition. Our game representation assigns one shared move to each mutable fact and preserves query values and intervention distances. Pairing source and target certificates characterizes all minimum contingencies for the six transitions among true, false, and undefined. When inserting or deleting a move resolves a draw, a directed tree
traces the newly winning and losing positions along alternating paths to the
source of that move. For the fixed one-rule win--move program, deciding
whether a contingency of size at most $k$ exists is $\mathrm{NP}$-complete
for each transition, even on bounded-degree graphs. Computing the minimum
size is $\mathrm{FP}^{\mathrm{NP}[\log]}$-complete, and counting minimum
contingencies is $\#\mathrm{P}$-hard under polynomial-time Turing reductions. When only one of true and false is reachable, the prime implicants of undefinedness satisfied at the observation cover every undefined state, and their supports induce nonempty subtrees of a common tree, we compute a minimum resolving intervention in polynomial time from the game instance. From the explicit
tree and these implicants, we construct an equally large family of disjoint
supports, certifying optimality.

\par
\endgroup
\medskip

\section{Introduction}\label{sec:introduction}
When a database query returns an unexpected answer, one may ask how the answer was derived, why an expected answer is absent, or which input facts are responsible for the observed answer. Lineage and provenance describe successful derivations \cite{buneman2001and,green2007provenance}, whereas why-not provenance can identify failed derivations that explain why an expected answer is absent \cite{lee2019pug}. Database causality asks a different question: whether inserting or deleting a designated input fact changes the query answer, possibly after changes to other input facts that preserve the observed answer \cite{meliou2010causality,meliou2010complexity}. We study this question for Datalog programs with recursion through negation under the well-founded semantics (WFS)~\cite{van1991well}. The answer to a Boolean query can then be true, false, or undefined, even when the input database is completely specified. Changing a true answer to undefined and changing it to false can therefore require different updates.

Deletion-based causality partitions the database into \emph{exogenous facts}, which remain fixed, and \emph{endogenous facts}, which may be deleted \cite{meliou2010causality}. An endogenous fact is an actual cause of a true answer if there exists a set of other endogenous facts, called a \emph{contingency}, whose deletion preserves the answer but whose deletion together with the designated fact makes the answer false. For Datalog with negation, we keep the program fixed and distinguish fixed input facts from mutable facts, which we may insert as well as delete. For an undefined answer, we seek a set of insertions and deletions, an \emph{intervention}, that yields a prescribed value, true or false. For a designated fact, a \emph{contingency} consists of changes to other facts that preserve the observed answer and make the designated insertion or deletion yield a prescribed different answer. We seek sets of minimum cardinality in both cases. For contingencies, we consider all six ordered transitions between distinct truth values.
\begin{example}\label{example_intro_1}
Consider the Datalog program $\Pi_{\mathsf{ex}}$ for an approval policy, where $\neg$ denotes default negation:
\begin{equation*}
\begin{aligned}
 \Approve(x)&\leftarrow\Eligible(x),\neg\Blocked(x),\\
 \Blocked(x)&\leftarrow\Violation(x),\\
 \Blocked(x)&\leftarrow\HighRisk(x),\neg\Cleared(x),\\
 \Cleared(x)&\leftarrow\Clearance(x),\\
 \Cleared(x)&\leftarrow\Audit(x),\neg\Blocked(x).
\end{aligned}
\end{equation*}
For an application $a$, let
\[
\begin{aligned}
\Univ=\{\Eligible(a),\Violation(a),\HighRisk(a),\Audit(a),\Clearance(a)\}
\end{aligned}
\]
be the set of facts that may be inserted or deleted (the mutable facts). There are no fixed input facts. Let $E=\{\Eligible(a),\allowbreak\HighRisk(a),\allowbreak\Audit(a)\}$ be the observed state. A state $F\subseteq\Univ$ specifies which mutable facts are present. Toggling $J\subseteq\Univ$ produces the symmetric difference $E\toggle J=(E\setminus J)\cup(J\setminus E)$.

For each state $F$, we evaluate $\Pi_{\mathsf{ex}} \cup F$ under the WFS. We write $\True$, $\False$, and $\Undef$ for true, false, and undefined, respectively.
At $E$, both $\Violation(a)$ and $\Clearance(a)$ are absent. The remaining rules for $\Blocked(a)$ and $\Cleared(a)$ each require the default negation of the other atom. Each atom remains possible if the other is false, but neither can be derived while the other is undecided. Both atoms are therefore undefined under WFS, as is $\Approve(a)$ because $\Eligible(a)$ is present. The table gives the values at $E$ and after each possible single change.
\begin{center}
\small
\setlength{\tabcolsep}{4.5pt}
\renewcommand{\arraystretch}{1.12}
\begin{tabular}{|l|cc|c|}
\hline
Change at $E$ & $\Blocked(a)$ & $\Cleared(a)$ & $\Approve(a)$\\
\hline
None                    & $\Undef$ & $\Undef$ & $\Undef$\\
\hline
Insert $\Clearance(a)$  & $\False$ & $\True$  & $\True$\\
Delete $\HighRisk(a)$   & $\False$ & $\True$  & $\True$\\
Insert $\Violation(a)$  & $\True$  & $\False$ & $\False$\\
Delete $\Audit(a)$      & $\True$  & $\False$ & $\False$\\
Delete $\Eligible(a)$   & $\Undef$ & $\Undef$ & $\False$\\
\hline
\end{tabular}
\end{center}
A designated fact is an \emph{actual cause} for a transition if it has a contingency for that transition. If a smallest such contingency has size $k$, the responsibility of the designated fact is $1/(1+k)$~\cite{chockler2004responsibility,meliou2010complexity}. A fact has higher responsibility when fewer other changes are needed. A fact whose change alone gives the specified transition is a \emph{counterfactual cause}. Each fact changed in the table therefore has responsibility $1$ for the indicated transition between $\True,\False$ and $\Undef$.

At $E\cup\{\Clearance(a)\}$, deleting $\Clearance(a)$ makes $\Approve(a)$ undefined. To make the same deletion yield false, first delete $\Audit(a)$. The query answer $\Approve(a)$ remains true until $\Clearance(a)$ is deleted. Thus, $\{\Audit(a)\}$ is a minimum contingency of size 1 for the transition $\True\longrightarrow\False$. At this observed state, $\Clearance(a)$ has responsibility \(1\) for $\True\longrightarrow\Undef$ and \(1/2\) for $\True\longrightarrow\False$.
\end{example}

Viegas Damásio et al. use conjunctions of literals to characterize the well-founded truth value of an atom after an update~\cite{viegas2013justifications}. For stratified Datalog, where negation does not occur along a cycle of predicate dependencies, the answers are Boolean. Conditions on which facts are present or absent before and after the designated change determine the minimum contingency size~\cite{thapa2026causalexplanationsstratifieddatalog}. We extend this question to recursive negation and characterize minimum contingencies for all six transitions.

Classical game normal forms represent the true answers of a Datalog query under the well-founded semantics as winning positions~\cite{flum1997total,flum2000games}. For causality, the game must preserve the query value at every mutable state and the number of facts changed by each intervention. Treating different occurrences of the same input fact as independently mutable can alter the minimum contingency size. We therefore construct a game in which each mutable fact has its own move, shared by all occurrences of that fact. Example~\ref{example_intro_2} introduces the one-rule win--move program used in the construction.
\begin{example}\label{example_intro_2}
Consider the one-rule program $\PiP_{\Win}$ with binary input relation $\Move$, where  $\Move(x,y)$ permits a move from $x$ to $y$. The query $\Win(x)$ asks whether the player to move at $x$ can force a win. The rule is
\begin{equation}
\Win(x)\leftarrow\Move(x,y),\neg\Win(y).
\label{eq:win-intro}
\end{equation}
The two players move on a directed graph that may contain self-loops. Starting at $x$, the players take turns following one outgoing move. A player with no move loses, and an infinite play is drawn. Moving to a losing position gives the player who made the move a winning strategy. Under optimal play, $\Win(x)$ is true, false, or undefined according as the player to move can force a win, the opponent can force a win, or neither can force a win~\cite{galanaki2008infinite,flum1997total,flum2000games}. A vertex is terminal if it has no outgoing move. We call the fixed moves exogenous. Let $\Move(a,s)$ and $\Move(b,b)$ be exogenous, let $\ell$ be terminal, and let
$\Univ_{\mathsf g}=\{\Move(s,a),\Move(s,b),\Move(s,\ell)\}$ be the mutable moves, with observed state $E_{\mathsf g}=\{\Move(s,a)\}$.
At the observed state, play from $s$ must follow the cycle $s\longrightarrow a\longrightarrow s$, which makes $\Win(s)$ undefined. Inserting $\Move(s,\ell)$ yields $\Undef\longrightarrow\True$, deleting $\Move(s,a)$ yields $\Undef\longrightarrow\False$, and inserting $\Move(s,b)$ preserves undefinedness.
\end{example}

\noindent\textbf{Our contributions.}
\begin{enumerate}
\item \textbf{Game representation.} For a fixed Datalog program with default negation and safe rules (each variable occurs in a positive body atom), we construct a game in polynomial time from the database, query, and mutable facts. All occurrences of a mutable fact use the same move, and distinct mutable facts use distinct moves. We preserve the query value at every state and map each insertion, deletion, and minimum contingency to its counterpart in the game (Section~\ref{sec:causal-normal-form}, Theorem~\ref{thm:causal-normal-form}).

\item \textbf{Minimum contingencies.} We characterize all minimum contingencies for the six transitions by pairing certificates for the observed and target answers. We use each certificate to specify move conditions that establish the corresponding value. For the designated move, we check the source condition against its observed presence or absence and the target condition against its status after the toggle. We require the two certificates to impose consistent conditions on all other mutable moves (Section~\ref{sec:certificates}, Theorems~\ref{thm:structured-completeness} and~\ref{thm:requirement-normal-form}).

\item \textbf{Propagation of one update.} We prove that whenever inserting or deleting one move resolves a draw at the query, the positions that become winning or losing admit a directed tree of moves in the updated game. Paths in the tree lead to the source of the updated move and alternate between winning and losing positions (Section~\ref{sec:certificates}, Theorem~\ref{thm:resolution-tree}).

\item \textbf{Complexity bounds.} We prove that deciding whether a contingency of size at most a given bound exists is NP-complete for each transition, and that resolving an undefined answer within the bound is also NP-complete. We establish both results for the single rule of Example~\ref{example_intro_2} when the graph of all possible moves has in-degree at most three and out-degree at most two. On unrestricted graphs, we also classify minimum-size computation and prove hardness of counting minimum contingencies (Section~\ref{sec:complexity}, Theorems~\ref{thm:np-completeness} and~\ref{thm:exact-complexity}).

\item \textbf{Tractable interventions.} We consider inclusion-minimal conjunctions of fact literals that hold at an undefined observation and force undefinedness. The facts in each conjunction form its support. Our algorithm applies when only one of true and false is reachable, every undefined state satisfies one of the conjunctions, and their supports induce nonempty subtrees of a common tree on the mutable facts. We compute a minimum resolving intervention in polynomial time from the game instance and target alone. With the tree and conjunctions, we construct an equally large family of pairwise disjoint supports, certifying optimality (Section~\ref{sec:minmax}, Theorem~\ref{thm:arboreal-minmax}).
\end{enumerate}
\section*{Related Work}
Meliou et al.~\cite{meliou2010complexity} define causes of conjunctive-query answers by deleting tuples designated as endogenous (eligible for deletion), and causes of non-answers by inserting tuples from a given set. For an initially true Boolean query, resilience is the minimum number of endogenous tuples whose deletion makes the query false. Freire et al.~\cite{freire2015complexity} classify resilience for self-join-free Boolean conjunctive queries and correct the earlier dichotomy for responsibility. Their resilience dichotomy extends to binary conjunctive queries in which at most one relation symbol occurs twice and the others occur once~\cite[Theorem~37]{freire2020selfjoins}. Makhija and Gatterbauer give integer programs for both problems under set and bag semantics~\cite{makhija2023unified}. For positive Datalog and a true answer, an endogenous fact is an actual cause under deletions exactly when it belongs to an inclusion-minimal set of endogenous facts that, together with the fixed facts, suffices for the answer~\cite{bertossi2017causes,thapa2026causalityminimalsupportsrecursive}.

Why-provenance identifies witnesses for an answer, and semiring provenance describes how input tuples contribute to a derivation~\cite{buneman2001and,green2007provenance}. For clique-frontier-guarded stratified Datalog, Amarilli et al. construct a circuit with cycles and one input per fact \cite{amarilli2019evaluating}. For bounded program body size and instance treewidth, the construction takes time linear in the product of the program and instance sizes~\cite[Theorem~34]{amarilli2019evaluating}. K{\"o}hler, Lud{\"a}scher, and Zinn explain wins, losses, and draws by the moves available in a game~\cite{kohler2013first}. Lee, Lud{\"a}scher, and Glavic compute provenance graphs for nonrecursive Datalog with negation~\cite{lee2019pug}. Bowers, Xia, and Lud{\"a}scher further distinguish moves according to their values and lengths~\cite{bowers2024game}. We (cf. Theorem~\ref{thm:resolution-tree}) relate the newly won and lost positions by a tree rooted at the source of the updated move.

Viegas Dam\'{a}sio, Analyti, and Antoniou give provenance formulas for all three well-founded values when facts may be inserted or deleted and rules may be deleted~\cite[Definition~8 and Theorem~3]{viegas2013justifications}. Cabalar and Fandinno identify rules that enable or inhibit a well-founded derivation~\cite{cabalarFandinno2017}. Bogaerts et al. use justifications for why- and why-not provenance under WFS~\cite{BogaertsEtAl2026}. Hellerstein associates a tuple with the admissible choices under which it holds when an input has several possible outcomes~\cite{hellerstein2026determinationprovenanceambiguityalgebra}. For a designated fact, we pair terms obtained from game certificates before and after its change to recover all minimum contingencies in the original database.

Let $\FPNPlog$ denote the functions computable in polynomial time with $O(\log(n+2))$ adaptive queries to an NP oracle on inputs of length $n$. Computing responsibility is $\FPNPlog$-complete in binary causal models under the original Halpern--Pearl definition~\cite{chockler2008specification} and, in data complexity, for a fixed Boolean conjunctive query~\cite[Theorem~40]{salimiBertossi2015repairs}. For positive Boolean Datalog, cause recognition and deciding whether responsibility exceeds a given threshold are NP-complete in data complexity~\cite{bertossi2017causes}. Bienvenu, Figueira, and Lafourcade study responsibility measures based on the sizes of minimal supports for unions of conjunctive queries with negation~\cite{bienvenu2026responsibility}. Comer and Tannen minimize the number of insertions and deletions that produce a missing answer~\cite{ComerTannen2026}. For a fixed semipositive Datalog query they compute a minimum repair in polynomial time. For unions of conjunctive queries with negation, computing the minimum repair size in combined complexity is complete for $\mathsf{OptP}[\log n]$, the class of minima over accepting branches of a nondeterministic polynomial-time machine with $O(\log n)$ bits in the minimum value on inputs of length $n$. Our lower bounds hold for each of the six transitions of a single fixed recursive program.

\section{Preliminaries}\label{sec:preliminaries}
\paragraph*{Datalog with Negation.}
Let $\EDB$ and $\IDB$ be disjoint finite schemas of extensional and intensional predicates, respectively, and let every predicate have a fixed arity in $\Nat=\{0,1,\ldots\}$. A term is a constant or a variable. An atom over $\EDB\cup\IDB$ has the form $R(t_1,\ldots,t_r)$, where $R$ has arity $r\in \Nat$. A literal is an atom $A$ or the default-negated atom $\neg A$; the two forms are positive and negative, respectively. An atom is ground if it contains no variables; a fact is a ground atom. A normal Datalog rule has the form
\begin{equation*}
     H\leftarrow B_1,\ldots,B_m,\neg C_1,\ldots,\neg C_n,
\end{equation*}
where $m,n\in\Nat$, the head $H$ is an intensional atom, and the body
consists of the positive atoms $B_1,\ldots,B_m$ and the negative literals
$\neg C_1,\ldots,\neg C_n$, with each $B_i$ and $C_j$ an atom over
$\EDB\cup\IDB$. A rule is safe if every variable in the head or in a negative literal occurs in a positive body atom. We consider finite programs of safe normal rules. A program is stratified if the predicates admit levels in $\Nat$ such that each rule head has level at least that of each positive body predicate and strictly greater than that of each negated body predicate.

\paragraph*{Well-Founded Semantics.}
A normal Datalog program is ground if no rule contains variables. For a finite ground normal program $P$, let $\herbrand(P)$ denote the Herbrand base of $P$, the set of ground atoms over the chosen signature and finite domain. The well-founded model of $P$ is the partial interpretation
$\WFM(P)=(T_P,F_P)$, where the disjoint sets $T_P,F_P\subseteq\herbrand(P)$ contain the atoms
assigned true and false, respectively \cite{van1991well}. For $A\in\herbrand(P)$, we write
$\WFM(P)(A)=\True$ if $A\in T_P$, $\WFM(P)(A)=\False$ if
$A\in F_P$, and $\WFM(P)(A)=\Undef$ otherwise.  For a default-negated literal $\neg A$, the truth value is $\True$, $\False$, or $\Undef$ according as the truth value of $A$ is
$\False$, $\True$, or $\Undef$, respectively.

For $X\subseteq\herbrand(P)$, the reduct $P^X$ discards each rule with a literal $\neg A$ such that $A\in X$ and removes the negative literals from the surviving rules. Let $g_P(X)$ be the least model of the resulting positive program. Starting from $\varnothing$, repeatedly add the head of each rule whose body atoms are present, until no further atom can be added. For $i\in\Nat$, the alternating fixpoint construction~\cite{vanGelder1993alternating} uses
\begin{equation}\label{eq:wfs-bounds}
 K_0=\varnothing,\qquad U_i=g_P(K_i),\qquad K_{i+1}=g_P(U_i).
\end{equation}
If $X\subseteq Y$, then $P^Y\subseteq P^X$ and $g_P(Y)\subseteq g_P(X)$. Thus $g_P\circ g_P$ is monotone. Since $K_0=\varnothing\subseteq K_1$, iterating $g_P\circ g_P$ gives $K_i\subseteq K_{i+1}$; applying $g_P$ gives $U_{i+1}\subseteq U_i$. Also $K_0\subseteq U_0$, and
\[
 K_i\subseteq U_i
 \ \Longrightarrow\ K_{i+1}=g_P(U_i)\subseteq U_i
 \ \Longrightarrow\ K_{i+1}=g_P(U_i)\subseteq g_P(K_{i+1})=U_{i+1}.
\]
Both sequences stabilize on the finite base, with
\[
 T_P=\bigcup_{i\in\Nat}K_i\quad\text{and}\quad F_P=\herbrand(P)\setminus\bigcap_{i\in\Nat}U_i.
\]

\paragraph*{Intervention Instances.}
\begin{definition}\label{def:instance}
An \emph{intervention instance} is a tuple
$I=(\PiP,\allowbreak\Dx,\allowbreak\Univ,\allowbreak E,\allowbreak q)$, where $\PiP$ is a finite safe normal Datalog
program over $\EDB\cup\IDB$,
$\Dx$ is a finite set of ground $\EDB$ atoms, $\Univ$ is a finite set of mutable ground $\EDB$ atoms disjoint from $\Dx$,
$E\subseteq\Univ$ is the observed state, consisting of the mutable facts present in the database,
and $q$ is a ground $\IDB$ atom. 
\end{definition}

For a syntactic object $X$, let $\Const(X)$ denote the set of constants
occurring in $X$. For an instance $I=(\PiP,\Dx,\Univ,E,q)$, let
$C_I=\Const(\PiP)\cup\Const(\Dx)\cup\Const(\Univ)\cup\Const(q)$. Let
$A_I=C_I$ if $C_I\neq\varnothing$, and let $A_I=\{a_0\}$ otherwise,
where $a_0$ is a fresh constant. We write $\Ground(\PiP,A_I)$ for the set
of all ground rule instances obtained by mapping the variables of $\PiP$ to
constants in $A_I$, and use Herbrand semantics over $A_I$.
The base contains all ground atoms over $\EDB\cup\IDB$ and $A_I$, including atoms occurring in no ground rule. When $I$ is
clear, for every database $D$ of ground $\EDB$-atoms over $A_I$,
$\PiP\cup D$ abbreviates $\Ground(\PiP,A_I)\cup D$.

Let $I=(\PiP,\Dx,\Univ,E,q)$ be an intervention instance for the rest of the section. A \emph{state} is a set $F\subseteq\Univ$. The database at $F$ is
$D_F=\Dx\cup F$: the facts in $\Dx$ are fixed, the facts in $F$ are
present, and all other ground $\EDB$-atoms over $A_I$ are false. Let $\Bthree=\{\True,\False,\Undef\}$. The response function $\Ans_I\colon 2^{\Univ}\longrightarrow\Bthree$ maps each state $F$ to the WFS value of $q$ under $\PiP$ on $D_F$, \ie $\Ans_I(F)=\WFM(\PiP\cup D_F)(q)$. For every $F\subseteq\Univ$, let $I[F]=(\PiP,\Dx,\Univ,F,q)$ denote the intervention instance with observed state $F$. For states $F,F'\subseteq\Univ$, let $F\toggle F'=(F\setminus F')\cup(F'\setminus F)$ denote their symmetric difference. Their \emph{Hamming distance} is $\Dist(F,F')=\lvert F\toggle F'\rvert$. An intervention $J\subseteq\Univ$ at $E$ produces the state $E\toggle J$ and the intervention instance $I[E\toggle J]$.

We omit the subscript $I$ from $\Ans_I$ when the instance is clear. The sets $\Dx$ and $\Univ$ play the roles of exogenous and endogenous facts in database causality~\cite{meliou2010complexity,bertossi2017causes}. For first-order queries on partial instances, Bogaerts et al. identify inclusion-minimal sets of signed facts whose joint reversal changes a true result under supervaluation~\cite[Definition~4.1]{BogaertsJakubowskiVandenBussche2024}. We prescribe both the source and the target WFS value for a designated fact.

\paragraph*{Causes and Minimum Interventions.}

Let $E$ be the observed state, $b=\Ans(E)$, $c\in\Bthree\setminus\{b\}$, and $\tau\in\Univ$. A \emph{$b\to c$ contingency} for $\tau$ at $E$ is a set $\Gamma\subseteq\Univ\setminus\{\tau\}$ such that, for $F=E\toggle\Gamma$,
$\Ans(F)=b$ and $\Ans(F\toggle\{\tau\})=c$. Since $\tau\notin\Gamma$, the designated toggle deletes $\tau$ if $\tau\in E$ and inserts $\tau$ otherwise. We take minima in $\Nat_{\Inf}=\Nat\cup\{\Inf\}$, with $n<\Inf$ and $n+\Inf=\Inf$ for $n\in\Nat$, and $\min\varnothing=\Inf$.

\begin{definition}
\label{def:typed-cause}
\label{eq:kappa}
The \emph{minimum $b\to c$ contingency size} of $\tau$ at $E$ is
\[
 \Kappa_E^{b\to c}(\tau)=\min\{|\Gamma| \mathrel{\mid} \Gamma\text{ is a $b\longrightarrow c$ contingency for }\tau\,\}.
\]
The corresponding responsibility is
\begin{equation}
\Resp_E^{b\longrightarrow c}(\tau)
=
\begin{cases}
\dfrac{1}{1+\Kappa_E^{b\longrightarrow c}(\tau)},
  & \text{if }\Kappa_E^{b\longrightarrow c}(\tau)<\Inf,\\[2mm]
0, & \text{otherwise}.
\end{cases}
\label{eq:responsibility}
\end{equation}
The fact $\tau$ is a \emph{$b\longrightarrow c$ actual cause} at $E$ if
$\Kappa_E^{b\longrightarrow c}(\tau)<\Inf$.
\end{definition}
Let $a_1,a_2,q$ be nullary, $\Univ=\{a_1,a_2\}$, and $\Dx=E=\varnothing$. For the rules $q\leftarrow\neg q$ and $q\leftarrow a_1,a_2$, inserting either fact alone leaves $q$ undefined. Inserting $a_1$ after $a_2$ makes $q$ true. Thus, $\{a_2\}$ is a minimum $\Undef\to\True$ contingency for $a_1$, whose responsibility is $1/2$.

If $\forall F\subseteq\Univ:\ \Ans(F)\in\{\True,\False\}$, then at any $F$ satisfying $\Ans(F)=\Ans(E)$ a toggle changes the answer exactly when it yields the other Boolean value. Definition~\ref{def:typed-cause} therefore gives the Boolean contingency condition for insertions and deletions in stratified programs~\cite{thapa2026causalexplanationsstratifieddatalog}, since stratified programs have total well-founded models~\cite[Theorem~6.1]{van1991well}. Suppose now that $\PiP$ is positive (all body literals are positive), $E=\Univ$, and $\Ans(E)=\True$. Let $\mathcal S$ consist of the inclusion-minimal $S\subseteq\Univ$ such that $\Ans(S)=\True$, the minimal endogenous supports relative to $\Dx$~\cite{bertossi2017causes,thapa2026causalityminimalsupportsrecursive}. For $\tau\in\Univ$ and $\Gamma\subseteq\Univ\setminus\{\tau\}$, deleting $\Gamma$ is a $\True\to\False$ contingency for $\tau$ exactly when
\[
\begin{aligned}
 &(\exists S\in\mathcal S:\ S\cap\Gamma=\varnothing)\wedge (\forall S\in\mathcal S:\ S\cap(\Gamma\cup\{\tau\})\ne\varnothing).
\end{aligned}
\]
Indeed, a positive query holds precisely when a support survives. If $\Dx$ alone entails the query, $\varnothing\in\mathcal S$ and the second conjunct fails.

\begin{definition}\label{def:radius}
Let $E\subseteq\Univ$ and $\Ans(E)=\Undef$. For
$c\in\{\True,\False\}$, a $c$-intervention is a set
$\Delta\subseteq\Univ$ such that $\Ans(E\toggle\Delta)=c$. 
The minimum numbers of changes that give value $c$ and that give either determinate value are, respectively,
\begin{enumerate}
 \item \label{eq:directional-radius} $\DirRad_c(E)=\min\{\,|\Delta|\mid\Delta\subseteq\Univ\ \wedge\
 \Ans(E\toggle\Delta)=c\,\}$, and
 \item \label{eq:radius} $\DetRad(E) =\min\{\DirRad_{\True}(E),\DirRad_{\False}(E)\}$.
\end{enumerate}
\end{definition}

\begin{figure}[htb]
\centering
{
\def\policyatom#1#2{\makebox[49pt]{$#1(a)$}\vrule width .4pt height 8.4pt depth 3.8pt\makebox[13pt]{$#2$}}%
\begin{tikzpicture}[x=1pt,y=1pt,font=\footnotesize,
  pol atom/.style={draw,line width=.45pt,inner sep=0pt,outer sep=.22pt},
  pol arrow/.style={->,>=stealth,line width=.55pt,shorten <=1.1pt,shorten >=1.1pt},
  pol heading/.style={anchor=west,inner sep=0pt,font=\sffamily\bfseries\footnotesize},
  pol score/.style={anchor=east,inner sep=0pt}]
\node[pol heading] at (0,63) {\textbf{(a)}\enspace Audit present};
\node[pol score] at (225,63) {$\Resp_{E'}^{\True\to\Undef}(\tau)=1$};
\node[pol atom] (polClearedA) at (32,45) {\policyatom{\Cleared}{\Undef}};
\node[pol atom] (polBlockedA) at (112.5,45) {\policyatom{\Blocked}{\Undef}};
\node[pol atom] (polApproveA) at (193,45) {\policyatom{\Approve}{\Undef}};
\draw[pol arrow] ([yshift=2.6pt]polClearedA.east) -- ([yshift=2.6pt]polBlockedA.west);
\draw[pol arrow] ([yshift=-2.6pt]polBlockedA.west) -- ([yshift=-2.6pt]polClearedA.east);
\draw[pol arrow] (polBlockedA.east) -- (polApproveA.west);
\node[pol heading] at (0,20) {\textbf{(b)}\enspace Audit deleted};
\node[pol score] at (225,20) {$\Resp_{E'}^{\True\to\False}(\tau)=1/2$};
\node[pol atom] (polClearedB) at (32,2) {\policyatom{\Cleared}{\False}};
\node[pol atom] (polBlockedB) at (112.5,2) {\policyatom{\Blocked}{\True}};
\node[pol atom] (polApproveB) at (193,2) {\policyatom{\Approve}{\False}};
\draw[pol arrow] (polClearedB.east) -- (polBlockedB.west);
\draw[pol arrow] (polBlockedB.east) -- (polApproveB.west);
\end{tikzpicture}}
\caption{ Let $q = \Approve(a)$, $\tau = \Clearance(a)$, and $E' = E \cup \{\tau\}$. Approval is true in both cases before clearance is withdrawn. The diagrams show the values after withdrawal. An arrow $P\to Q$ denotes the rule $Q\leftarrow\neg P$ after fixing the extensional facts in Example~\ref{ex:policy-program}. Deleting the audit removes the rule $\Cleared(a)\leftarrow\neg\Blocked(a)$.}
\label{fig:policy-contingency}
\Description{Each box gives a ground atom and its value after withdrawing clearance. With the audit present, cleared and blocked depend negatively on one another, and all three atoms are undefined. After deleting the audit, cleared is false, blocked is true, and approval is false. Approval was true before withdrawing clearance in either case. Responsibility is one for true to undefined and one half for true to false.}
\end{figure}
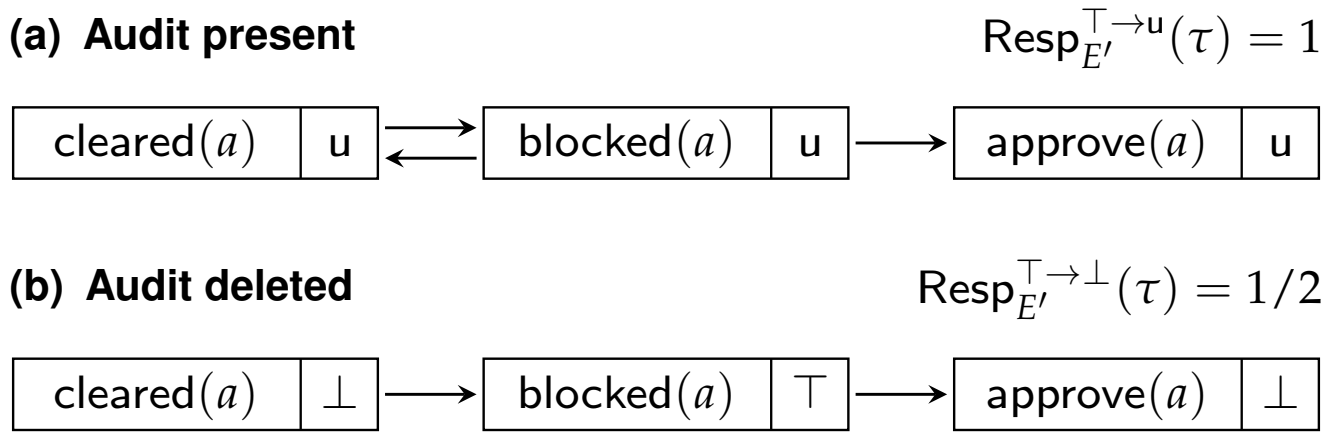

\begin{example}\label{ex:policy-program}
At $E\cup\{\Clearance(a)\}$ in Example~\ref{example_intro_1}, Figure~\ref{fig:policy-contingency} compares the value of $\Blocked(a)$ after deleting $\Clearance(a)$, with and without $\Audit(a)$.
\end{example}

If $\Ans(E)=\Undef$, then $\DetRad(E)=1+\min_{\tau\in\Univ,\,c\in\{\True,\False\}}\Kappa_E^{\Undef\to c}(\tau)$. A minimum contingency followed by the designated toggle reaches a determinate state. Conversely, removing any toggle from a smallest intervention yielding a determinate answer leaves the query undefined, giving a contingency for the removed fact. If no intervention resolves the query, both sides equal $\Inf$.

The equality need not hold for a prescribed target. Take nullary predicates $a,b,q$, mutable set $\Univ=\{a,b\}$, $\Dx=E=\varnothing$, query $q$, and the rules
\begin{equation}\label{eq:directional-counterexample}
\begin{aligned}
q&\leftarrow a,b,\\
q&\leftarrow \neg a,\neg b,\neg q.
\end{aligned}
\end{equation}
At $E$, only $q\leftarrow\neg q$ remains after evaluating $a,b$, so $q$ is undefined. At either singleton both bodies are false, whereas $\{a,b\}$ derives $q$. Thus $\DirRad_{\True}(E)=2$. No nonempty contingency preserves undefinedness and neither single insertion produces truth, so both $\Undef\to\True$ contingency sizes are $\Inf$.

\section{Game Normal Form}
\label{sec:causal-normal-form}
The classical win--move construction represents a specified answer relation of a Datalog query under WFS~\cite{galanaki2008infinite,flum1997total,flum2000games}. We require a single game position that distinguishes all three answers under each change to the input facts.

\begin{example}\label{ex:true-answer-equivalence}
Let $a,q$ be nullary predicates, $\Univ=\{a\}$ and $\Dx=E=\varnothing$. Consider
$P_1=\{q\leftarrow\neg q;\ q\leftarrow a\}$ and $
P_2=\{q\leftarrow a\}$. Both programs make $q$ true exactly when $a$ is present, but at the empty state $P_1$ assigns $\Undef$ to $q$, whereas $P_2$ assigns $\False$. Thus, $a$ is an $\Undef\longrightarrow\True$ cause only under $P_1$; causal preservation requires the complete three-valued response.
\end{example}

Let $I=(\PiP,\Dx,\Univ,E,q)$ and $J=(\PiP_{\Win},D^{\mathsf g},\Univ',E',\Win(s))$
be intervention instances. For a bijection
$\eta\colon\Univ\longrightarrow\Univ'$ and $F\subseteq\Univ$, let
$\eta[F]=\{\eta(\tau)\mid\tau\in F\}$.
\begin{definition}\label{def:response-isometry}
The bijection $\eta$ preserves the observation and the query values if
\[
E'=\eta[E]\ \wedge\ \bigl(\forall F\subseteq\Univ:\ \Ans_I(F)=\Ans_J(\eta[F])\bigr).
\]
\end{definition}

Any bijection of mutable facts preserves Hamming distances because, for all
$F,F'\subseteq\Univ$, $\eta[F\toggle F']
  =\eta[F]\toggle\eta[F'] $ and $ \bigl|\eta[F]\toggle\eta[F']\bigr| =\lvert F\toggle F'\rvert$.

\paragraph*{Game Positions and Moves.}
We identify every move atom $\Move(u,v)$ with the pair $(u,v)$. For the target instance $J$, let $\Mx=\{(u,v)\mid\Move(u,v)\in D^{\mathsf g}\}$, $\Mu=\{(u,v)\mid\Move(u,v)\in\Univ'\}$, and $\Pot=\Mx\cup\Mu$. Let $V$ be the finite vertex set, including the query position, all terminal positions, and both endpoints of each possible move. For a directed graph $A=(V,M)$, let $\Succ_A(v)=\{w\in V\mid(v,w)\in M\}$ and let
\[
\Val_A(v)=\WFM(\Ground(\PiP_{\Win},V)\cup M)(\Win(v)).
\]
Grounding over $V$ includes the isolated vertices. For a state $F\subseteq\Mu$, let $M_F=\Mx\cup F$. We abbreviate $\Succ_{(V,M_F)}$ and $\Val_{(V,M_F)}$ by $\Succ_F$ and $\Val_F$, respectively. We omit the subscript when the graph is understood. For a query position $q\in V$, the query is $\Win(q)$. A position is winning, losing, or drawn when the value is $\True$, $\False$, or $\Undef$, respectively.

In the game constructed from $I$, every mutable move has a terminal target and is the only possible move from its source:
\[
\forall(p,z)\in\Mu:\quad
  \Succ_{(V,\Pot)}(p)=\{z\}\ \wedge\
  \Succ_{(V,\Pot)}(z)=\varnothing.
\]
For each $\tau\in\Univ$, let $p_\tau$ denote its input position and $\eta(\tau)$ the mutable move from $p_\tau$ to a terminal. All occurrences of $\tau$ use $p_\tau$. For $F\subseteq\Univ$, let $G_F=(V,\Mx\cup\eta[F])$ be the game at state $F$.

\paragraph*{Truth and Falsity Indicators.}
A Boolean query is total if it has a value in $\{\True,\False\}$ at every state. Apply the total-game normal form to the $\True$ and $\False$ answer relations of the source program~\cite[Cor.~15, Thm.~16]{flum2000games}. Let $t,f$ be their distinguished positions. Both are determinate at every state, and for $F\subseteq\Univ$,
\[
\begin{aligned}
\Val_{G_F}(t)=\True&\iff\Ans_I(F)=\True,\\
\Val_{G_F}(f)=\True&\iff\Ans_I(F)=\False.
\end{aligned}
\]

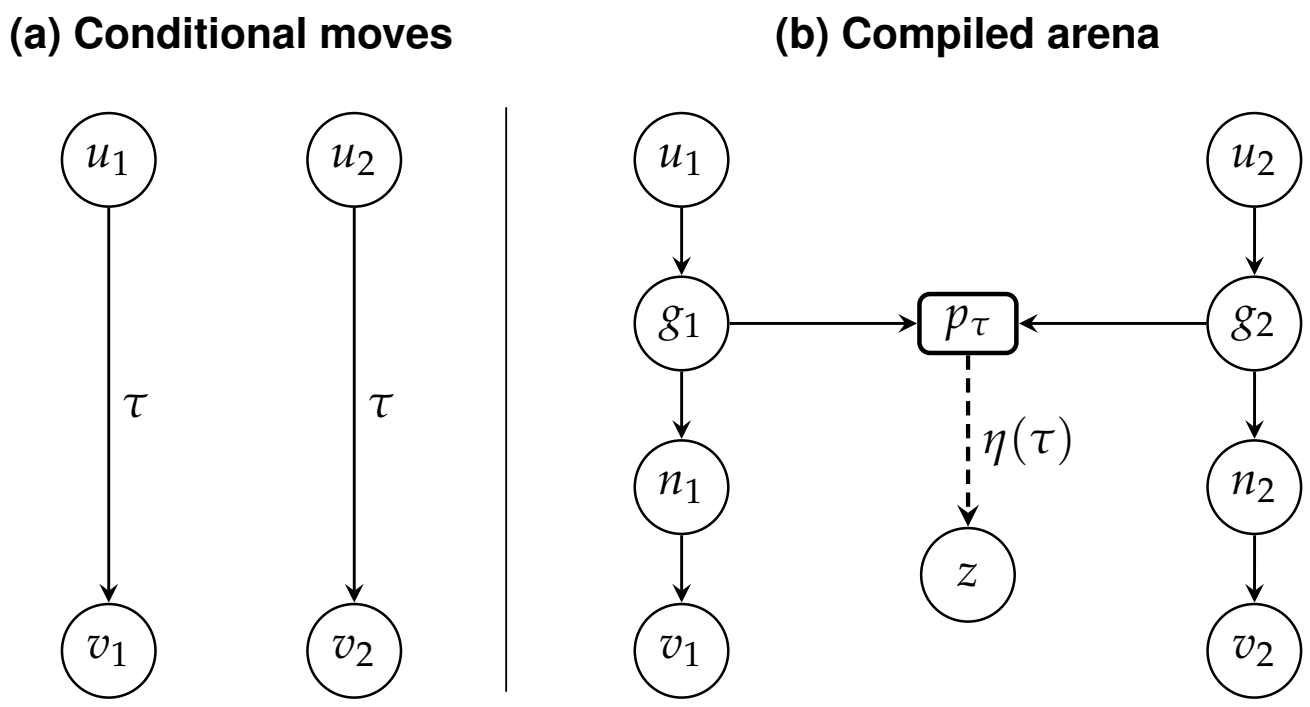
\begin{figure}[tb]
\centering
\begin{tikzpicture}[
  x=1pt,y=1pt,
  vertex/.style={circle,draw,inner sep=0pt,minimum size=16pt,
                 line width=.48pt,font=\footnotesize},
  input/.style={rectangle,rounded corners=2pt,draw,inner xsep=4pt,
                inner ysep=2pt,line width=.85pt,font=\footnotesize},
  move/.style={->,>=stealth,line width=.55pt},
  mutable/.style={move,dash pattern=on 2.7pt off 1.6pt,line width=.8pt},
  heading/.style={font=\sffamily\bfseries\footnotesize,inner sep=0pt},
  edgelabel/.style={font=\footnotesize,inner sep=2pt}]
\node[heading] at (33,111) {(a) Conditional moves};
\node[heading] at (159,111) {(b) Compiled arena};
\draw[line width=.35pt] (80,-1) -- (80,99);
\node[vertex] (su1) at (12,90) {$u_1$};
\node[vertex] (su2) at (54,90) {$u_2$};
\node[vertex] (sv1) at (12,6) {$v_1$};
\node[vertex] (sv2) at (54,6) {$v_2$};
\draw[move] (su1) -- node[edgelabel,right] {$\tau$} (sv1);
\draw[move] (su2) -- node[edgelabel,right] {$\tau$} (sv2);
\node[vertex] (u1) at (110,90) {$u_1$};
\node[vertex] (u2) at (208,90) {$u_2$};
\node[vertex] (g1) at (110,62) {$g_1$};
\node[vertex] (g2) at (208,62) {$g_2$};
\node[vertex] (n1) at (110,34) {$n_1$};
\node[vertex] (n2) at (208,34) {$n_2$};
\node[vertex] (v1) at (110,6) {$v_1$};
\node[vertex] (v2) at (208,6) {$v_2$};
\node[input] (ptau) at (159,62) {$p_\tau$};
\node[vertex] (z) at (159,19) {$z$};
\draw[move] (u1) -- (g1);
\draw[move] (g1) -- (n1);
\draw[move] (n1) -- (v1);
\draw[move] (u2) -- (g2);
\draw[move] (g2) -- (n2);
\draw[move] (n2) -- (v2);
\draw[move] (g1) -- (ptau);
\draw[move] (g2) -- (ptau);
\draw[mutable] (ptau) -- node[edgelabel,right] {$\eta(\tau)$} (z);
\end{tikzpicture}
\caption{Two moves with the same condition $\tau$. Both gates read $p_\tau$. If the dashed move to terminal $z$ is present, $\Val(g_i)=\Val(v_i)$ for $i\in\{1,2\}$. If it is absent, both gates are winning.}
\Description{The two source moves have the same condition. In the compiled arena the paths from u1 through g1 and n1 to v1 and from u2 through g2 and n2 to v2 flank the input position. Both gates have a fixed move to this position, whose sole possible move to a terminal is mutable.}
\label{fig:shared-guard}
\end{figure}

The move conditions are defined by nonrecursive semipositive programs. Their intensional predicates admit a dependency order, and only extensional atoms may occur negated. The source variables range over $A_I$. The game construction uses $\widehat A_I=A_I\cup\{a_{\mathrm{aux}}\}$, with $a_{\mathrm{aux}}$ fresh. Let two fresh constant symbols denote a chosen element of $A_I$ and $a_{\mathrm{aux}}$, respectively. Unary extensional relations distinguish $A_I$ and $\widehat A_I$. Grounding the move rules gives one circuit for all states.
Let $e=(u,v)$ be a possible move with Boolean condition $\varphi$ and circuit output position $c_\varphi$. Replace $e$ by a move from $u$ to a fresh position $g_e$ with
\begin{equation}
\Val(g_e)=
\begin{cases}
\Val(v), & \Val(c_\varphi)=\True,\\
\True,   & \Val(c_\varphi)=\False.
\end{cases}
\label{eq:conditional-move-gate}
\end{equation}
A false condition therefore gives the opponent a win if the move is chosen. In Figure~\ref{fig:shared-guard}, both moves have condition $\tau$ and both use $p_\tau$.

\paragraph*{Query Position.}
A fixed five-position component reconstructs the WFS value from $t$ and
$f$. Let $\bar t$ have the sole move to $t$, let $\bar f$ have the
sole move to $f$, let $d$ have a self-loop, let $h$ move to $d$ and
$\bar f$, and let $s$ move to $\bar t$ and $h$. The following table gives the values at the five indicated positions.
\begin{equation}
\setlength{\arraycolsep}{4.5pt}
\renewcommand{\arraystretch}{1.10}
\begin{array}{|c|cc|cc|c|}\hline
\Ans_I(F) & t & f & \bar t & h & s\\ \hline
\True  & \True  & \False & \False & \Undef & \True\\
\False & \False & \True  & \True  & \True  & \False\\
\Undef & \False & \False & \True  & \Undef & \Undef\\ \hline
\end{array}
\label{eq:three-value-combiner}
\end{equation}
Since $d$ has only a self-loop, it is drawn. Thus $h$ wins exactly when $\bar f$ loses, and is drawn otherwise. The two successors $\bar t,h$ of $s$ then give $\Val_{G_F}(s)=\Ans_I(F)$.

\paragraph*{Compilation algorithm.}
Given $I=(\PiP,\Dx,\Univ,E,q)$, we construct
$\NF(I)=(\PiP_{\Win},D^{\mathsf g},\Univ^{\mathsf g},E^{\mathsf g},\Win(s))$ as follows.
\begin{enumerate}
\item Construct the two total games for $q$ on the auxiliary domain and take their disjoint union, with distinguished positions $t$ and $f$.
\item Ground the move-defining rules over $\widehat A_I$. For each ground head, take the disjunction of the conjunctions of body literals in the defining ground rules. A single successor implements negation; edges to the complements of the inputs implement disjunction, and De Morgan's law gives conjunction. All occurrences of the translated atom for $\tau\in\Univ$ read the same input position; this includes the unary representation of a nullary fact. The translated inputs for $\Dx$ are true. Those for original input atoms outside $\Dx\cup\Univ$ are false. The two domain predicates have the fixed extensions $A_I$ and $\widehat A_I$.
\item Give each $\tau\in\Univ$ a distinct input position $p_\tau$ and optional edge $\eta(\tau)$ to a terminal, including facts unused by the circuits. For each conditional move $u\to v$ with output $c_\varphi$, introduce fresh $g,n$ and the fixed moves $u\to g$, $g\to c_\varphi$, $g\to n$, $n\to v$. Add $\bar t,\bar f,d,h,s$ and their fixed moves as in the construction of the query position. Let $D^{\mathsf g}$ contain all fixed moves and let $\Univ^{\mathsf g}=\eta[\Univ]$, $E^{\mathsf g}=\eta[E]$.
\end{enumerate}
\begin{theorem}\label{thm:causal-normal-form}
For every fixed safe normal Datalog program $\PiP$, the construction maps each intervention instance $I=(\PiP,\Dx,\Univ,E,q)$ to $\NF(I)$ in polynomial time, and the bijection $\eta\colon\Univ\longrightarrow\Univ^{\mathsf g}$ preserves the observation and the query values in the sense of Definition~\ref{def:response-isometry}. For every $\tau\in\Univ$, $\eta(\tau)$ is the only mutable move controlled by $\tau$, and no auxiliary move is mutable.
\end{theorem}

\begin{proof}[Proof sketch]
Let $F\subseteq\Univ$. The only possible move from $p_\tau$ leads to a terminal, so $p_\tau$ wins exactly when $\tau\in F$. Induction on the Boolean circuit gates gives the value of each move condition. For an enabled move $e=(u,v)$, $g_e$ has the value of $v$; for a disabled move, $g_e$ wins. Expanding and contracting finite derivations of wins and losses therefore preserves both values at each original position (Lemma~\ref{lem:conditional-edge-expansion}). The remaining positions are drawn. Equation~\eqref{eq:three-value-combiner} gives $\Val_{G_F}(s)=\Ans_I(F)$. The input moves are distinct and $E^{\mathsf g}=\eta[E]$. For fixed $\PiP$, the move rules have bounded arity and a bounded number of variables. Enumerating their ground instances and adding a constant number of positions per gate, wire, and possible move takes polynomial time.
\end{proof}

Let $F\subseteq\Univ$, let $\Ans_I(F)=b$, choose
$c\in\Bthree\setminus\{b\}$, and let $\tau\in\Univ$. The map
$\Gamma\longmapsto\eta[\Gamma]$ is a cardinality-preserving bijection between the $b\longrightarrow c$ contingencies
for $\tau$ in $I[F]$ and those for $\eta(\tau)$ in
$\NF(I)[\eta[F]]$. For $b=\Undef$, the bijection also maps the interventions reaching $c$ bijectively onto those reaching $c$ in the game. Hence, minimum sizes and responsibilities agree, and the bijection maps minimizers onto minimizers.

\begin{example}\label{ex:running-guarded-arena}
In Example~\ref{example_intro_2}, inserting $\Move(s,\ell)$ or deleting $\Move(s,a)$ resolves the query in one move. Both moves leave $s$. In the compiled arena, the corresponding mutable edges leave distinct input positions.
\end{example}

\begin{example}\label{thm:guard-necessity}
Let $m\geq1$. Take nullary intensional predicates $q,b_1,\ldots,b_m$ and a nullary input $a$. The program has rules
\[
 q\leftarrow\neg q,\qquad q\leftarrow b_1,\ldots,b_m\quad\text{and}\quad
 b_i\leftarrow a\quad(1\le i\le m).
\]
Replace the occurrence of $a$ in rule $b_i\leftarrow a$ by a distinct input $a_i$. Initially all inputs are absent. Encode absence of $a$ by $\varnothing$ and presence of $a$ by $\{a_1,\ldots,a_m\}$. The responses agree at both encoded states, but changing $a$ changes all $m$ copies. If the copies are independently mutable, inserting $a_i$ changes $q$ from $\Undef$ to $\True$ exactly after the other $m-1$ copies have been inserted. The minimum contingency size of $a_i$ is $m-1$, giving responsibility $1/m$. In the source, inserting $a$ alone realizes the same transition. For $m>1$, the two encodings therefore give different Hamming distances and responsibilities.
\end{example}

\section{Certificates for game positions}\label{sec:certificates}
In this section, the arena is any finite directed graph with disjoint sets of fixed and mutable moves, state $F\subseteq\Mu$, and query position $q\in V$.
Thus, $M_F=\Mx\cup F$ and $\Ans(F)=\Val_F(q)$. We compute the winning and losing positions by retrograde analysis~\cite[Alg.~1]{kohler2013first}. For $i\in\Nat$, the stages are
\begin{enumerate}
  \item \label{eq:stage-zero} $\Wset_0=\varnothing,
 \quad \Lset_0=\{v\in V\mid\Succ_F(v)=\varnothing\}$,
  \item \label{eq:stages} $\Wset_{i+1}=\Wset_i\cup\{v\in V\mid\Succ_F(v)\cap\Lset_i\neq\varnothing\},\quad \Lset_{i+1}=\Lset_i\cup\{v\in V\mid\Succ_F(v)\subseteq\Wset_i\}$.
\end{enumerate}
Let $\Wset_F=\bigcup_{i\in\Nat}\Wset_i$,
$\Lset_F=\bigcup_{i\in\Nat}\Lset_i$ and $\Dset_F=V\setminus(\Wset_F\cup\Lset_F)$.
The rank of $v\in\Wset_F\cup\Lset_F$ is its first stage,
$\rho_F(v)=\min\{i\in\Nat\mid v\in\Wset_i\cup\Lset_i\}$.
We call $\rho_F$ the canonical rank and $(\Wset_F,\Lset_F,\Dset_F)$ the canonical partition.

\begin{definition}\label{def:rank-certificate}
Let $\Rank=(\Wp,\Wm,\rhoF)$, where $\Wp,\Wm\subseteq V$ are disjoint and $\rhoF:\Wp\cup\Wm\to\Nat$. The rank conditions at $F$ are
\begin{enumerate}
\item \label{eq:rank-win}
$\forall v\in\Wp\;\exists w\in\Succ_F(v)\cap\Wm:\ 
\rhoF(w)<\rhoF(v)$;
\item \label{eq:rank-lose}
$\forall v\in\Wm,\quad
\Succ_F(v)\subseteq
\{w\in\Wp \mid \rhoF(w)<\rhoF(v)\}$.
\end{enumerate}
A triple satisfying both conditions certifies a win at $q$ if $q\in\Wp$, and a loss at $q$ if $q\in\Wm$.
\end{definition}

\begin{definition}\label{def:draw-certificate}
A certificate for a draw at $q$ in state $F$ is
$\Draw=(Z,\Rank)$, where
$\Rank=(\Wp,\Wm,\rhoF)$ satisfies the rank conditions at $F$,
$q\in Z\subseteq V$ and $Z\cap(\Wp\cup\Wm)=\varnothing$,
and
\begin{enumerate}
\item \label{eq:draw-recurrent}
$\forall z\in Z:\ \Succ_F(z)\cap Z\neq\varnothing$,
\item \label{eq:draw-boundary}
$\forall z\in Z:\ \Succ_F(z)\setminus Z\subseteq\Wp.$
\end{enumerate}
\end{definition}

\begin{theorem}\label{thm:certificate-completeness}
For every state $F$ of a finite arena and every $q\in V$,
\begin{enumerate}
 \item \label{eq:true-certificate} $\Val_F(q)=\True \Longleftrightarrow
 q$ has a certificate for a win at $F$,
 \item \label{eq:false-certificate} $\Val_F(q)=\False \Longleftrightarrow
 q$ has a certificate for a loss at $F$,
 \item \label{eq:undefined-certificate} $\Val_F(q)=\Undef \Longleftrightarrow
 q$ has a certificate for a draw at $F$.
\end{enumerate}

\end{theorem}
\begin{proof}[Proof sketch]
A winning position has a lower-ranked losing successor. Every move from a losing position leads to a lower-ranked winning successor. By choosing a lower-ranked losing successor on each turn, the winning player forces the opponent to reach a terminal position. In $\Dset_F$, either player can stay in the set until the opponent leaves it for a winning position. Conversely, induction on the rank shows that a certificate puts $\Wp$ in $\Wset_F$ and $\Wm$ in $\Lset_F$. The two conditions on $Z$ then let either player avoid a loss from $q\in Z$. Appendix~\ref{app:certificates} derives the three regions from the alternating fixpoint.
\end{proof}
\begin{corollary}\label{cor:canonical-partition}\label{eq:canonical-partition}
For every state $F$ of a finite arena,
\begin{enumerate}
 \item $\Wset_F=\{v\in V\mid\Val_F(v)=\True\},$
 \item $\Lset_F=\{v\in V\mid\Val_F(v)=\False\}$ and
 \item $\Dset_F=\{v\in V\mid\Val_F(v)=\Undef\}.$
\end{enumerate}

The set $\Dset_F$ is the greatest $Z\subseteq V$ satisfying
$\forall z\in Z:\ (\Succ_F(z)\cap Z\ne\varnothing)\wedge(\Succ_F(z)\setminus Z\subseteq\Wset_F)$.

For every $v\in\Wset_F\cup\Lset_F$, the canonical rank satisfies
$\rho_F(v)\leq |V|-1$.
\end{corollary}

Appendix~\ref{app:certificates} proves the corollary by relating the stages to the alternating fixpoint.

\subsection{Conjunctions of literals}

A Boolean term over $\Mu$ is a conjunction of positive and negative literals on distinct mutable moves. We represent it by the pair $\Req=(\Req^+,\Req^-)$ of disjoint sets of moves occurring positively and negatively. The empty conjunction is true.
For $F,E\subseteq\Mu$, satisfaction and cost are given by
\begin{enumerate}
    \item $ F\models\Req\ \Longleftrightarrow\ (\Req^+\subseteq F)\wedge(\Req^-\cap F=\varnothing),$
    \item $\Cost_E(\Req)=|\Req^+\setminus E|+|\Req^-\cap E|.$
\end{enumerate}

We write $\Req\preceq\Req'$ if $(\Req^+\subseteq\Req'^+)\wedge(\Req^-\subseteq\Req'^-)$, and $\Req\prec\Req'$ if $(\Req\preceq\Req')\wedge(\Req\ne\Req')$. Thus $F\models\Req'$ implies $F\models\Req$ whenever $\Req\preceq\Req'$.

To specify a certificate independently of the state, choose a tuple $C=(\Wp,\Wm,\rhoF,\omega)$ with disjoint
$\Wp,\Wm\subseteq V$, $\rhoF:\Wp\cup\Wm\to\{0,\ldots,|V|-1\}$, and
$\omega:\Wp\to\Wm$ satisfying
\begin{equation}
\forall v\in\Wp:\quad
(v,\omega(v))\in\Pot \;\wedge\;
\rhoF(\omega(v))<\rhoF(v),
\label{eq:rank-requirement-witness}
\end{equation}
\begin{equation}
\Mx\cap(\Wm\times V)
\subseteq\{(v,w)\in\Wm\times\Wp\mid \rhoF(w)<\rhoF(v)\}.
\label{eq:rank-requirement-fixed}
\end{equation}
The tuple $C$ determines the term $\ReqOf(C)=(\Req_C^+,\Req_C^-)$ with
\begin{enumerate}
 \item \label{eq:rank-positive-requirement} $\Req_C^{+}=\{(v,\omega(v))\mid v\in\Wp\ \wedge\ (v,\omega(v))\in\Mu\},$
 \item \label{eq:rank-negative-requirement} $\Req_C^{-}=(\Mu\cap(\Wm\times V))\setminus
      \{(v,w)\in\Wm\times\Wp\mid\rhoF(w)<\rhoF(v)\}.$
\end{enumerate}
For a draw, take $D=(Z,C,\zeta)$, where $C$ satisfies equations \eqref{eq:rank-requirement-witness}--\eqref{eq:rank-requirement-fixed},
$q\in Z\subseteq V\setminus(\Wp\cup\Wm)$, $\zeta:Z\to Z$, and
\begin{equation}
 \forall z\in Z:\ (z,\zeta(z))\in\Pot,
 \qquad
 \Mx\cap\bigl(Z\times(V\setminus Z)\bigr)\subseteq Z\times\Wp .
\label{eq:draw-requirement-fixed}
\end{equation}
The term $\ReqOf(D)=(\Req_D^+,\Req_D^-)$ is given by
\begin{enumerate}
 \item $\Req_D^+=\Req_C^+\cup\{(z,\zeta(z))\mid z\in Z\ \wedge\ (z,\zeta(z))\in\Mu\},$
 \item \label{eq:draw-negative-requirement} $\Req_D^-=\Req_C^-\cup
 \{(z,w)\in\Mu\mid z\in Z\ \mathbin{\wedge}\ w\notin Z\cup\Wp\}.$
\end{enumerate}
The positive and negative moves of $C$ have disjoint sources $\Wp$ and $\Wm$. The additional moves of $D$ start in $Z$ and have disjoint target sets $Z$ and $V\setminus(Z\cup\Wp)$. Thus, each term is consistent.

Let $\Certs_{\True}(q)$ and $\Certs_{\False}(q)$ consist of the tuples $C$ with $q\in\Wp$ and $q\in\Wm$, respectively, and $\Certs_{\Undef}(q)$ of the tuples $D$ satisfying~\eqref{eq:draw-requirement-fixed} and the stated conditions on $Z,C,\zeta$. For $b\in\Bthree$, let
\[
\begin{aligned}
 \RawReqs_b(q)&=\{\ReqOf(C)\mid C\in\Certs_b(q)\},\\
 \Reqs_b(q)&=\{\Req\in\RawReqs_b(q)\mid
    \nexists\Req'\in\RawReqs_b(q):\ \Req'\prec\Req\}.
\end{aligned}
\]
Both families are finite since $\Mu$ is finite.

\begin{theorem}\label{thm:structured-completeness}
$\forall F\subseteq\Mu\ \forall b\in\Bthree:\,\,
(\Ans(F)=b)\ \Longleftrightarrow\ 
(\exists\Req\in\Reqs_b(q): F\models\Req)$.
\end{theorem}

\begin{proof}[Proof sketch]
For $\Ans(F)=b$, take the canonical sets and ranks. Choose a lower-ranked losing successor at each winning position. If $b=\Undef$, choose an internal successor at each position of $Z=\Dset_F$. These choices give a term satisfied at $F$ and hence a minimal certificate term obtained by deleting zero or more literals. Conversely, a state satisfying the term contains the selected moves and excludes the forbidden ones. The resulting certificate gives $\Ans(F)=b$ by Theorem~\ref{thm:certificate-completeness} (Appendix~\ref{app:requirement-normal-form}).
\end{proof}

\subsection{Minimum contingencies}
For a term $\Req$ and $\tau\in\Mu$, we define $\Flip_\tau(\Req)$ by reversing the sign of $\tau$:
\begin{enumerate}
 \item $\Flip_\tau(\Req)^+
 =(\Req^+\setminus\{\tau\})\cup(\Req^-\cap\{\tau\})$,
 \item \label{eq:requirement-flip} $\Flip_\tau(\Req)^-
 =(\Req^-\setminus\{\tau\})\cup(\Req^+\cap\{\tau\})$.
\end{enumerate}
By Lemma~\ref{lem:requirement-flip}, $F\models\Flip_\tau(\Req)\iff F\toggle\{\tau\}\models\Req$.
Two terms $\Req_1,\Req_2$ are compatible if
$(\Req_1^+\cup\Req_2^+)\cap(\Req_1^-\cup\Req_2^-)=\varnothing$.
For compatible terms, their conjunction is
$\Req_1\ReqJoin\Req_2=(\Req_1^+\cup\Req_2^+,\Req_1^-\cup\Req_2^-)$.
A term $\Req$ agrees with $E$ at $\tau$ if
\[
 (\tau\in\Req^+\Rightarrow\tau\in E)\ \wedge (\tau\in\Req^-\Rightarrow\tau\notin E).
\]
For distinct $b,c\in\Bthree$, let $\Pairs_{E,\tau}^{b\to c}$ consist of pairs
$(\Req_b,\Req_c)\in\Reqs_b(q)\times\Reqs_c(q)$ such that $\Req_b$ and $\Flip_\tau(\Req_c)$ are compatible and their conjunction agrees with $E$ at $\tau$. Call such pairs admissible.

\begin{theorem}\label{thm:requirement-normal-form}
 Let $E\subseteq\Mu$ be the observed state, $b=\Ans(E)$,
$c\in\Bthree\setminus\{b\}$, and $\tau\in\Mu$. Then,
\begin{equation}
 \Kappa_E^{b\longrightarrow c}(\tau)
 =\min_{(\Req_b,\Req_c)\in\Pairs_{E,\tau}^{b\longrightarrow c}}
   \Cost_E\bigl(\Req_b\ReqJoin\Flip_\tau(\Req_c)\bigr).
\label{eq:requirement-factorization}
\end{equation}
The minimum contingencies are precisely the sets
$(\Req^+\setminus E)\cup(\Req^-\cap E)$, where
$\Req=\Req_b\ReqJoin\Flip_\tau(\Req_c)$ and $(\Req_b,\Req_c)$ attains the minimum in equation~\eqref{eq:requirement-factorization}.
\end{theorem}
\begin{proof}[Proof sketch]
A contingency $\Gamma$ gives an admissible pair by applying Theorem~\ref{thm:structured-completeness} at $E\toggle\Gamma$ and $E\toggle\Gamma\toggle\{\tau\}$. For the resulting term $\Req$, the set $D=(\Req^+\setminus E)\cup(\Req^-\cap E)$ is contained in $\Gamma$. Conversely, for any admissible pair, $E\toggle D$ satisfies $\Req$ and $\tau\notin D$. Theorem~\ref{thm:structured-completeness} gives the two prescribed answers, so $D$ is a contingency of size $\Cost_E(\Req)$. For minimum $\Gamma$, containment forces $D=\Gamma$ (Appendix~\ref{app:requirement-normal-form}).
\end{proof}
For a mutable input fact $t\in\Univ$ in the original instance with observation $E_0$, apply the formula with $\tau=\eta(t)$ and $E=\eta[E_0]$. If $\Req=\Req_b\ReqJoin\Flip_\tau(\Req_c)$ is obtained from a minimizing pair, then the ordered pair of insertion and deletion sets in the original database is
\[
 \bigl(\eta^{-1}[\Req^+\setminus E],\ \eta^{-1}[\Req^-\cap E]\bigr)
\]
by Theorem~\ref{thm:causal-normal-form}.

\begin{example}\label{ex:running-requirements}
In the arena of Example~\ref{example_intro_2}, the fixed moves are
$a\to s$ and $b\to b$, and $\ell$ is terminal. Table~\ref{tab:contingency-endpoints}
lists the three mutable moves. At the observation $E'=\{(s,a),(s,b)\}$,
we ask when deleting $\tau=(s,a)$ makes $\Win(s)$ false.

For the source term $\Req_{\Undef}$, take the draw region $Z=\{s,a,b\}$
with witnesses $s\to a$, $a\to s$, and $b\to b$ and the empty triple $(\varnothing,\varnothing,\varnothing)$.
The internal witness requires $\tau$ and the boundary forbids $(s,\ell)$.
Both literals are necessary: the empty state loses at $s$, and
$\{\tau,(s,\ell)\}$ wins. For the target term $\Req_{\False}$,
take $s$ as a losing vertex of rank zero. The term forbids all three moves.
Each is necessary since either $(s,a)$ or $(s,b)$ alone gives a draw,
and $(s,\ell)$ alone gives a win.

Flipping the sign of $\tau$ in $\Req_{\False}$ makes the two terms
compatible. Their conjunction $\Req=\Req_{\Undef}\ReqJoin\Flip_\tau(\Req_{\False})$
agrees with $E'$ at $\tau$. Only the row for $(s,b)$ disagrees with $E'$.
Deleting that move gives the state $F=\{\tau\}$, where the cycle through
$a$ preserves the draw. Deleting $\tau$ then leaves $s$ terminal.
Conversely, any contingency must delete $(s,b)$: if that move remains,
the player at $s$ can choose $b$, where the self-loop forces a draw. A contingency
must also leave $(s,\ell)$ absent, since a move to $\ell$ wins.
Thus $\{(s,b)\}$ is the unique contingency and
$\Kappa_{E'}^{\Undef\to\False}(\tau)=1$.

At the original observation $E_{\mathsf g}=\{\tau\}$ the same conjunction has
cost zero. Inserting $(s,\ell)$ there also gives a transition with empty
contingency, from $\Undef$ to $\True$, witnessed by the move from $s$ of
rank one to $\ell$ of rank zero.
\end{example}

\begin{table}[htb]
\centering
\caption{A minimum contingency for deleting $\tau=(s,a)$.
In a term, $+$ means required, $-$ forbidden, and $*$ unrestricted.
In a state, $1$ means present and $0$ absent. A bold zero marks the move deleted in each step.}
\label{tab:contingency-endpoints}
{\small
\setlength{\tabcolsep}{4pt}
\renewcommand{\arraystretch}{1.13}
\begin{tabular}{|l|ccc|ccc|}
\hline
\raisebox{-.78\ht\strutbox}[0pt][0pt]{Move} & \multicolumn{3}{c|}{Terms} & \multicolumn{3}{c|}{State}\\
\cline{2-7}
& $\Req_{\Undef}$ & $\Req_{\False}$ & $\Req$ & $E'$ & $F$ & $F\toggle\{\tau\}$ \\
\hline
$(s,a)=\tau$ & $+$ & $-$ & $+$ & $1$ & $1$ & $\mathbf{0}$ \\
$(s,b)$      & $*$ & $-$ & $-$ & $1$ & $\mathbf{0}$ & $0$ \\
$(s,\ell)$   & $-$ & $-$ & $-$ & $0$ & $0$ & $0$ \\
\hline
$\Win(s)$ & \multicolumn{3}{c|}{} & $\Undef$ & $\Undef$ & $\False$ \\
\hline
\end{tabular}}
\end{table}

\subsection{Propagation from the updated move}
We next describe how a single updated move determines the new value of the query vertex.
Let
$G=(V,M)$ and $G'=(V,M')$ satisfy $M\toggle M'=\{\tau\}$, where $\tau=(x,y)$.  Let $(\Wset,\Lset,\Dset)$ and
$(\Wset',\Lset',\Dset')$ denote the respective canonical partitions, let
$\rho'$ be the canonical rank on $\Wset'\cup\Lset'$, and assume
$q\in\Dset\cap(\Wset'\cup\Lset')$.  We use
$\Delta\Wset=\Wset'\setminus\Wset$ and $\Delta\Lset=\Lset'\setminus\Lset$.

For $S\subseteq V$ and $x\in S$, a spanning in-arborescence of $S$ rooted at $x$ is a directed spanning tree in which every vertex of $S$ has a unique directed path to $x$.

\begin{theorem}\label{thm:resolution-tree}
Let $S=\Delta\Wset\cup\Delta\Lset$. There exists a parent map
$\Parent:S\setminus\{x\}\to S$ such that, for every $v\in S\setminus\{x\}$,
$(v,\Parent(v))\in M'\ \wedge\ \rho'(\Parent(v))<\rho'(v)$. The parent map satisfies
\begin{enumerate}
    \item $\Parent(v)\in\Delta\Lset$ for $v\in\Delta\Wset\setminus\{x\}$, and
    \item $\Parent(v)\in\Delta\Wset$ for $v\in\Delta\Lset\setminus\{x\}$.
\end{enumerate}
The edges $(v,\Parent(v))$ form a spanning in-arborescence of $S$ rooted at $x$.
\end{theorem}

\begin{proof}[Proof sketch]
For $v\ne x$, the successor sets coincide. If $v\in\Delta\Wset$, a lower-ranked losing successor in $G'$ cannot have lost in $G$, since its move would already have won at $v$. If $v\in\Delta\Lset$, all successors win in $G'$ at smaller ranks. At least one did not win in $G$, since otherwise $v$ would already have lost. Choose such a successor as the parent in each case. Its rank is strictly smaller. Thus every parent path terminates at $x$, the only vertex where the construction can stop (Appendix~\ref{app:resolution-tree}).
\end{proof}

\begin{corollary}\label{cor:resolution-root-parity}
Under the assumption of Theorem~\ref{thm:resolution-tree}, for an update of the move $(x,y)$,
\begin{itemize}
    \item If $M'=M\cup\{(x,y)\}$, then $x\in\Delta\Wset$ and
$y\in\Lset\cap\Lset'$. 
\item If $M'=M\setminus\{(x,y)\}$, then
$x\in\Delta\Lset$ and $y\notin\Wset$.
\end{itemize}
For any parent map supplied by Theorem~\ref{thm:resolution-tree}, the unique parent path
$q=v_0,v_1,\ldots,v_m=x$
alternates between $\Delta\Wset$ and $\Delta\Lset$. Thus, $q\in\Wset'$ exactly
when either the update is an insertion and $m$ is even, or the update is a deletion and $m$ is odd; otherwise,
$q\in\Lset'$.
\end{corollary}
Appendix~\ref{app:resolution-tree} identifies the value at $x$ under insertion and deletion and proves the parity assertion by induction on the path length.

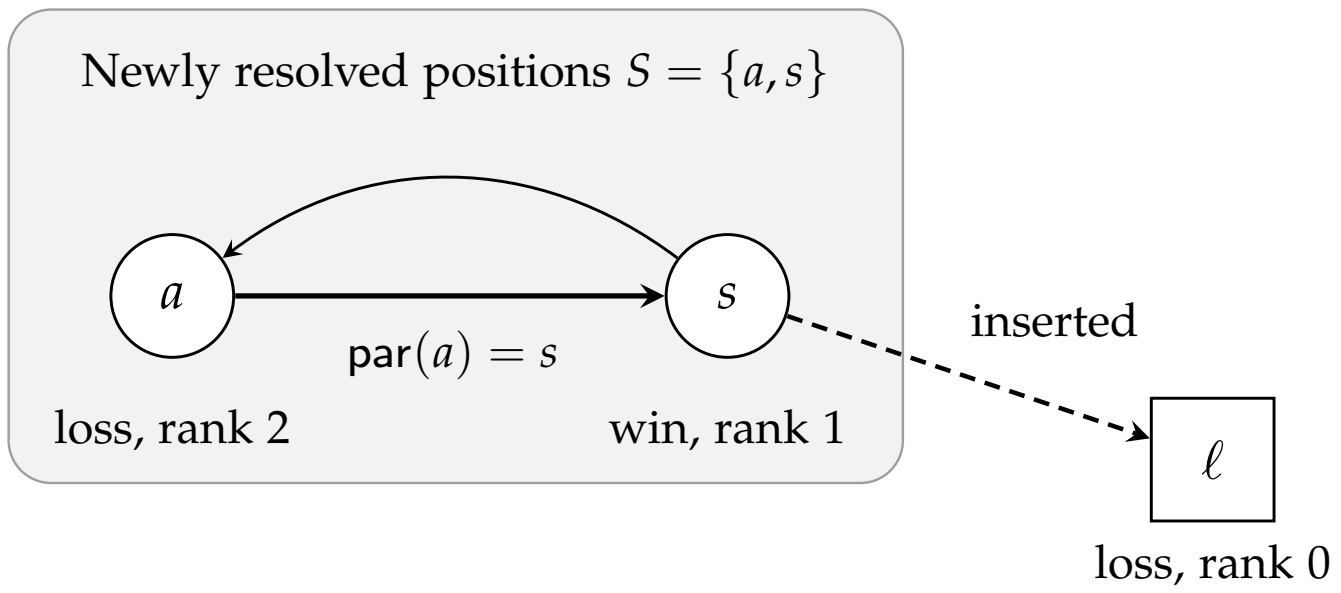
\begin{figure}[tb]
\centering
\begin{tikzpicture}[x=1pt,y=1pt,
  position/.style={circle,draw,fill=white,minimum size=21pt,inner sep=0pt,
                   line width=.55pt,font=\small},
  move/.style={->,>=stealth,line width=.55pt},
  parentmove/.style={move,line width=1pt},
  inserted/.style={move,line width=.85pt,dash pattern=on 3pt off 1.9pt},
  annotation/.style={font=\footnotesize,inner sep=1pt}]
\path[rounded corners=8pt,fill=black!5,draw=black!38,line width=.45pt]
  (0,10) rectangle (153,91);
\node[font=\footnotesize] at (76.5,80) {Newly resolved positions $S=\{a,s\}$};
\node[position] (a) at (28,42) {$a$};
\node[position] (s) at (123,42) {$s$};
\node[position,rectangle] (ell) at (206,14) {$\ell$};
\draw[move] (s) to[out=143,in=37] (a);
\draw[parentmove] (a) -- (s);
\draw[inserted] (s) -- (ell);
\node[annotation] at (76,32) {$\Parent(a)=s$};
\node[annotation] at (179,38) {inserted};
\node[annotation] at (28,19) {loss, rank $2$};
\node[annotation] at (123,19) {win, rank $1$};
\node[annotation] at (206,-4) {loss, rank $0$};
\end{tikzpicture}
\caption{In Example~\ref{example_intro_2}, inserting the dashed move
$s\to\ell$ resolves the drawn cycle. The bold edge $a\to s$ is the
parent tree on $S$, rooted at $s$. The labels give the new outcomes
and ranks.}
\Description{The shaded region contains a and s, which formed a drawn two-cycle.
The inserted move goes from s to the terminal ell outside the region.
After insertion, ell loses at rank zero, s wins at rank one, and a loses
at rank two. The bold parent edge goes from a to s.}
\label{fig:propagation-example}
\end{figure}

At the observation $E_{\mathsf g}$ of Example~\ref{example_intro_2}, inserting $s\longrightarrow\ell$ gives
$\Delta\Wset=\{s\}$, $\Delta\Lset=\{a\}$, and $\Parent(a)=s$.
Deleting $s\longrightarrow a$ gives
$\Delta\Wset=\{a\}$, $\Delta\Lset=\{s\}$, and $\Parent(a)=s$.
Figure~\ref{fig:propagation-example} takes $q=a$ for the insertion. The new move to the terminal $\ell$ makes $s$ winning, so $a$, whose only move leads to $s$, loses.

\section{Complexity}\label{sec:complexity}

We establish data complexity for a fixed program $\PiP$. The lower bounds hold for the one-rule program $\PiP_{\Win}$ of Example~\ref{example_intro_2}. In each construction, the query at a designated vertex $v$ is $\Win(v)$.

\subsection{Decision problems}\label{subsec:decision-problems}

Let $b,c\in\Bthree$ be distinct. Both decision problems take an instance $I=(\PiP,\Dx,\Univ,E,q)$ of Definition~\ref{def:instance} and a threshold $k\in\Nat$. The program is fixed, the relations are listed explicitly, and $k$ is encoded in binary. For $\Cause_{b\to c}[\PiP]$, the input also specifies a ground extensional atom $\tau$. Both problems reject malformed encodings. On well-formed inputs, the decision procedures are as follows.
\begin{itemize}
\item $\Cause_{b\to c}[\PiP]$ rejects if $(\tau\notin\Univ)\vee(\Ans(E)\ne b)$. Otherwise it accepts iff $\Kappa_E^{b\to c}(\tau)\le k$.
\item $\DetRadius[\PiP]$ rejects if $\Ans(E)\ne\Undef$. Otherwise it accepts iff $\DetRad(E)\le k$.
\end{itemize}

Calautti et al. prove that deciding whether a specified set of facts is exactly the support of a proof tree is NP-complete in data complexity, already for linear Datalog~\cite{calautti2024complexity}. Our decision problem asks for a bounded contingency for a designated fact and an ordered WFS transition.

A vertex cover of an undirected graph is a set of vertices meeting each edge. The reduction from Vertex Cover uses an indexed set system $\Hyp=(X,(S_j)_{j\in\Idx})$ with finite nonempty sets $X,\Idx$, $\varnothing\ne S_j\subseteq X$ for $j\in\Idx$, and $\bigcup_{j\in\Idx}S_j=X$. A cover is an index set $C\subseteq\Idx$ with $\bigcup_{j\in C}S_j=X$. The minimum cover size is $\Cov(\Hyp)$.

In the first construction, each $j\in\Idx$ has positions $s_j,d_j$, a fixed self-loop at $d_j$, and a mutable move $\alpha_j=(s_j,d_j)$. There are fixed moves $q\to e_x$ for $x\in X$ and $e_x\to s_j$ for $x\in S_j$. These are all the moves. The observed state is $E_{\Hyp}^{u}=\{\alpha_j\mid j\in\Idx\}$, the set of all mutable moves. Deleting $\alpha_j$ selects index $j$. For $C\subseteq\Idx$, let $\Delta_C=\{\alpha_j\mid j\in C\}$. After deleting $\Delta_C$, $s_j$ loses for $j\in C$ and draws otherwise. Hence, $e_x$ wins for $x\in\bigcup_{j\in C}S_j$ and draws otherwise. Since $q$ moves to each $e_x$,
\[
 \Ans(E_{\Hyp}^{u}\toggle\Delta_C)=
 \begin{cases}\False,&\bigcup_{j\in C}S_j=X,\\
 \Undef,&\text{otherwise.}\end{cases}
\]
Thus,
$\DetRad(E_{\Hyp}^{u})=\DirRad_{\False}(E_{\Hyp}^{u})=\Cov(\Hyp)$, whereas
$\DirRad_{\True}(E_{\Hyp}^{u})=\Inf$.
Let $\ast\notin X$ and $j_\ast\notin\Idx$, let $S_{j_\ast}=\{\ast\}$, and let $\Hyp^+=\bigl(X\cup\{\ast\},(S_j)_{j\in\Idx\cup\{j_\ast\}}\bigr)$.
Let $\tau=\alpha_{j_\ast}$ be the move added for $j_\ast$ and use the observed
state $E_{\Hyp^+}^{u}=E_{\Hyp}^{u}\cup\{\tau\}$.  Let $q$ be the
original query vertex, with one additional move $q\to e_\ast$, and let $p$ have the sole move to $q$.

Figure~\ref{fig:recurrent-selector} gives the construction for a graph with one edge $x$ and endpoints $1,2$, so $X=S_1=S_2=\{x\}$.
Deleting $\alpha_1$ makes $s_1$ terminal and $e_x$ winning. While $\tau$ is present, $e_\ast$ is drawn, so $q$ is drawn.
Deleting $\tau$ now makes $s_{j_\ast}$ terminal and $e_\ast$ winning. Both successors of $q$ win, so $q$ loses. If $\tau$ alone is deleted, $e_x$ remains drawn and $q$ remains drawn. By symmetry, $\{\alpha_1\}$ and $\{\alpha_2\}$ are exactly the minimum contingencies for $\tau$ at $q$.

\begin{figure}[htbp]
\centering
\begin{tikzpicture}[
  x=1pt,y=1pt,
  position/.style={circle,draw,line width=.5pt,minimum size=17pt,
    inner sep=1pt,outer sep=.4pt,font=\small},
  fixed/.style={->,>=stealth,line width=.6pt},
  mutable/.style={fixed,dash pattern=on 3pt off 2pt},
  edge label/.style={font=\small,inner sep=1.6pt},
  group label/.style={font=\footnotesize,inner sep=0pt}
]
\node[position] (p)  at (75,104) {$p$};
\node[position,line width=.85pt] (q)  at (113,104) {$q$};
\node[position] (ex) at (53,79) {$e_x$};
\node[position] (es) at (163,79) {$e_\ast$};
\node[position] (s1) at (23,52) {$s_1$};
\node[position] (s2) at (83,52) {$s_2$};
\node[position] (ss) at (163,52) {$s_{j_\ast}$};
\node[position] (d1) at (23,18) {$d_1$};
\node[position] (d2) at (83,18) {$d_2$};
\node[position] (ds) at (163,18) {$d_{j_\ast}$};
\draw[fixed] (p) -- (q);
\draw[fixed] (q) -- (ex);
\draw[fixed] (q) -- (es);
\draw[fixed] (ex) -- (s1);
\draw[fixed] (ex) -- (s2);
\draw[fixed] (es) -- (ss);
\draw[mutable] (s1) -- node[edge label,right] {$\alpha_1$} (d1);
\draw[mutable] (s2) -- node[edge label,right] {$\alpha_2$} (d2);
\draw[mutable] (ss) -- node[edge label,right] {$\tau$} (ds);
\draw[fixed] (d1) to[out=34,in=-34,looseness=5] (d1);
\draw[fixed] (d2) to[out=34,in=-34,looseness=5] (d2);
\draw[fixed] (ds) to[out=34,in=-34,looseness=5] (ds);
\draw[line width=.45pt] (6,4) -- (6,1) -- (108,1) -- (108,4);
\draw[line width=.45pt] (132,4) -- (132,1) -- (194,1) -- (194,4);
\node[group label,below=3pt] at (57,1) {Edge $x=\{1,2\}$};
\node[group label,below=3pt] at (163,1) {Added singleton $\{\ast\}$};
\end{tikzpicture}

\smallskip
{\small\setlength{\tabcolsep}{5pt}\renewcommand{\arraystretch}{1.1}%
\begin{tabular}{|c|cc|cc|}
\hline
$C$ & $e_x$ & $e_\ast$ & $q$ & $p$ \\ \hline
$\varnothing$ & $\Undef\to\Undef$ & $\Undef\to\True$ &
  $\Undef\to\Undef$ & $\Undef\to\Undef$ \\
$\{1\},\,\{2\},\,\{1,2\}$ & $\True\to\True$ & $\Undef\to\True$ &
  $\Undef\to\False$ & $\Undef\to\True$ \\ \hline
\end{tabular}}\par
\caption{The game for an edge with endpoints 1 and 2. The dashed moves are mutable. For each set $C$ of deleted endpoint indices, the table gives position values before and after deleting $\tau$.}
\Description{The query q has two successors. The original element leads to
two index positions and the added singleton leads to one. Each index position has a
mutable move to a draw vertex with a fixed self-loop. Before deleting tau,
the singleton element and the query are drawn. Deleting tau makes the
singleton element winning. The query then loses and its predecessor wins
exactly when at least one original mutable move has been deleted.}
\label{fig:recurrent-selector}
\end{figure}

In the second construction, replace each $d_j$ by a terminal $\ell_j$ and let $\alpha_j=(s_j,\ell_j)$ be present at the observation.  Let
$E_{\Hyp}^{t}=\{\alpha_j\mid j\in\Idx\}$.  After deleting the moves
indexed by $C$, $s_j$ is losing exactly for $j\in C$, $e_x$ is
winning exactly when $x\in\bigcup_{j\in C}S_j$, and $q$ loses
exactly when $C$ covers $X$.  For $\Hyp^+$ let $\tau=(s_{j_\ast},\ell_{j_\ast})$ and $E_{\Hyp^+}^{t}=E_{\Hyp}^{t}\cup\{\tau\}$, let $q$ be the query vertex with a move to each $e_x$ for $x\in X\cup\{\ast\}$, and let $p$ have the sole move to $q$.

In the third construction the mutable moves are $\Univ_{\Hyp}^{o}=\{\beta_j\mid j\in\Idx\}\cup\{\tau\}$, with $\tau$ distinct from the $\beta_j$, and $E_{\Hyp}^{o}=\varnothing$. We write $\Delta_C=\{\beta_j\mid j\in C\}$. A move from $r$ to a vertex with a fixed self-loop is enabled at $F\subseteq\Univ_{\Hyp}^{o}$ exactly when
\[
 (\tau\in F)\ \wedge\ (\forall x\in X\ \exists j\in\Idx:\ (x\in S_j)\wedge(\beta_j\in F)).
\]
The conditional-move construction of Eq.~\eqref{eq:conditional-move-gate} implements this test. Give $p$ the sole move to $r$. All auxiliary moves in the three constructions are fixed.

\begin{table}[t]
\centering
\caption{Transitions at the root and at the one-move predecessor $p$ when $\Delta_C$ changes the moves indexed by a cover $C$. The root is $q$ in the first two constructions and $r$ in the third.}
\label{tab:six-transitions}
\small
\setlength{\tabcolsep}{5pt}
\renewcommand{\arraystretch}{1.13}
\begin{tabular}{|l l|cc|}
\hline
Construction & Toggle of $\tau$ & Root & $p$ \\
\hline
Self-loops & Delete & $\Undef\to\False$ & $\Undef\to\True$ \\
Terminals & Delete & $\True\to\False$ & $\False\to\True$ \\
\hline
Conditional move & Insert & $\False\to\Undef$ & $\True\to\Undef$ \\
\hline
\end{tabular}
\end{table}

For each construction and query position in Table~\ref{tab:six-transitions}, let $E$ be the observation and $b\to c$ the indicated transition. For $C\subseteq\Idx$,
\[
\begin{aligned}
  \Ans(E\toggle\Delta_C)&=b,\\
  \Ans(E\toggle\Delta_C\toggle\{\tau\})=c
    &\iff \bigcup_{j\in C}S_j=X.
\end{aligned}
\]
The mutable moves other than $\tau$ are indexed by $\Idx$. Thus $C\mapsto\Delta_C$ is a cardinality-preserving bijection from covers to $b\to c$ contingencies for $\tau$, and $\Kappa_E^{b\to c}(\tau)=\Cov(\Hyp)$.

\begin{theorem}\label{thm:np-completeness}
For $\PiP_{\Win}$ and every pair of distinct $b,c\in\Bthree$,
$\Cause_{b\longrightarrow c}[\PiP_{\Win}]$ is NP-complete in data
complexity. The problem $\DetRadius[\PiP_{\Win}]$ is NP-complete in data complexity. The lower bounds use only mutable move
atoms as mutable facts and hold at in-degree at most $3$ and out-degree at most
$2$ in the union of all fixed and mutable moves, including all auxiliary positions.
\end{theorem}
\begin{proof}[Proof sketch]
A set of mutable moves is a polynomial-size witness. Verify its cardinality and the WFS values in the definition of the decision problem (Proposition~\ref{prop:np-membership}).

For hardness, take a graph $H$ of maximum degree three with at least one edge and no isolated vertices. Vertex Cover remains NP-complete on these graphs~\cite{garey1974some}. Let $X=E(H)$ and $S_v=\{e\in E(H)\mid v\in e\}$. In each row of Table~\ref{tab:six-transitions}, the bijection $C\mapsto\Delta_C$ sends a vertex cover to a contingency of the same size. In the first construction without the additional singleton it sends a vertex cover to an intervention of the same size that makes the query false.

For the degree bounds, each edge of $H$ has two endpoints and each vertex meets at most three edges. Replace the moves from $q$ to the element positions by one move to a binary conjunction tree. At each internal node, two fresh positions $a,b$ read child values at $u,v$ by the moves $a\to u$, $a\to v$, and $b\to a$. The output $b$ wins if both children win, loses if either child loses, and draws otherwise. Thus $q$ retains its value. In the construction with a conditional move, use binary gates for the cover test. To use an input $w$ twice, add $t\to w$, $w_1\to t$, and $w_2\to t$. Then $\Val(w_1)=\Val(w_2)=\Val(w)$ by~\eqref{eq:single-successor-table}. Repeating this construction uses $O(|V(H)|+|E(H)|)$ positions and fixed moves in total. Appendix~\ref{app:selector-degree} verifies the degrees and the correspondence for every state.
\end{proof}

\subsection{Optimization and counting}\label{subsec:optimization-counting}

We write $\SharpP$ for the functions counting accepting computations of nondeterministic polynomial-time machines. A polynomial-time metric reduction from $f$ to $g$ has the form $f(x)=d(x,g(h(x)))$, where $h$ and $d$ are computable in polynomial time. A polynomial-time Turing reduction permits adaptive oracle calls.

A well-formed input encodes an intervention instance and, for either function with a candidate, a ground extensional atom $\tau$. Let $n=|\Univ|$. On well-formed inputs, the functions are
\begin{align}
 \widehat\Kappa_E^{b\longrightarrow c}(\tau)
 &=\begin{cases}
   \min\{n,\Kappa_E^{b\longrightarrow c}(\tau)\},
     &(\tau\in\Univ)\wedge(\Ans(E)=b),\\
   n,&\text{otherwise,}
  \end{cases}
\label{eq:total-kappa}
\\
 \widehat\DetRad(E)
 &=\begin{cases}
   \min\{n+1,\DetRad(E)\},&\Ans(E)=\Undef,\\
   n+1,&\text{otherwise.}
  \end{cases}
\label{eq:total-radius}
\end{align}
For $\tau\in\Univ$, a contingency omits $\tau$ and has size at most $n-1$. An intervention has size at most $n$. Thus $n$ and $n+1$ distinguish infeasibility in the respective functions.  When $(\tau\in\Univ)\wedge(\Ans(E)=b)$, let $\CountMin_E^{b\to c}(\tau)$ count the minimum $b\to c$ contingencies for $\tau$, with count zero if the contingency family is empty. In the remaining cases set the count to zero. The three functions return zero on malformed encodings.

\begin{theorem}\label{thm:exact-complexity}
For the fixed program $\PiP_{\Win}$ and each transition in \Cref{thm:np-completeness}, computing
$\widehat\Kappa_E^{b\longrightarrow c}(\tau)$ is $\FPNPlog$-complete under
polynomial-time metric reductions.  Computing $\widehat\DetRad(E)$ is
$\FPNPlog$-complete under the same reductions.  Computing
$\CountMin_E^{b\longrightarrow c}(\tau)$ is
$\SharpP$-hard under polynomial-time Turing reductions.
\end{theorem}

\begin{proof}[Proof sketch]
For either optimum, test feasibility at the greatest possible witness size. If feasible, binary search for the least accepting threshold uses $O(\log(n+2))$ NP queries. Equations~\eqref{eq:total-kappa} and~\eqref{eq:total-radius} specify the remaining outputs on well-formed inputs.

Maximum clique size is $\FPNPlog$-complete under polynomial-time metric reductions~\cite{krentel1986complexity}. Let $G$ be the input graph and $H$ its complement. The complements of cliques of $G$ are exactly the vertex covers of $H$. If $H$ has an edge, the constructions in Section~\ref{subsec:decision-problems} give either optimum as the minimum vertex-cover size of $H$. Subtracting this size from $|V(G)|$ recovers the maximum clique size. If $H$ is edgeless, the maximum is $|V(G)|$.

For the counting lower bound, attach a private leaf $v'$ to each vertex $v$ of a bipartite graph $G=(V,A)$. The $|V|$ disjoint leaf edges require at least $|V|$ vertices, and $V$ attains the bound. A minimum cover therefore selects exactly one of $v,v'$ for each $v$. Hence the covers $C$ of $G$ correspond bijectively to the minimum covers $K(C)\mathrel{=}C\cup\{v'\mid v\in V\setminus C\}$ of the enlarged graph. The bijection $C\mapsto\Delta_{K(C)}$ gives the required count for each transition. Counting vertex covers of bipartite graphs is $\SharpP$-hard~\cite{provanBall1983}. Appendix~\ref{app:complexity} gives the reductions, including the empty graph.
\end{proof}

\section{Subtree Min--Max Theorem}\label{sec:minmax}
We give conditions under which reaching a prescribed answer amounts to changing at least one fact in each of a family of sets.

Let $q$ be the query vertex of a finite arena, $\Mu$ its mutable moves, and $E\subseteq\Mu$ its observed state. For $b\in\Bthree$, we define $f_b:2^{\Mu}\to\{0,1\}$ by $f_b(F)=1\iff\Ans(F)=b$. A term $\Req$ is an \emph{implicant} of $f_b$ if $\forall F\subseteq\Mu:\ (F\models\Req\Rightarrow f_b(F)=1)$. An implicant $\Req$ is \emph{prime} if no $\Req'\prec\Req$ is an implicant~\cite{darwiche2022computation}. Section~\ref{sec:certificates} takes minimality only among terms obtained from certificates.

For example, let the fixed moves on $q,x,y,t$ be $(q,x),(q,y),(y,x)$ and the only mutable move be $e=(x,t)$. If $e$ is absent, $x$ loses and $y$ wins. If $e$ is present, $t$ loses, $x$ wins, and $y$ loses. Thus $q$ wins in both states and the empty term is the unique prime implicant of $f_{\True}$. Each term obtained from a certificate for a win at $q$ specifies a losing successor. Neither successor loses in both states, so neither choice gives the empty term. By Theorem~\ref{thm:structured-completeness}, each state satisfies a term obtained from a certificate. With only one mutable move, these terms are $(\varnothing,\{e\})$ and $(\{e\},\varnothing)$.

Assume $\Ans(E)=\Undef$. We write $\mathcal O_E^{\min}$ for the prime implicants of $f_{\Undef}$ satisfied by $E$. This family is nonempty: $(E,\Mu\setminus E)$ implies $f_{\Undef}$ and has an inclusion-minimal subterm with the same property. The \emph{support} of a term $\Req$ is $B(\Req)=\Req^+\cup\Req^-$. The support hypergraph is $\mathcal H_E=(\Mu,(B(\Req))_{\Req\in\mathcal O_E^{\min}})$. In a Datalog instance, the bijection of Theorem~\ref{thm:causal-normal-form} preserves satisfaction, subterm inclusion and the query value at every state. It therefore maps these prime implicants and their supports bijectively to those of the corresponding game.

For a finite indexed hypergraph $\mathcal H=(X,(S_j)_{j\in J})$, a \emph{transversal} is a set $C\subseteq X$ such that $\forall j\in J:\ C\cap S_j\ne\varnothing$. A \emph{packing} is a family of pairwise disjoint hyperedges. Let $\Trans(\mathcal H)$ be the minimum transversal size and $\nu(\mathcal H)$ the maximum packing size. A hypergraph is a \emph{subtree hypergraph} if there is a tree on its vertices in which each hyperedge induces a nonempty connected subgraph~\cite{vandenHeuvelJohnson2008}. For such hypergraphs the transversal and packing numbers coincide~\cite[Cor.~66.5a]{schrijver2003optimization}.

\begin{definition}\label{def:subtree-representation}
Let $c\in\{\True,\False\}$ and let $T$ be a finite tree on vertex set $\Mu$. For $B\subseteq\Mu$, write $T[B]$ for the subgraph induced by $B$. Consider the following conditions on $E$, $c$, and $T$:
\begin{enumerate}
\item \label{item:one-value} $\forall F\subseteq\Mu:\ \Ans(F)\in\{\Undef,c\}$;
\item \label{item:observed-coverage} $\forall F\subseteq\Mu:\ \bigl(\Ans(F)=\Undef\iff\exists\Req\in\mathcal O_E^{\min}: F\models\Req\bigr)$;
\item \label{item:connected-supports} $\forall\Req\in\mathcal O_E^{\min}:\ B(\Req)\ne\varnothing\ \wedge\ T[B(\Req)]\text{ is connected}$.
\end{enumerate}
\end{definition}

Clause~\ref{item:observed-coverage} holds exactly when restoring one changed fact to its observed value preserves undefinedness:
\begin{equation}\label{eq:coverage-restoration}
 \forall\Delta\subseteq\Mu\ \forall e\in\Delta:\quad
 \Ans(E\toggle\Delta)=\Undef\ \Rightarrow \Ans\bigl(E\toggle(\Delta\setminus\{e\})\bigr)=\Undef.
\end{equation}
Given $T$ and the listed supports, we can test Clause~\ref{item:connected-supports} in polynomial time by checking that each support $B$ is nonempty and that a search in $T[B]$ reaches all vertices of $B$ from one of them. Equation~\eqref{eq:coverage-restoration} gives a polynomial-size witness to failure of Clause~\ref{item:observed-coverage}. Let $\mathsf{coNP}$ denote the class of complements of problems in $\mathsf{NP}$. For the fixed win--move rule, testing Clause~\ref{item:observed-coverage} from the game, query, and undefined observation is $\mathsf{coNP}$-complete under polynomial-time many-one reductions (Proposition~\ref{prop:recognizing-coverage}).

\begin{theorem}\label{thm:arboreal-minmax}
If $E$, $c$, and $T$ satisfy Definition~\ref{def:subtree-representation}, then
\begin{equation}
 \DirRad_c(E)=\Trans(\mathcal H_E)=\nu(\mathcal H_E).
\label{eq:arboreal-minmax}
\end{equation}
Given $(V,\Mx,\Mu,E,q)$ and $c$, we can compute a minimum $c$-intervention in polynomial time whenever such a tree $T$ exists. Given also $T$ and the list $\mathcal O_E^{\min}$, we can compute in polynomial time a maximum subfamily with pairwise disjoint supports and a certificate for $c$ at $q$ after the intervention.
\end{theorem}

\begin{proof}[Proof sketch]
Let $\Delta\subseteq\Mu$ and $F=E\toggle\Delta$. For $\Req\in\mathcal O_E^{\min}$, $E\models\Req$ gives $F\models\Req\iff\Delta\cap B(\Req)=\varnothing$. Clauses~\ref{item:one-value} and~\ref{item:observed-coverage} therefore give
\[
 \Ans(F)=c\iff
 \forall\Req\in\mathcal O_E^{\min}:\ \Delta\cap B(\Req)\ne\varnothing .
\]
Thus, the query tests whether $\Delta$ is a transversal. By Clause~\ref{item:connected-supports}, the supports induce nonempty subtrees of $T$. The algorithm of van den Heuvel and Johnson~\cite{vandenHeuvelJohnson2008} computes a minimum transversal using this test alone. Given $T$ and the supports explicitly, root $T$. Among the remaining supports choose one whose vertex nearest the root is as deep as possible. Select that vertex for the intervention and that support for the packing. The supports containing the selected vertex are precisely those meeting the selected support (Lemma~\ref{lem:subtree-konig}). Removing them and repeating gives a transversal and an equally large packing. Retrograde evaluation at the resulting state gives the certificate.
\end{proof}

Example~\ref{example_intro_1} violates Clause~1 because both determinate answers are reachable. Clause~2 can fail even under Clause~1. Consider the rules $q\leftarrow a,\neg b,\neg q$ and $q\leftarrow b,\neg a,\neg q$ with mutable facts $a,b$ and $c=\False$. At either singleton state the only remaining rule is $q\leftarrow\neg q$, giving $\Undef$. At the empty and full states both rule bodies are false, giving $\False$. At $E=\{a\}$, deleting either literal of $a\wedge\neg b$ admits the empty or full state. This is the unique prime implicant of $f_{\Undef}$ satisfied by $E$. At $\{b\}$ the query is undefined and this term is false.

\begin{figure}[htb]
\centering
{\setlength{\unitlength}{1pt}%
\begin{picture}(202,48)
\linethickness{0.45pt}
\put(28,36){\line(1,0){38}}
\put(82,36){\line(1,0){38}}
\put(136,36){\line(1,0){38}}
\put(20,36){\circle{16}}
\put(66,28){\framebox(16,16){$b$}}
\put(120,28){\framebox(16,16){$c$}}
\put(182,36){\circle{16}}
\put(20,36){\makebox(0,0){$a$}}
\put(182,36){\makebox(0,0){$d$}}
\linethickness{0.8pt}
\put(20,17){\line(0,1){5}}
\put(20,17){\line(1,0){54}}
\put(74,17){\line(0,1){5}}
\put(128,17){\line(0,1){5}}
\put(128,17){\line(1,0){54}}
\put(182,17){\line(0,1){5}}
\put(47,6){\makebox(0,0){\small $\{a,b\}$}}
\put(155,6){\makebox(0,0){\small $\{c,d\}$}}
\end{picture}}
\caption{A minimum intervention on a path. Squares mark the deleted facts. The brackets mark two disjoint supports that certify optimality.}
\label{fig:tree-edge-supports}
\Description{Four vertices form the path a, b, c, d. Squares denote the deleted facts at b and c. Separate brackets join a with b and c with d, identifying a matching of size two.}
\end{figure}
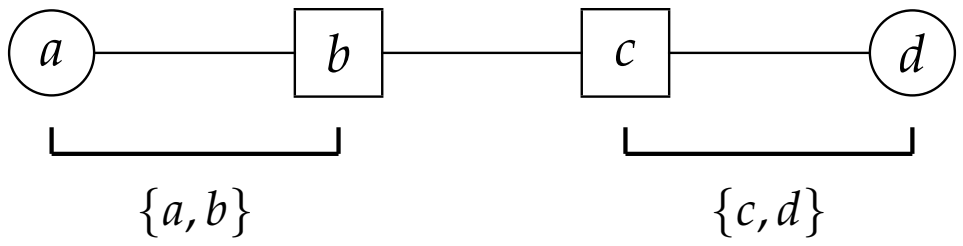

\emph{When the input lists the supports.} The exogenous facts $\mathsf{Edge}(i)$ and $\mathsf{In}(i,a)$ list the sets and their members, respectively. The mutable facts are $\mathsf{On}(a)$, one per vertex $a$. The following fixed program $\PiP_{\mathsf{hyp}}$ has query $q$.
\begin{equation}\label{eq:hypergraph-program}
\begin{aligned}
 \mathsf{Bad}(i)&\leftarrow\mathsf{Edge}(i),\mathsf{In}(i,a),\neg\mathsf{On}(a),\\
 \mathsf{Draw}(i)&\leftarrow\mathsf{Edge}(i),\neg\mathsf{Bad}(i),\neg\mathsf{Draw}(i),\\
 q&\leftarrow\mathsf{Draw}(i).
\end{aligned}
\end{equation}
\ Let the input list a nonempty family $(B_i)_{i\in I}$ of distinct, nonempty, inclusion-incomparable sets, and let $E$ contain every mutable fact. Identify each vertex $a$ with $\mathsf{On}(a)$. A missing member of $B_i$ derives $\mathsf{Bad}(i)$, making $\mathsf{Draw}(i)$ false. If $B_i\subseteq F$, the rule for $d=\mathsf{Draw}(i)$ reduces to $d\leftarrow\neg d$, making $d$ undefined. Hence,
\[
 \Ans(F)=\Undef\ \iff\ \exists i\in I:\ B_i\subseteq F,
 \qquad \Ans(F)\in\{\Undef,\False\}.
\]
Every term satisfied at $E$ is positive. A positive term on $P$ forces undefinedness iff its least satisfying state $P$ contains a listed $B_i$. Incomparability therefore gives exactly the $B_i$ as supports of the prime implicants satisfied by $E$ (Lemma~\ref{lem:sperner-realization}). For the path $a$--$b$--$c$--$d$ in Figure~\ref{fig:tree-edge-supports}, deleting $\mathsf{On}(b)$ and $\mathsf{On}(c)$ meets every support and resolves $q$ to false. Each deletion meets at most one of the disjoint supports $\{a,b\}$ and $\{c,d\}$, so two deletions are necessary. More generally, for the edges of a nontrivial tree, a minimum vertex cover gives the intervention and a maximum matching certifies the minimum size by Eq.~\eqref{eq:arboreal-minmax}.

\emph{Size of the explicit family.} Take the arena with a conditional move from Section~\ref{sec:complexity} on $n\geq 1$ disjoint edges $\{u_i,v_i\}$, $1\leq i\leq n$. Let $\beta_w$ denote the mutable move indexed by endpoint $w$, and let $E$ contain $\tau$ and all $2n$ such moves. The move from the root to the draw vertex is enabled precisely when
\[
\tau\in F\ \wedge\ \bigl(\forall i\in\{1,\ldots,n\}:\ 
   (\beta_{u_i}\in F\ \vee\ \beta_{v_i}\in F)\bigr).
\]
The root is undefined in those states and false otherwise. Every term satisfied by $E$ is positive. If such a term omits $\tau$ or both moves $\beta_{u_i},\beta_{v_i}$ for an edge, the state consisting of the required moves makes the root false. Conversely, requiring $\tau$ and one of $\beta_{u_i},\beta_{v_i}$ for each $i$ suffices for undefinedness. If both occur for an edge, either can be omitted. The prime implicants of $f_{\Undef}$ satisfied by $E$ therefore have supports of the form 
$\{\tau,\beta_{w_1},\ldots,\beta_{w_n}\}$, where  $w_i\in\{u_i,v_i\}$ for each $ 1\leq i\leq n$.
There are $2^n$ supports on $2n+1$ mutable moves, each connected in the star centred at $\tau$. Deleting $\tau$ meets every support and resolves the root to false.

The \emph{incidence graph} has one vertex per hypergraph vertex and hyperedge, with one edge per incidence. A tree decomposition consists of vertex sets (bags) indexed by a tree. Each graph vertex occurs in a nonempty connected set of bags, and both ends of each edge occur in a common bag. The \emph{treewidth} $\operatorname{tw}(G)$ of a graph $G$ is the minimum, over tree decompositions, of the largest bag size minus one. \ifdefined\balance\par\newpage\noindent\fi For a hypergraph $H$, the \emph{incidence treewidth} is the treewidth of the incidence graph of $H$.

\begin{proposition}\label{prop:subtree-strictness}
There are instances satisfying Definition~\ref{def:subtree-representation} for a tree $T$ for which no path on $\Mu$ satisfies Clause~3. The incidence treewidth of the support hypergraphs satisfying that definition is unbounded. Both statements hold for the fixed program $\PiP_{\mathsf{hyp}}$.
\end{proposition}

\begin{proof}[Proof sketch]
By Lemma~\ref{lem:sperner-realization}, $\PiP_{\mathsf{hyp}}$ realizes the triples consisting of the centre of a star and two of its $n$ leaves. For $n=3$, each omitted leaf would have to be an endpoint of a path on four vertices. For $n\geq3$, deleting the centre from the incidence graph leaves a subdivision of the complete graph $K_n$, giving treewidth at least $n-1$.
\end{proof}

\begin{corollary}\label{thm:packing-failure}
The program $\PiP_{\mathsf{hyp}}$ realizes instances satisfying Clauses~1 and~2 of Definition~\ref{def:subtree-representation} for which $\Trans(\mathcal H_E)/\nu(\mathcal H_E)$ is unbounded.
\end{corollary}
\begin{proof}[Proof sketch]
For $n\geq 2$, take as input to $\PiP_{\mathsf{hyp}}$ the family of all $k$-subsets of $[n]=\{1,\ldots,n\}$, where $k=\lfloor n/2\rfloor+1$. By Lemma~\ref{lem:sperner-realization}, these subsets are precisely the supports. They intersect pairwise, and a set meets them all exactly when its complement has size less than $k$. Thus the ratio is $n-k+1=\lceil n/2\rceil$ (Appendix~\ref{app:arboreal}).
\end{proof}

\section{Conclusion and Outlook}\label{sec:conclusion}
We have extended contingency-based causality to Boolean Datalog queries with recursion through negation under the well-founded semantics. For each of the six transitions among true, false, and undefined, we have characterized the minimum contingencies that make a designated fact pivotal. For the same fact at the same observation, the minimum contingency size can depend on the target answer. Our game representation preserves query values and intervention distances. When one update resolves a draw, a path of newly winning and newly losing positions connects the query to the source of the changed move.

We have shown that deciding whether a contingency of a given size or smaller exists is NP-complete for each transition, even for the single win--move rule. We also give a polynomial-time algorithm for a prescribed determinate answer when it is the only reachable determinate value, the prime implicants of undefinedness that hold at the observation cover every undefined state, and their supports induce nonempty subtrees of a common tree. The algorithm takes the game, query, observation, and target as input. Given the tree and the implicants, we also compute a family of pairwise disjoint supports whose cardinality certifies the minimum.

Two questions remain. First, we proved that deciding coverage from the game is $\mathsf{coNP}$-complete (Proposition~\ref{prop:recognizing-coverage}). Which syntactic restrictions on recursive rules guarantee coverage at every undefined observation? By the same proposition, it suffices to guarantee that restoring any changed fact to its observed value preserves undefinedness. Second, when both true and false are reachable, meeting each support under the coverage assumption gives a determinate answer. Which conditions on the game allow us to find a minimum intervention for a \emph{prescribed} determinate answer in polynomial time?

\vspace{1cm}
\phantomsection
\addcontentsline{toc}{section}{References}
\bibliographystyle{plainnat}
\bibliography{references}
\vspace{2cm}
\appendixproofstrue
\appendix
\section*{Appendix}

\section{Minimum Intervention Size}\label{app:boundary-distance}
\newenvironment{proofsteps}{%
 \setlength{\leftmargini}{1em}%
 \setlength{\leftmarginii}{1em}%
 \setlength{\leftmarginiii}{1em}%
 \begin{enumerate}
 \setlength{\itemsep}{0.35\baselineskip}%
 \setlength{\parsep}{0pt}%
 \setlength{\topsep}{0.35\baselineskip}%
}{\end{enumerate}}
\newenvironment{proofcases}{%
 \setlength{\leftmargini}{1em}%
 \setlength{\leftmarginii}{1em}%
 \setlength{\leftmarginiii}{1em}%
 \begin{itemize}
 \setlength{\itemsep}{0.35\baselineskip}%
 \setlength{\parsep}{0pt}%
 \setlength{\topsep}{0.35\baselineskip}%
}{\end{itemize}}

\begin{proposition}\label{prop:boundary-distance}
Let $\Univ$ and $\Ans$ be the mutable facts and response function of an intervention instance. If $E\subseteq\Univ$ and $\Ans(E)=\Undef$, then
\begin{equation}
 \DetRad(E)=1+\min_{\substack{\tau\in\Univ\\ c\in\{\True,\False\}}}
 \Kappa_E^{\Undef\to c}(\tau).
\label{eq:boundary-distance}
\end{equation}
For each prescribed $c\in\{\True,\False\}$, the equality
$\DirRad_c(E)=1+\min_{\tau\in\Univ}\Kappa_E^{\Undef\to c}(\tau)$ can fail.
\end{proposition}

\begin{proof}
Let $r=\DetRad(E)$ and let $k$ be the minimum on the right of~\eqref{eq:boundary-distance}. We prove both inequalities in $\Nat_{\Inf}$.

If $k<\Inf$, finiteness of $\Univ$ gives a fact $\tau$, a target $c$, and a contingency $\Gamma$ of size $k$. By Definition~\ref{def:typed-cause}, $\tau\notin\Gamma$ and $\Ans(E\toggle\Gamma\toggle\{\tau\})=c$.
Thus, $\Gamma\cup\{\tau\}$ resolves the query, and $r\leq |\Gamma\cup\{\tau\}|=1+k$. For $k=\Inf$, the same inequality holds since $1+\Inf=\Inf$.

If $r<\Inf$, choose $\Delta\subseteq\Univ$ with $|\Delta|=r$ and $\Ans(E\toggle\Delta)=c\in\{\True,\False\}$. Since $\Ans(E)=\Undef$, $\Delta\ne\varnothing$. Take $\tau\in\Delta$ and let $\Gamma=\Delta\setminus\{\tau\}$. Since $|\Gamma|=r-1$, minimality of $r$ gives $\Ans(E\toggle\Gamma)=\Undef$. Also, $E\toggle\Gamma\toggle\{\tau\}=E\toggle\Delta$. Thus, $\Gamma$ is a $\Undef\to c$ contingency for $\tau$, and $k\leq |\Gamma|=r-1$. For $r=\Inf$, the inequality $1+k\leq r$ holds for every $k\in\Nat_{\Inf}$. Together the inequalities give~\eqref{eq:boundary-distance}, including $\Univ=\varnothing$.

For the directional claim, take the nullary program in~\eqref{eq:directional-counterexample} with $\Univ=\{a,b\}$ and $\Dx=E=\varnothing$. Let $P_F$ be the program on $q$ after evaluating the extensional literals at $F\subseteq\Univ$. For $X\subseteq\{q\}$,
\[
 q\in g_{P_F}(X)\iff (F=\Univ)\vee\bigl(F=\varnothing\wedge q\notin X\bigr).
\]
The reduct has a fact $q$ precisely in these two cases. Thus its alternating bounds stabilize at $(\varnothing,\{q\})$ for $F=\varnothing$, at $(\varnothing,\varnothing)$ for $|F|=1$, and at $(\{q\},\{q\})$ for $F=\Univ$. Hence $\DirRad_{\True}(E)=2$. For either $\tau$, only $\Gamma=\varnothing$ preserves the undefined answer, and $\Ans(\{\tau\})=\False$. Therefore $\min_{\tau\in\Univ}\Kappa_E^{\Undef\to\True}(\tau)=\Inf$.

For $c=\False$, take a fresh nullary query $q'$ with the sole defining rule $q'\leftarrow\neg q$. Since $q'$ occurs in no body, the rule preserves the value of $q$ at every state and assigns its complement to $q'$. Thus $q'$ is undefined at $\varnothing$, true at either singleton, and false at $\{a,b\}$. For this query, $\DirRad_{\False}(E)=2$. Only the empty contingency preserves undefinedness, and neither single insertion gives $\False$. Hence $\min_{\tau\in\Univ}\Kappa_E^{\Undef\to\False}(\tau)=\Inf$.
\end{proof}

\section{Game Normal Form}
\label{app:causal-normal-form}

\subsection{Game positions and Boolean circuits}
Let $\PiP$ be a safe normal Datalog program treated as constant in data
complexity, and let $Q$ be an intensional predicate of the declared
schema. Let $q=Q(\bar a)$ be a ground goal, $A$ a finite nonempty source
domain containing $\Const(\PiP)\cup\Const(q)$, and $D$ an extensional
database over $A$. We use a unary domain predicate with extension
$A$. The games may use one additional domain element and fixed structural
facts, as specified in the proof.

\begin{lemma}\label{lem:two-total-games}
The pair $(\PiP,Q)$ determines two total game-normal-form programs.
On every $D$ over $A$, their respective roots $t_{\bar a}$ and $f_{\bar a}$ satisfy
\begin{align}
 \Val(t_{\bar a})=\True
 &\Longleftrightarrow \WFM(\PiP\cup D)(q)=\True,
 \label{eq:true-total-game}\\
 \Val(f_{\bar a})=\True
 &\Longleftrightarrow \WFM(\PiP\cup D)(q)=\False.
 \label{eq:false-total-game}
\end{align}
Both programs define moves by nonrecursive semipositive rules
over the augmented extensional signature.
\end{lemma}

\begin{proof}
\begin{proofsteps}
\item \emph{Grounding domain.}
Let $A$ be the specified finite nonempty source domain, so $\Const(\PiP)\cup\Const(D)\cup\Const(q)\subseteq A$. Both the source grounding and its Herbrand base use $A$. In the construction for an intervention instance $I$, take $A=A_I$ for all its states.
If $Q$ has no defining rule, add $Q(\bar x)\leftarrow Q(\bar x)$, with distinct variables in $\bar x$ (the tuple is empty for nullary $Q$). Each ground instance is satisfied by every interpretation. Thus the models of each positive reduct, its least model, and the alternating bounds are unchanged. The queried predicate now occurs in a rule head, as required by the cited normal form.

Choose $a_*\in A$ and $a_{\rm aux}\notin A$. On $\widehat A=A\cup\{a_{\rm aux}\}$, interpret two fresh constant symbols $c_0,c_1$ by $a_*,a_{\rm aux}$, respectively. We ground over elements of $\widehat A$, so $c_0$ denotes $a_*$ even if a source constant already names that element. The interpretations of $c_0,c_1$ are distinct also for $|A|=1$. Use a fresh unary extensional predicate $\mathsf{adom}$ with extension $A$. For each variable $x$ of a source rule, add $\mathsf{adom}(x)$ to the rule body. Replace each nullary predicate $P$ by a fresh unary predicate $P'$ and each occurrence of $P$ by $P'(c_0)$, including occurrences in the facts and the goal. Interpret any intensional predicate without defining rules as the empty extensional relation.

Let $j$ map source ground atoms to their translations over $\widehat A$. We write $j(q)=\widehat Q(\widehat{\bar a})$: for positive-arity $Q$ take $(\widehat Q,\widehat{\bar a})=(Q,\bar a)$, and for nullary $Q$ take $(Q',(c_0))$. The fixed signature needs only the constants of $\PiP$ and $c_0,c_1$; the entries of $\bar a$ are elements of $A$, including those absent from $\PiP$. Let $g,\widehat g$ be the reduct operators of the source and padded programs, including their extensional facts. For each set $\widehat X$ of ground atoms over the padded signature and $\widehat A$, let $X=j^{-1}(\widehat X)$, $L=g(X)$, and $\widehat L=\widehat g(\widehat X)$. The two reducts retain corresponding rules under substitutions over $A$: the atom of each negative body literal lies in $X$ iff its translation lies in $\widehat X$. Let $D_{\rm dom}=\{\mathsf{adom}(a)\mid a\in A\}$. The following inclusions give agreement at the least models:
\begin{enumerate}
\item $L\subseteq j^{-1}(\widehat L)$. Every translated input fact and every atom in $D_{\rm dom}$ lies in $\widehat L$. Whenever $j^{-1}(\widehat L)$ satisfies the body of an original reduct rule, $\widehat L$ satisfies its translated body including the domain tests. It contains the translated head. Thus $j^{-1}(\widehat L)$ is a model of the original reduct.
\item $\widehat L\subseteq j[L]\cup D_{\rm dom}$. This extension contains the padded facts. A substitution using $a_{\rm aux}$ for a source variable fails its domain test. Every remaining reduct rule translates an original rule, whose head belongs to $L$ whenever its body holds in $L$. Hence the extension models the padded reduct.
\end{enumerate}
Leastness in the respective reducts proves the inclusions. The fresh domain predicate lies outside the image of $j$, so 
$j^{-1}(\widehat g(\widehat X))=g(j^{-1}(\widehat X))$. For the lower and upper bounds in~\eqref{eq:wfs-bounds}, $j^{-1}(\widehat K_0)=\varnothing=K_0$, and
\[
 j^{-1}(\widehat K_i)=K_i
 \ \Longrightarrow\ j^{-1}(\widehat U_i)=g(K_i)=U_i
 \ \Longrightarrow\ j^{-1}(\widehat K_{i+1})=g(U_i)=K_{i+1}.
\]
Induction and stabilization give equal values at $q$ and $j(q)$.

Let $\widehat D=j[D]\cup D_{\rm dom}$ and let $\gamma$ be the reduct operator on ground intensional atoms after evaluating the extensional literals in $\widehat D$. Only input facts have extensional heads, so the least model of every reduct of the padded program has extensional part $\widehat D$. For a set $Y$ of ground intensional atoms, both reducts test negative intensional literals against $Y$, giving $\widehat g(\widehat D\cup Y)=\widehat D\cup\gamma(Y)$. Applying the identity twice identifies the fixed points of $\gamma\circ\gamma$ with those of $\widehat g\circ\widehat g$ under $Y\mapsto\widehat D\cup Y$. The map preserves and reflects inclusion, so the least fixed points give the same intensional lower bound in~\eqref{eq:wfs-bounds}. Applying the identity once more gives the same intensional upper bound.

\item \emph{Two total games.}
For $b\in\{\True,\False\}$, let $Q^b$ be the relation of tuples at which $\widehat Q$ has value $b$ in the padded program. Corollary~15 of \cite{flum2000games} applies separately to the true-answer and false-answer queries of a well-founded Datalog program on finite structures. For each $b$ it gives a game-normal-form program whose true answer relation is $Q^b$. Theorem~16 of the same paper preserves that true answer relation and makes the program total. The win--move rule is the only recursive rule. The move definitions are nonrecursive and semipositive. The domain $\widehat A$ is finite, and every predicate in the translation has positive arity. For $k$ distinct position labels, choose $d$ with $2^d\ge k$ and assign distinct tuples in $\{c_0,c_1\}^d$. The number $k$ depends only on the program; the tuples distinguish labels since $c_0,c_1$ have distinct interpretations (cf.~\cite[Sec.~4.2, fn.~7]{flum2000games}). A fresh unary extensional predicate $\mathsf{dom}$ of extension $\widehat A$ supplies any positive domain literals required for safety in the move rules. Equality and inequality are interpreted on $\widehat A$. Use the tagged tuples as vertices, so the ground arena has a binary move relation.

\item \emph{Answer position.}
Each answer rule has the form $\mathsf{answer}(\bar v)\leftarrow\Win(\bar t)$, with every head variable occurring in $\bar t$. For the selected tuple $\widehat{\bar a}$, let $X_b=\{\bar t\theta\mid\bar v\theta=\widehat{\bar a}\}$, where the pair $(\bar v,\bar t)$ ranges over the answer rules for $Q^b$ and $\theta$ ranges over their ground substitutions on $\widehat A$. In particular, body variables absent from $\bar v$ range over the whole domain. Use fresh positions $\bar x$ for $x\in X_b$ and a root $r_b$ with moves $r_b\to\bar x$ and $\bar x\to x$. We index the roots by head tuples and distinguish the roots from the intermediate vertices by fixed tags. We count the root and intermediate labels in $k$ and pad shorter argument tuples with $c_0$ to a common arity before adjoining their length-$d$ tags. Equality with the components of the root tuple selects the corresponding ground rule instances. Since every $x\in X_b$ is determinate, the win--move rule gives
\[
\Val(r_b)=\True\iff \exists x\in X_b:\ \Val(x)=\True.
\]
If the condition fails, every successor of $r_b$ is winning, so $r_b$ is losing even when $X_b=\varnothing$. The existential condition is exactly membership of $\widehat{\bar a}$ in the answer relation of the total program for $Q^b$. Using disjoint copies of the two games and setting $t_{\bar a}=r_{\True}$ and $f_{\bar a}=r_{\False}$ proves~\eqref{eq:true-total-game} and~\eqref{eq:false-total-game}. The rule sets and tags depend only on $\PiP$ and $Q$.
\end{proofsteps}
\end{proof}

\begin{lemma}\label{lem:three-value-combiner}
For the selected ground tuple, abbreviate $t=t_{\bar a}$ and
$f=f_{\bar a}$ for the roots from \Cref{lem:two-total-games}.  Add positions
$\bar t,\bar f,d,h,s$ and moves
\[
 \bar t\to t,\qquad
 \bar f\to f,\qquad
 d\to d,\qquad
 h\to d,\qquad
 h\to\bar f,\qquad
 s\to\bar t,\qquad
 s\to h.
\]
Then, for every database $D$, $\Val(s)=\WFM(\PiP\cup D)(q)$.
\end{lemma}

\begin{proof}
Let $a=\WFM(\PiP\cup D)(q)$. If $v$ is the sole successor of $\bar v$, a move to a loss wins, a forced move to a win loses, and a forced move to a draw preserves the draw. Thus,
\begin{equation}
 \Val(\bar v)=
 \begin{cases}
  \False,&\Val(v)=\True,\\
  \True,&\Val(v)=\False,\\
  \Undef,&\Val(v)=\Undef.
 \end{cases}
\label{eq:single-successor-table}
\end{equation}
The self-loop makes $d$ nonterminal, hence $d\notin\Wset_0\cup\Lset_0$. If $d\notin\Wset_i\cup\Lset_i$, the sole successor of $d$ is neither losing nor winning at stage $i$. Both recurrence clauses exclude $d$ at stage $i+1$. Thus, $d$ remains drawn.  Equations~\eqref{eq:true-total-game}--
\eqref{eq:false-total-game}, determinacy of $t,f$,
Eq.~\eqref{eq:single-successor-table}, and the win--move rule give
\begin{equation*}
 \setlength{\arraycolsep}{3.4pt}
\renewcommand{\arraystretch}{1.10}
\begin{array}{|c|cc|ccc|c|}\hline
 a&\Val(t)&\Val(f)&\Val(\bar t)&\Val(\bar f)&\Val(h)&\Val(s)\\ \hline
 \True &\True &\False&\False&\True &\Undef&\True\\
 \False&\False&\True &\True &\False&\True &\False\\
 \Undef&\False&\False&\True &\True &\Undef&\Undef\\ \hline
\end{array}
\end{equation*}
The three exhaustive rows give $\Val(s)=a$.
\end{proof}

For a semipositive program $R$ over a grounding domain $A$ and an
extensional state $D$, let $R(D)$ denote the least model of the positive
ground program obtained by evaluating every EDB body atom of
$\Ground(R,A)$ against $D$.

\begin{lemma}\label{lem:input-circuit}
Let $R$ be a fixed nonrecursive semipositive Datalog program.
Given a finite domain $A$ listed explicitly and a ground output atom
$\beta$ of $R$ over $A$, we can construct in polynomial time an acyclic
Boolean circuit $C_\beta$ with one input per ground extensional fact.
For every extensional state $D$, $C_\beta(D)=1$ iff $\beta\in R(D)$.
\end{lemma}

\begin{proof}
\begin{proofsteps}
\item \emph{Construction.} Order the intensional predicates so that each intensional body predicate precedes the head predicate. Such an order exists because $R$ is nonrecursive. Ground the rules over $A$, discard instances with a false ground equality or inequality, and remove true tests from the remaining bodies.  For a ground rule instance, make one conjunction gate whose
inputs are the positive body atoms and the complements of the negative
extensional body atoms; an empty body uses the true constant.  For every ground intensional atom, including one with no defining instance, make one disjunction gate over the conjunction gates of all rule
instances with the chosen head; an empty family uses the false constant.  Process the
intensional predicates in topological order, reusing the already constructed
output gate whenever an earlier intensional atom occurs in a body.

\item \emph{Correctness.} For a ground rule $r$ surviving the equality tests, write $r^+$ and $r^-$ for the atoms of its positive and negative body literals. Let $C_\alpha$ be the gate for a ground atom $\alpha$ and let $\mathcal G_\beta$ be the surviving rules with head $\beta$. By construction,
\[
 C_\beta(D)=1\iff
 \exists r\in\mathcal G_\beta:\;
 \bigl(\forall\alpha\in r^+: C_\alpha(D)=1\bigr)
 \wedge\bigl(\forall\alpha\in r^-: C_\alpha(D)=0\bigr).
\]
For an extensional $\alpha$, $C_\alpha(D)=1\iff\alpha\in D$. At the first intensional predicate, only these input gates occur on the right; the equivalence is exactly the condition for deriving $\beta$ in $R(D)$. At each later predicate the positive intensional body atoms have earlier predicates, so the induction hypothesis replaces their gate tests by membership in $R(D)$. All negative tests are extensional. The same equivalence gives $C_\beta(D)=1\iff\beta\in R(D)$, including $\mathcal G_\beta=\varnothing$ and an empty rule body.

\item \emph{Size and shared inputs.} Let $r$ be the maximum number of variables in a rule of $R$, with $r=0$ for the empty program.  Every rule
has at most $|A|^r$ ground instances.  The program and all predicate arities
are treated as constants; hence, the number of gates and wires is polynomial in $|A|$. Each ground rule contains a bounded number of input or program constants. Enumerating the rules, indexing their gates and writing their labels therefore takes polynomial time in the input length.
Creating one input gate for each ground extensional fact and reusing it proves
the shared-input assertion.
\end{proofsteps}
\end{proof}

\begin{lemma}\label{lem:boolean-game-circuit}
Every acyclic Boolean circuit $C$ over $\wedge,\vee,\neg,0,1$
admits an acyclic game with a total output position $c$. For every
input state $D$, $c$ wins iff $C(D)=1$. Each ground input fact $\alpha$ has one shared position $p_\alpha$
whose sole possible move leads to a terminal and is enabled iff $\alpha$
is present.
\end{lemma}

\begin{proof}
\begin{proofsteps}
\item \emph{Base positions.} Let $z$ be terminal; thus, $\Val(z)=\False$.  Let $D$ be the set of present input facts. For an input $\alpha$, the
position $p_\alpha$ has the move $p_\alpha\to z$ when $\alpha$ is present
and has no move otherwise.  Therefore,
\begin{equation}
 \Val(p_\alpha)=
 \begin{cases}
  \True,&\alpha\in D,\\
  \False,&\alpha\notin D.
 \end{cases}
\label{eq:input-position-value}
\end{equation}

\item \emph{Gate constructions.} If $a$ represents a total Boolean value, a new position $\bar a$ with the
sole move to $a$ represents the complement by
Eq.~\eqref{eq:single-successor-table}.  If $a_1,\ldots,a_m$ represent total
values, create their complements and a position $o$ with moves to all
$\bar a_i$.  Then,
\[
\begin{aligned}
 \Val(o)=\True
 &\iff \exists i\in\{1,\ldots,m\}:\ \Val(\bar a_i)=\False\\
 &\iff \exists i\in\{1,\ldots,m\}:\ \Val(a_i)=\True.
\end{aligned}
\]
If no $a_i$ is true, every successor of $o$ wins and $o$ loses, also when $m=0$. For a conjunction, use the position with sole move to the disjunction of the $\bar a_i$. It wins iff $\neg\exists i\in\{1,\ldots,m\}:\Val(a_i)=\False$, equivalently iff $\forall i\in\{1,\ldots,m\}:\Val(a_i)=\True$. This includes the empty conjunction. The false constant is $z$; the true constant has sole move to $z$.

\item \emph{Structural induction.} In topological order, the input and constant positions give the base case. If the input positions of a gate represent their Boolean values, the case for the corresponding connective gives the output position the value of the gate. Induction therefore covers every gate, including the circuit output. All occurrences of $\alpha$ use the same $p_\alpha$.
\end{proofsteps}
\end{proof}

\begin{lemma}\label{lem:conditional-edge-expansion}
Let $H_D$ be a finite game with Boolean move conditions
$\varphi_e(D)$, using $1$ for unconditional moves. On positions disjoint
from $H_D$, represent the conditions by an acyclic circuit game as in
Lemma~\ref{lem:boolean-game-circuit}, with output $c_e$ for $\varphi_e$
and possibly shared inputs and subcircuits. We obtain $K_D$ by replacing
each possible move $e=(u,v)$ with the fixed moves
\begin{equation}
 u\to g_e\to n_e\to v \qquad\text{and}\qquad g_e\to c_e,
 \label{eq:edge-expansion}
\end{equation}
where $g_e,n_e$ are fresh for each $e$. Then
$\Val_{K_D}(w)=\Val_{H_D}(w)$ for every original position $w$.
\end{lemma}

\begin{proof}
For the given database $D$, let $H=H_D$ and $K=K_D$. In a finite arena $A$, write $A\vdash u^+$ and $A\vdash u^-$ for finite derivations by the rules
\[
 \frac{A\vdash v^-}{A\vdash u^+}\ (v\in\Succ_A(u))
 \quad\text{and}\quad
 \frac{(A\vdash v^+)_{v\in\Succ_A(u)}}{A\vdash u^-}.
\]
The second rule has no premises at a terminal. These judgments give exactly the winning and losing positions. Indeed, induction on the stages in Section~\ref{sec:certificates} gives a derivation for each position when it first enters a stage: a terminal uses the empty rule, a win uses one earlier loss, and a nonterminal loss uses the earlier wins at its successors. Conversely, induct on derivation height. A leaf is a terminal loss at stage zero. For a win with a premise at stage $i$, use stage $i+1$. For a nonterminal loss, take $i$ to be the maximum of the finitely many premise stages; monotonicity places all successors in stage $i$, hence the position is losing at stage $i+1$.

The circuit vertices have only circuit successors. A derivation rooted there therefore uses the same moves before and after expansion. In particular, $K\vdash c_e^+$ for enabled $e$ and $K\vdash c_e^-$ for disabled $e$, by Lemma~\ref{lem:boolean-game-circuit}. The two judgments are disjoint since winning and losing positions are disjoint. For an enabled $e=(u,v)$, the two rules applied at $g_e,n_e$ therefore give
\[
 K\vdash g_e^+\iff K\vdash n_e^-\iff K\vdash v^+
  \quad\text{and}\quad
 K\vdash g_e^-\iff K\vdash n_e^+\iff K\vdash v^-.
\]
For a disabled $e$, its losing successor $c_e$ gives $K\vdash g_e^+$. This proves~\eqref{eq:conditional-move-gate}, including the draw case.

\emph{From $H$ to $K$.} Induct on the height of a derivation at an original position $u$.
\begin{enumerate}
\item For $H\vdash u^+$, choose the premise $H\vdash v^-$ at its enabled edge $e=(u,v)$. The induction hypothesis gives $K\vdash v^-$. The move $n_e\to v$ gives $K\vdash n_e^+$. Both $n_e$ and $c_e$ thus have winning derivations, giving $K\vdash g_e^-$ and then $K\vdash u^+$.
\item For $H\vdash u^-$, consider each possible edge $e=(u,v)$. If $e$ is enabled, its premise gives $K\vdash v^+$ by induction, hence $K\vdash n_e^-$ and $K\vdash g_e^+$. If $e$ is disabled, $K\vdash c_e^-$ gives $K\vdash g_e^+$. These are all successors of $u$ in $K$, so $K\vdash u^-$, also when there are none.
\end{enumerate}
\emph{From $K$ to $H$.} We induct on the height of a derivation at an original position $u$. The premises at an original successor give proper subderivations.
\begin{enumerate}
\item For $K\vdash u^+$, the losing premise is $g_e^-$ for a possible $e=(u,v)$. Its premises give $c_e^+$ and $n_e^+$, so $e$ is enabled and the sole successor of $n_e$ supplies $v^-$. Induction gives $H\vdash v^-$, hence $H\vdash u^+$.
\item For $K\vdash u^-$, each enabled $e=(u,v)$ supplies a winning premise $g_e^+$. Its losing child cannot be $c_e$, since $K\vdash c_e^+$. It is $n_e^-$, whose premise is $v^+$. By induction $H\vdash v^+$ for each enabled successor. Applying the loss rule in $H$ gives $H\vdash u^-$, including when no edge is enabled.
\end{enumerate}
The two translations preserve wins and losses at each original position. Their complement consists of the draws.
\end{proof}

\subsection{Compilation and preservation}
\begin{proof}[Proof of Theorem~\ref{thm:causal-normal-form}]
Let $I=(\PiP,\Dx,\Univ,E,q)$. Use fresh names for all auxiliary positions and predicates, except where an input or subcircuit is explicitly shared.
\begin{proofsteps}
\item \emph{Construction.}
Use the domain relation of Lemma~\ref{lem:two-total-games} with extension $A_I$. Choose the interpretations of $c_0,c_1$ once for $I$. For each $F\subseteq\Univ$, let $H_F$ be the disjoint union of the two games from that lemma on $\Dx\cup F$, with distinguished positions $t$ and $f$. They are total, with
\begin{align}
 \Val_{H_F}(t)=&\True\iff\Ans_I(F)=\True,\nonumber \\
 \Val_{H_F}(f)=&\True\iff\Ans_I(F)=\False.
\label{eq:conditional-game-response}
\end{align}

For each possible goal predicate in the fixed schema, the move-defining programs are determined by $\PiP$ and that predicate. They are nonrecursive and semipositive.  By \Cref{lem:input-circuit,lem:boolean-game-circuit}, construct total circuits for the ground move conditions, sharing inputs across both games. For every $\tau\in\Univ$, reserve a distinct input position $p_\tau$, including when no condition reads $\tau$. Its sole possible move is $\eta(\tau)=(p_\tau,z)$, where $z$ is terminal. Let $j$ be the ground-atom translation in Lemma~\ref{lem:two-total-games}. Let $D_0$ consist exactly of the facts of the two fixed domain relations. At a padded extensional input $\beta$, use
\[
 \begin{cases}
 p_\tau,&\beta=j(\tau),\ \tau\in\Univ,\\
 \text{the true constant},&\beta\in j[\Dx]\cup D_0,\\
 \text{the false constant},&\text{otherwise}.
 \end{cases}
\]
In particular, if $\tau=P$ is nullary, $j(\tau)=P'(c_0)$ reads $p_\tau$ and all $P'(a)$ with $a\ne c_0$ are false. Declare exactly the moves $\eta(\tau)$ mutable, and let $\Univ^{\mathsf g}=\{\eta(\tau)\mid\tau\in\Univ\}$ and $E^{\mathsf g}=\eta[E]$.
Expand the moves of $H_F$ by~\Cref{lem:conditional-edge-expansion}, using the same possible moves and circuits at every state. Add the five fresh positions and fixed moves of Lemma~\ref{lem:three-value-combiner} to the resulting graph. Apart from the designated input moves $\eta(\tau)$, all moves are fixed.

\item \emph{Response preservation.} Let $F\subseteq\Univ$.  Eq.~\eqref{eq:input-position-value} shows that
the shared input position $p_\tau$ is winning exactly when $\tau\in F$. By Lemmas~\ref{lem:input-circuit} and~\ref{lem:boolean-game-circuit}, the circuit outputs give precisely the enabled moves of $H_F$. Lemma~\ref{lem:conditional-edge-expansion} therefore preserves~\eqref{eq:conditional-game-response} at $t,f$. Adding the five new positions does not change any outgoing move of either game. The five-position construction of Lemma~\ref{lem:three-value-combiner} therefore gives $\Val_{G_F}(s)=\Ans_I(F)$. The grounding domain of $\NF(I)$ contains $s$ and the endpoints of every fixed or mutable move. Any additional constructed vertex is isolated at every state. Restricting to this domain preserves the successors of every retained vertex, hence, by induction in the stage recurrence~\ref{eq:stage-zero}--\ref{eq:stages}, its winning and losing stages. Thus
\begin{equation}
 \Ans_{\NF(I)}(\eta[F])
 =\WFM(\PiP_{\Win}\cup D^{\mathsf g}\cup\eta[F])(\Win(s))
 =\Ans_I(F).
\label{eq:normal-form-response}
\end{equation}

The distinct input positions make $\eta$ injective, and $\Univ^{\mathsf g}=\eta[\Univ]$ makes it surjective. Together with $E^{\mathsf g}=\eta[E]$ and~\eqref{eq:normal-form-response}, the bijection satisfies Definition~\ref{def:response-isometry}.

\item \emph{Size.} Let $n=|A_I|$ and $N=n+1$, the size of the auxiliary domain. Let $r$ be the maximum number
of variables in a rule used by the two total-game normal forms, and let
$a$ be the maximum extensional arity.  There are $O(N^a)$ ground input
positions and $O(N^r)$ ground instances of each rule. For a tuple position of arity $h$, a fixed tag has $N^h$ possible positions and at most $N^{2h}$ ordered pairs of positions. The finitely many tags and the arities depend only on the programs for the fixed schema. Grounding an answer rule for the selected tuple likewise takes at most $N^r$ instances. Since $n\geq1$, $N\leq 2n$.  The constructions
in \Cref{lem:input-circuit,lem:boolean-game-circuit,lem:conditional-edge-expansion}
add a constant number of positions and moves
per circuit gate, wire, and game move.  Therefore, for a constant
$c$ depending only on $\PiP$ and the declared schemas, the number of positions and moves is $O(n^c+|\Dx|+|\Univ|)$. Number the positions consecutively. Each endpoint then uses $O(\log(n+2))$ bits, since the schema is fixed and $|\Dx|+|\Univ|$ is polynomial in $n$. Ground rule enumeration and writing the moves and the map $\eta$ take polynomial time in the input length.

\item \emph{Causal preservation.} For $F,F'\subseteq\Univ$, bijectivity of $\eta$ gives
\begin{equation*}
 \eta[F\toggle F']
 =
 \eta[F]\toggle \eta[F']
 \quad\text{and}\quad
 |\eta[F]\toggle \eta[F']|
 =
 |F\toggle F'|.
\end{equation*}
Let $F_0\subseteq\Univ$, let distinct $b,c\in\Bthree$ satisfy
$\Ans_I(F_0)=b$, let $\tau\in\Univ$, and let
$\Gamma\subseteq\Univ\setminus\{\tau\}$.  By
Eq.~\eqref{eq:normal-form-response},
\begin{equation*}
\begin{aligned}
 \Ans_I(F_0\toggle\Gamma)=b
 &\Longleftrightarrow
 \Ans_{\NF(I)}
   (\eta[F_0]\toggle \eta[\Gamma])=b,\\
 \Ans_I(F_0\toggle\Gamma\toggle\{\tau\})=c
 &\Longleftrightarrow
 \Ans_{\NF(I)}
   (\eta[F_0]\toggle \eta[\Gamma]
      \toggle\{\eta(\tau)\})=c.
\end{aligned}
\end{equation*}
The map $\Gamma\mapsto \eta[\Gamma]$ is a cardinality-preserving bijection
between the admissible contingencies in $I[F_0]$ and
$\NF(I)[\eta[F_0]]$.  It follows that the minimum sizes agree, including $\min\varnothing=\Inf$, and that the bijection maps minimizers onto minimizers. Responsibilities agree by~\eqref{eq:responsibility}. If $b=\Undef$, the same two identities give
\[
 \forall\Delta\subseteq\Univ:\quad
 \Ans_I(F_0\toggle\Delta)=c
 \iff \Ans_{\NF(I)}(\eta[F_0]\toggle\eta[\Delta])=c.
\]
Since $|\eta[\Delta]|=|\Delta|$, the minimum intervention sizes for each determinate target agree, as do their minimum and the corresponding families of minimizers.
\end{proofsteps}
\end{proof}

\subsection{Independently Mutable Copies}\label{app:guard-necessity}

\begin{proof}[Proof of Example~\ref{thm:guard-necessity}]
Let $m\geq1$, $\Univ_m=\{a_1,\ldots,a_m\}$, and $E_m=\varnothing$. Let $F\subseteq\Univ_m$. In every positive reduct of the program on copies, the only rule for $b_i$ is $b_i\leftarrow a_i$, so its least model contains $b_i$ iff $a_i\in F$. The positive rule for $q$ thus derives $q$ iff $F=\Univ_m$. In the source it derives $q$ iff $a$ is present. After evaluating these positive rules, write $P_{m,F}$ for the remaining program on $q$, including a fact $q$ when $F=\Univ_m$. For $X\subseteq\{q\}$,
\[
 q\in g_{P_{m,F}}(X)\iff (F=\Univ_m)\vee(q\notin X).
\]
If $F=\Univ_m$, both alternating bounds stabilize at $\{q\}$; otherwise they are $\varnothing$ and $\{q\}$. Thus $\Ans(F)=\True$ iff $F=\Univ_m$ and $\Ans(F)=\Undef$ otherwise. The source has the same response on its two states as the case $m=1$ with $a_1=a$. Its two states have distance one and the toggle of $a$ has responsibility one. The encoding $\iota(\varnothing)=\varnothing$, $\iota(\{a\})=\Univ_m$ preserves the two answers but gives distance $m$.

For $1\le i\le m$ and $\Gamma\subseteq \Univ_m\setminus\{a_i\}$, $\Ans(\Gamma)=\Undef$, and
\[
  \Ans(\Gamma\cup\{a_i\})=\True
  \iff \Gamma\cup\{a_i\}=\Univ_m
  \iff \Gamma=\Univ_m\setminus\{a_i\}.
\]
The unique contingency has size $m-1$ and gives responsibility $1/m$.

\end{proof}

\section{Certificates for game positions}
\label{app:certificates}

Let $F\subseteq\Mu$ be a state of the finite arena on $V$.

\begin{lemma}\label{lem:stage-partition}
For $i\in\Nat$,
$(\Wset_i\subseteq\Wset_{i+1})\wedge(\Lset_i\subseteq\Lset_{i+1})\wedge(\Wset_i\cap\Lset_i=\varnothing)$.
The player to move can force a win from each $v\in\Wset_F$, and the opponent can force a win from each $v\in\Lset_F$.
\end{lemma}

\begin{proof}
The two inclusions follow from the unions in the recurrence. We prove the game outcomes by induction on $i$. At stage zero the winning set is empty and every vertex in $\Lset_0$ is terminal. Assume the outcomes hold at stage $i$. From $v\in\Wset_{i+1}\setminus\Wset_i$ a player can move to $w\in\Lset_i$, where the induction hypothesis gives the opponent of the player at $w$ a winning strategy. At $v\in\Lset_{i+1}\setminus\Lset_i$, every move hands the opponent a position in $\Wset_i$ and hence a winning strategy. Previously classified vertices retain their strategies. The two sets at each stage are disjoint: strategies forcing a win for opposite players would determine a single play won by both. Taking unions proves the claims for $\Wset_F$ and $\Lset_F$.
\end{proof}

\begin{lemma}\label{lem:canonical-ranks}
The triple $(\Wset_F,\Lset_F,\rho_F)$ satisfies the rank conditions of Definition~\ref{def:rank-certificate}.
\end{lemma}

\begin{proof}
By Lemma~\ref{lem:stage-partition}, $\Wset_i\cap\Lset_j\subseteq\Wset_k\cap\Lset_k=\varnothing$ for $i,j\in\Nat$ and $k=\max(i,j)$. Thus $\Wset_F\cap\Lset_F=\varnothing$. For $v\in\Wset_F\cup\Lset_F$ and $i\in\Nat$, first entry in the stage recurrence of Section~\ref{sec:certificates} gives
\[
\begin{aligned}
 (v\in\Wset_F)\wedge(\rho_F(v)=i+1)
  &\ \Longrightarrow\ \exists w\in\Succ_F(v)\cap\Lset_i:\ \rho_F(w)\le i,\\
 (v\in\Lset_F)\wedge(\rho_F(v)=0)
  &\ \Longrightarrow\ \Succ_F(v)=\varnothing,\\
 (v\in\Lset_F)\wedge(\rho_F(v)=i+1)
  &\ \Longrightarrow\ \Succ_F(v)\subseteq\Wset_i.
\end{aligned}
\]
In the last case $\forall w\in\Succ_F(v):\ \rho_F(w)\le i$. Since $\Wset_0=\varnothing$, the three cases exhaust the domain of $\rho_F$ and give clauses~\ref{eq:rank-win} and~\ref{eq:rank-lose} of Definition~\ref{def:rank-certificate}.
\end{proof}

\begin{lemma}\label{lem:rank-soundness}
Let $\Rank=(\Wp,\Wm,\rhoF)$ satisfy the rank conditions at $F$.
Then $\Wp\subseteq\Wset_F$ and $\Wm\subseteq\Lset_F$.
Every vertex in $\Wp$
is winning and every vertex in $\Wm$ is losing.
\end{lemma}

\begin{proof}
For $j\in\Nat$, let $P_j=\{v\in\Wp\mid\rhoF(v)\le j\}$ and $N_j=\{v\in\Wm\mid\rhoF(v)\le j\}$. We prove $P_j\subseteq\Wset_j$ and $N_j\subseteq\Lset_j$ by induction. At $j=0$, a smaller rank is impossible. Thus, clause~\ref{eq:rank-win} of Definition~\ref{def:rank-certificate} gives $P_0=\varnothing$, and clause~\ref{eq:rank-lose} gives $\Succ_F(v)=\varnothing$ for $v\in N_0$. Thus, $N_0\subseteq\Lset_0$. Assume the inclusions at $j$. The rank conditions give
\[
\begin{aligned}
 v\in P_{j+1}\setminus P_j
   &\ \Longrightarrow\ \Succ_F(v)\cap N_j\ne\varnothing
    \ \Longrightarrow\ \Succ_F(v)\cap\Lset_j\ne\varnothing
    \ \Longrightarrow\ v\in\Wset_{j+1},\\
 v\in N_{j+1}\setminus N_j
   &\ \Longrightarrow\ \Succ_F(v)\subseteq P_j
    \ \Longrightarrow\ \Succ_F(v)\subseteq\Wset_j
    \ \Longrightarrow\ v\in\Lset_{j+1}.
\end{aligned}
\]
Monotonicity retains the vertices of $P_j,N_j$. Taking unions over $j$ proves the inclusions. Lemma~\ref{lem:stage-partition} gives the two game outcomes.
\end{proof}

\begin{lemma}\label{lem:draw-structure}
The set $\Dset_F$ satisfies
\begin{align}
 \forall v\in\Dset_F:\quad&
 \Succ_F(v)\cap\Dset_F\neq\varnothing,
\label{eq:canonical-recurrence}\\
 \forall v\in\Dset_F:\quad&
 \Succ_F(v)\setminus\Dset_F\subseteq\Wset_F.
\label{eq:canonical-boundary}
\end{align}
\end{lemma}

\begin{proof}
Let $v\in\Dset_F$. For $w\in\Succ_F(v)\cap\Lset_F$, the stage rule would give $v\in\Wset_{\rho_F(w)+1}$, contradicting $v\in\Dset_F$. Hence $\Succ_F(v)\setminus\Dset_F\subseteq\Wset_F$. If $\Succ_F(v)\cap\Dset_F=\varnothing$, all successors would belong to $\Wset_F$. In that case let $i=\max(\{0\}\cup\{\rho_F(w)\mid w\in\Succ_F(v)\})$. Finiteness and the stage recurrence in Section~\ref{sec:certificates} give $\Succ_F(v)\subseteq\Wset_i$ and $v\in\Lset_{i+1}$, again a contradiction. The maximum also covers $\Succ_F(v)=\varnothing$.
\end{proof}

\begin{lemma}\label{lem:draw-soundness}
If $(Z,\Rank)$ certifies a draw, then $Z\subseteq\Dset_F$.
\end{lemma}

\begin{proof}
Let $Y\subseteq V$. We show that $Y\subseteq\Dset_F$ whenever
\[
 \forall z\in Y:\quad (\Succ_F(z)\cap Y\ne\varnothing)
 \ \wedge\ (\Succ_F(z)\setminus Y\subseteq\Wset_F).
\]
Otherwise choose $z\in Y\cap(\Wset_F\cup\Lset_F)$ of minimum canonical rank. If $z\in\Wset_F$, Lemma~\ref{lem:canonical-ranks} gives a losing successor of smaller rank. This successor cannot be outside $Y$, because all exits are winning. If $z\in\Lset_F$, choose an internal successor; by the same lemma it is winning at smaller rank. Both cases contradict minimality. Now take $Y=Z$: Definition~\ref{def:draw-certificate} gives the internal successor and places every exit in $\Wp\subseteq\Wset_F$ by Lemma~\ref{lem:rank-soundness}.
\end{proof}

\begin{proof}[Proof of Corollary~\ref{cor:canonical-partition}]
\begin{proofsteps}
\item \emph{Well-founded values.}
Identify $\Win(v)$ with $v\in V$. For $X\subseteq V$, let
\[
 g(X)=\{v\in V\mid \exists w\in\Succ_F(v):w\notin X\}.
\]
After evaluating the move facts, the reduct at $X$ derives $\Win(v)$ exactly when a rule for $v$ has a successor outside $X$. Thus $g(X)$ is the set of win atoms in the least model of that reduct. Equation~\eqref{eq:wfs-bounds} gives $K_0=\varnothing$, $U_i=g(K_i)$, $K_{i+1}=g(U_i)$ for $i\in\Nat$. The true and false sets are $\bigcup_i K_i$ and $V\setminus\bigcap_i U_i$, respectively. As in Section~\ref{sec:preliminaries}, $K_i\subseteq K_{i+1}$ and $U_{i+1}\subseteq U_i$. We show
\begin{equation}
 (\Wset_{2i},\Lset_{2i})=(K_i,V\setminus U_i)\qquad(i\in\Nat).
\label{eq:stage-sound-values}
\end{equation}
At $i=0$, $\Wset_0=K_0=\varnothing$ and $\Lset_0=\{v\in V\mid\Succ_F(v)=\varnothing\}=V\setminus g(\varnothing)=V\setminus U_0$. Suppose the equality holds at $i$. For $j\in\Nat$, the stage rules give
$\Wset_{j+1}=\Wset_j\cup g(V\setminus\Lset_j)$ and
$\Lset_{j+1}=\Lset_j\cup(V\setminus g(\Wset_j))$.
Substituting the induction hypothesis, $g(K_i)=U_i$, and $g(U_i)=K_{i+1}$ gives
\[
\begin{aligned}
 \Wset_{2i+1}&=K_i\cup g(U_i)=K_{i+1},&
 \Lset_{2i+1}&=(V\setminus U_i)\cup(V\setminus g(K_i))=V\setminus U_i,\\
 \Wset_{2i+2}&=K_{i+1}\cup g(U_i)=K_{i+1},&
 \Lset_{2i+2}&=(V\setminus U_i)\cup(V\setminus g(K_{i+1}))=V\setminus U_{i+1}.
\end{aligned}
\]
The unions reduce by $K_i\subseteq K_{i+1}$ and $U_{i+1}\subseteq U_i$.
Induction gives~\eqref{eq:stage-sound-values}. Taking unions over the even stages, since each odd stage is contained in the next even stage, gives
$\Wset_F=\bigcup_i K_i$ and $\Lset_F=V\setminus\bigcap_i U_i$.
These are the true and false sets of the alternating fixpoint, so precisely the vertices in their complement are undefined.

\item \emph{Rank bound.} If $\Wset_{i+1}=\Wset_i$ and $\Lset_{i+1}=\Lset_i$, then
the recurrence clauses yield the same pair at every later stage.  Let $r$ be a
canonical rank. If $\Lset_0=\varnothing$, every vertex has a successor and the recurrence gives $\Wset_1=\Lset_1=\varnothing$, contradicting the existence of $r$. The initial losing stage and stages $1,\ldots,r$ each add a vertex not ranked at any earlier stage; otherwise, the iteration would stabilize before
stage $r$.  Therefore, $r+1\leq|V|$ and $r\leq|V|-1$.

\item \emph{Greatest draw region.} The inclusion proved for arbitrary $Y$ in Lemma~\ref{lem:draw-soundness} places every set satisfying the two conditions in $\Dset_F$. Lemma~\ref{lem:draw-structure} shows that $\Dset_F$ itself satisfies them.
\end{proofsteps}
\end{proof}

\begin{proof}[Proof of Theorem~\ref{thm:certificate-completeness}]
Let $\Rank_F=(\Wset_F,\Lset_F,\rho_F)$. Lemma~\ref{lem:canonical-ranks} gives its rank conditions. For $b\in\{\True,\False\}$, take $X_b=\Wset_F$ if $b=\True$ and $X_b=\Lset_F$ if $b=\False$. Corollary~\ref{cor:canonical-partition} and Lemma~\ref{lem:rank-soundness} give, for any certificate triple $\Rank$,
\[
\begin{aligned}
 \Val_F(q)=b &\ \Longrightarrow\ q\in X_b\ \Longrightarrow\ \Rank_F\text{ certifies }b\text{ at }q,\\
 \Rank\text{ certifies }b\text{ at }q &\ \Longrightarrow\ q\in X_b\ \Longrightarrow\ \Val_F(q)=b.
\end{aligned}
\]
For $b=\Undef$, a draw certificate places $q$ in $Z\subseteq\Dset_F$ by Lemma~\ref{lem:draw-soundness}. Conversely, $\Val_F(q)=\Undef$ gives $q\in\Dset_F$, disjoint from $\Wset_F\cup\Lset_F$. Lemma~\ref{lem:draw-structure} verifies both successor conditions, so $(\Dset_F,\Rank_F)$ certifies a draw.
\end{proof}

\section{Terms and minimum contingencies}
\label{app:requirement-normal-form}

\begin{lemma}\label{lem:rank-requirements}
For every $F\subseteq\Mu$ and $b\in\{\True,\False\}$,
$\Ans(F)=b\ \Longleftrightarrow\ \exists\Req\in\RawReqs_b(q):F\models\Req$.
\end{lemma}

\begin{proof}
\emph{Soundness.} Let $C=(\Wp,\Wm,\rhoF,\omega)\in\Certs_b(q)$ and $F\models\ReqOf(C)$. For $v\in\Wp$, the selected move $(v,\omega(v))\in\Pot$ is either fixed or lies in $\Req_C^+\subseteq F$. Hence
\begin{equation}
 \omega(v)\in\Succ_F(v)\cap\Wm\ \wedge\ \rhoF(\omega(v))<\rhoF(v).
\label{eq:req-rank-win}
\end{equation}
For $v\in\Wm$ and $w\in\Succ_F(v)$, a fixed move $(v,w)$ satisfies~\eqref{eq:rank-requirement-fixed}. A mutable move belongs to $F$ and hence lies outside $\Req_C^-$. By the definition of $\Req_C^-$, in either case
\begin{equation}
 w\in\Wp\ \wedge\ \rhoF(w)<\rhoF(v).
\label{eq:req-mutable-lose}
\end{equation}
Thus the triple satisfies both rank conditions. As $C\in\Certs_b(q)$, membership of $q$ in the corresponding set certifies $\Ans(F)=b$ by Theorem~\ref{thm:certificate-completeness}.

\emph{Completeness.} If $\Ans(F)=b$, Corollary~\ref{cor:canonical-partition} places $q$ in $\Wset_F$ or $\Lset_F$ according to $b$. Take $\Wp=\Wset_F$, $\Wm=\Lset_F$ and $\rhoF=\rho_F$. For each $v\in\Wp$, Lemma~\ref{lem:canonical-ranks} supplies $\omega(v)\in\Wm$ with
\begin{equation}
 (v,\omega(v))\in M_F\ \wedge\ \rhoF(\omega(v))<\rhoF(v).
\label{eq:req-canonical-witness}
\end{equation}
For $C=(\Wp,\Wm,\rhoF,\omega)$, each mutable witness lies in $M_F\cap\Mu=F$, giving $\Req_C^+\subseteq F$. By Lemma~\ref{lem:canonical-ranks}, every move in $M_F$ from $\Wm$ leads to a lower-ranked vertex in $\Wp$. The fixed moves therefore satisfy~\eqref{eq:rank-requirement-fixed}, and $\Req_C^-\cap F=\varnothing$.
The canonical ranks are at most $|V|-1$ by Corollary~\ref{cor:canonical-partition}. Thus $C\in\Certs_b(q)$ and $F\models\ReqOf(C)$.
\end{proof}

\begin{lemma}\label{lem:draw-requirements}
For every $F\subseteq\Mu$,
$\quad\Ans(F)=\Undef\ \Longleftrightarrow\ \exists\Req\in\RawReqs_{\Undef}(q):F\models\Req$.
\end{lemma}

\begin{proof}
Let $D=(Z,C,\zeta)\in\Certs_{\Undef}(q)$, where $C=(\Wp,\Wm,\rhoF,\omega)$. Let $J=\{(z,\zeta(z))\mid z\in Z\}$ and $B=Z\times(V\setminus(Z\cup\Wp))$. By~\eqref{eq:draw-requirement-fixed}, $J\subseteq\Pot$ and $B\cap\Mx=\varnothing$. Since $M_F=\Mx\cup F$, $F\subseteq\Mu$, and $\Mx\cap\Mu=\varnothing$,
\[
\begin{aligned}
 J\cap\Mu\subseteq F &\iff J\subseteq M_F,\\
 (B\cap\Mu)\cap F=\varnothing &\iff B\cap M_F=\varnothing .
\end{aligned}
\]
Thus the definition of $\ReqOf(D)$ gives
\[
 F\models\ReqOf(D)\iff
 (F\models\ReqOf(C))\ \wedge\ (J\subseteq M_F)\ \wedge\ (B\cap M_F=\varnothing).
\]
If $F\models\ReqOf(D)$, \eqref{eq:req-rank-win}--\eqref{eq:req-mutable-lose} give the rank conditions. For each $z\in Z$, the second conjunct gives $\zeta(z)\in\Succ_F(z)\cap Z$, and the third gives $\Succ_F(z)\setminus Z\subseteq\Wp$. Since $q\in Z$ and $Z\cap(\Wp\cup\Wm)=\varnothing$, Theorem~\ref{thm:certificate-completeness} yields $\Ans(F)=\Undef$.

Conversely, if $\Ans(F)=\Undef$, take $Z=\Dset_F$, so $q\in Z$ by Corollary~\ref{cor:canonical-partition}. Take $\Wp=\Wset_F$, $\Wm=\Lset_F$ and $\rhoF=\rho_F$. Choose $\omega$ by~\eqref{eq:req-canonical-witness} and $\zeta(z)\in\Succ_F(z)\cap Z$ for $z\in Z$ by~\eqref{eq:canonical-recurrence}. Let $C=(\Wp,\Wm,\rhoF,\omega)$ and $D=(Z,C,\zeta)$, with $J$ and $B$ denoting the corresponding sets of moves. The construction in Lemma~\ref{lem:rank-requirements} gives~\eqref{eq:rank-requirement-witness}--\eqref{eq:rank-requirement-fixed} and $F\models\ReqOf(C)$ independently of $q$. The choice of $\zeta$ gives $J\subseteq M_F\subseteq\Pot$, and~\eqref{eq:canonical-boundary} gives $B\cap M_F=\varnothing$, hence $B\cap\Mx=\varnothing$. Thus $D\in\Certs_{\Undef}(q)$ and the equivalence gives $F\models\ReqOf(D)$.
\end{proof}

\begin{lemma}\label{lem:requirement-flip}
For every term $\Req$, state $F\subseteq\Mu$ and $\tau\in\Mu$, 
\[F\toggle\{\tau\}\models\Req\Longleftrightarrow F\models\Flip_\tau(\Req).\]
\end{lemma}

\begin{proof}
Let $F'=F\toggle\{\tau\}$ and $\Req'=\Flip_\tau(\Req)$. For $\sigma\in\Mu\setminus\{\tau\}$, membership in $F'$ and both signs in $\Req'$ agree with those in $F$ and $\Req$, respectively. If $\tau\in\Req^+$, the required tests are $\tau\in F'$ and $\tau\notin F$, respectively. If $\tau\in\Req^-$, they are $\tau\notin F'$ and $\tau\in F$. Both pairs are equivalent by the definition of $F'$. If $\tau\notin\Req^+\cup\Req^-$, neither term tests $\tau$. Thus $F'\models\Req\Leftrightarrow F\models\Req'$.
\end{proof}

\begin{proof}[Proof of Theorem~\ref{thm:structured-completeness}]
By Lemmas~\ref{lem:rank-requirements} and~\ref{lem:draw-requirements}, $\Ans(F)=b$ exactly when $F$ satisfies a term $\Req'\in\RawReqs_b(q)$. Among terms $\Req\in\RawReqs_b(q)$ with $\Req\preceq\Req'$, choose one with the fewest literals; the family is finite and contains $\Req'$. A strict predecessor would have fewer literals, so $\Req\in\Reqs_b(q)$. Since the literals of $\Req$ form a subset of those of $\Req'$, $F\models\Req'$ implies $F\models\Req$. Conversely, $\Reqs_b(q)\subseteq\RawReqs_b(q)$, proving both directions.
\end{proof}

\begin{proof}[Proof of Theorem~\ref{thm:requirement-normal-form}]
For a term $\Req$ let $B=\Req^+\cup\Req^-$ and $D=(\Req^+\setminus E)\cup(\Req^-\cap E)$. Hence $D\subseteq B$ and $|D|=\Cost_E(\Req)$. For $\Gamma\subseteq\Mu$,
\[
\begin{aligned}
 E\toggle\Gamma\models\Req
 &\iff (\Gamma\cap\Req^+=\Req^+\setminus E)\ \wedge\
          (\Gamma\cap\Req^-=\Req^-\cap E)\\
 &\iff \Gamma\cap B=D.
\end{aligned}
\]
On $\Req^+$, precisely the missing positive moves must change. On $\Req^-$, precisely the present negative moves must change. The last equivalence uses $\Req^+\cap\Req^-=\varnothing$.

\emph{From a contingency to a pair.}
Let $\Gamma$ be a contingency and $F=E\toggle\Gamma$. Theorem~\ref{thm:structured-completeness} supplies $\Req_b\in\Reqs_b(q)$ and $\Req_c\in\Reqs_c(q)$ with $F\models\Req_b$ and $F\toggle\{\tau\}\models\Req_c$. By Lemma~\ref{lem:requirement-flip}, $F\models\Flip_\tau(\Req_c)$. The two terms have a common model and hence are compatible. Since $\tau\notin\Gamma$, their conjunction $\Req$ agrees with $E$ at $\tau$. For this admissible pair, the equivalence gives
\begin{equation}
 D=\Gamma\cap B\subseteq\Gamma\quad\text{and}\quad\Cost_E(\Req)=|D|\le|\Gamma|.
\label{eq:req-first-bound}
\end{equation}

\emph{From a pair to a contingency.}
For an admissible pair, let $\Req=\Req_b\ReqJoin\Flip_\tau(\Req_c)$. Agreement at $\tau$ gives $\tau\notin D$. Since $D\cap B=D$, the state
\begin{equation}
 F^*=E\toggle D=(E\cup\Req^+)\setminus\Req^-
\label{eq:req-nearest-state}
\end{equation}
satisfies $\Req$. Thus $F^*\models\Req_b$ and $F^*\toggle\{\tau\}\models\Req_c$ by Lemma~\ref{lem:requirement-flip}. Theorem~\ref{thm:structured-completeness} gives $\Ans(F^*)=b$ and $\Ans(F^*\toggle\{\tau\})=c$. Hence $D$ is a contingency of size $\Cost_E(\Req)$.

The two constructions show that the families are empty together, giving $\Inf$ on both sides of~\eqref{eq:requirement-factorization}. Otherwise they are finite and nonempty. Inequality~\eqref{eq:req-first-bound} gives the pair minimum as a lower bound on contingency size. Applying~\eqref{eq:req-nearest-state} to a pair of least cost attains that bound. For a minimum $\Gamma$, the first construction gives $D\subseteq\Gamma$ and $|D|=|\Gamma|$, hence $D=\Gamma$. Conversely, the second construction realizes every minimum-cost pair.
\end{proof}

\section{Propagation from the Updated Move}\label{app:resolution-tree}

Let $G=(V,M)$ and $G'=(V,M')$ differ only in the membership of $(x,y)$. We write $(\Wset,\Lset,\Dset)$ and $(\Wset',\Lset',\Dset')$ for their respective canonical partitions. Let
$\rho'$ be the canonical rank on $\Wset'\cup\Lset'$, and let $A=\Wset'\setminus\Wset$ and $B=\Lset'\setminus\Lset$.
\begin{lemma}\label{lem:continuation}
For every $v\in V\setminus\{x\}$,
\begin{align}
 v\in A
 &\,\Longrightarrow\,
 \exists w\in\Succ_{G'}(v)\cap B:\ 
 \rho'(w)<\rho'(v),
\label{eq:continue-win}\\
 v\in B
 &\,\Longrightarrow\,
 \exists w\in\Succ_{G'}(v)\cap A:\ 
 \rho'(w)<\rho'(v).
\label{eq:continue-lose}
\end{align}
\end{lemma}

\begin{proof}
For the old game, the stage rules yield
\[
 v\in\Wset\iff\Succ_G(v)\cap\Lset\ne\varnothing\quad\text{and}\quad
 v\in\Lset\iff\Succ_G(v)\subseteq\Wset .
\]
Necessity follows from Lemma~\ref{lem:canonical-ranks}. A losing successor of rank $i$ puts $v$ in $\Wset_{i+1}$. If all successors are winning, let $i$ be the maximum of $0$ and their ranks. Then $\Succ_G(v)\subseteq\Wset_i$ gives $v\in\Lset_{i+1}$, including the terminal case.

For $v\ne x$, the changed move leaves a different vertex, hence
\begin{equation}
 \Succ_G(v)=\Succ_{G'}(v).
\label{eq:unchanged-successors}
\end{equation}
If $v\in A$, choose $w\in\Succ_{G'}(v)\cap\Lset'$ with $\rho'(w)<\rho'(v)$. Since $v\notin\Wset$, its old successors avoid $\Lset$. By~\eqref{eq:unchanged-successors}, $w\in\Lset'\setminus\Lset=B$, proving~\eqref{eq:continue-win}.

If $v\in B$, then $v\notin\Lset$, so $\Succ_G(v)\nsubseteq\Wset$. Choose $w\in\Succ_G(v)\setminus\Wset$. By~\eqref{eq:unchanged-successors} and the new losing rank condition, $w\in\Wset'$ and $\rho'(w)<\rho'(v)$. Thus $w\in A$, proving~\eqref{eq:continue-lose}.
\end{proof}

\begin{lemma}\label{lem:chain-end}
Every vertex $v\in A\cup B$ has in $(V,M')$ a simple path $v=v_0,v_1,\ldots,v_m=x$ whose ranks strictly decrease and whose vertices alternate between $A$ and $B$.
\end{lemma}

\begin{proof}
Let $v_0=v$. Whenever $v_i\ne x$, choose $v_{i+1}$ by~\eqref{eq:continue-win} if $v_i\in A$, or by~\eqref{eq:continue-lose} if $v_i\in B$. The sets alternate, $(v_i,v_{i+1})\in M'$, and $\rho'(v_{i+1})+1\le\rho'(v_i)$. Induction gives $\rho'(v_i)+i\le\rho'(v_0)$. At most $\rho'(v_0)$ steps are possible, and Lemma~\ref{lem:continuation} permits a next step at every vertex except $x$. Thus the path ends at $x$, with length zero if $v=x$. Strict decrease excludes repeated vertices.
\end{proof}

\begin{lemma}\label{lem:pivot-polarity}
Assume $A\cup B\neq\varnothing$.
If $M'=M\cup\{(x,y)\}$, then $x\in A$, $y\in\Lset\cap\Lset'$, and $\rho'(y)<\rho'(x)$.
If $M'=M\setminus\{(x,y)\}$, then $x\in B$ and $y\notin\Wset$.
\end{lemma}

\begin{proof}
Let $S=A\cup B\ne\varnothing$. By Lemma~\ref{lem:chain-end}, $x\in S$, and every $v\in S\setminus\{x\}$ has a nonempty decreasing path to $x$. Thus $\rho'(v)>\rho'(x)$ on $S\setminus\{x\}$. In particular, for a determinate $w$ with $\rho'(w)<\rho'(x)$,
\[
 (w\in\Wset'\Rightarrow w\in\Wset)\ \wedge\
 (w\in\Lset'\Rightarrow w\in\Lset).
\]
We use the old successor characterizations proved in Lemma~\ref{lem:continuation}.

\emph{Insertion.} If $x\in B$, every successor in $M'$ wins at a smaller rank. Hence $\Succ_G(x)\subseteq\Succ_{G'}(x)\subseteq\Wset$, so $x\in\Lset$, contrary to $x\in B$. Thus $x\in A$. Choose $z\in\Succ_{G'}(x)\cap\Lset'$ with $\rho'(z)<\rho'(x)$. Then $z\in\Lset$, but $x\notin\Wset$ implies $\Succ_G(x)\cap\Lset=\varnothing$. Thus $z\in\Succ_{G'}(x)\setminus\Succ_G(x)=\{y\}$.

\emph{Deletion.} If $x\in A$, a lower-ranked losing successor in $M'$ belongs to $\Lset$ and remains an old successor, giving $x\in\Wset$, a contradiction. Thus $x\in B$. Every remaining successor wins at a smaller new rank, so $\Succ_{G'}(x)\subseteq\Wset$. Since $x\notin\Lset$, $\Succ_G(x)\nsubseteq\Wset$. From $\Succ_G(x)=\Succ_{G'}(x)\cup\{y\}$ we obtain $y\notin\Wset$.
\end{proof}

\begin{proof}[Proof of Theorem~\ref{thm:resolution-tree}]
\begin{proofsteps}
\item \emph{Newly winning or losing vertices.} Since $q\in\Dset$ and $q\in\Wset'\cup\Lset'$, we have
\begin{equation}
 q\in
 \begin{cases}
 A,&q\in\Wset',\\
 B,&q\in\Lset'.
 \end{cases}
\label{eq:q-difference}
\end{equation}
Thus, $A\cup B\neq\varnothing$, and Lemma~\ref{lem:chain-end} places $x$ in $A\cup B$.

\item \emph{In-arborescence.} For each $v\in(A\cup B)\setminus\{x\}$, choose one successor
$\Parent(v)$ supplied by Eq.~\eqref{eq:continue-win} or
Eq.~\eqref{eq:continue-lose}.  The choice gives $(v,\Parent(v))\in M'$ and $\rho'(\Parent(v))<\rho'(v)$, together with the parent conditions in Theorem~\ref{thm:resolution-tree}.  Let $T$ contain the chosen moves.
Every vertex other than $x$ has exactly one outgoing move in $T$.  The root
$x$ has none.  For any $v\in A\cup B$, iteration gives $v,\Parent(v),\Parent^2(v),\ldots$.
At each step the rank decreases in $\Nat$, so the sequence reaches a vertex at which $\Parent$ is undefined.  The domain of $\Parent$ is $(A\cup B)\setminus\{x\}$; hence, the terminal vertex is $x$.  Therefore,
every vertex has a unique directed path in $T$ to $x$. The underlying graph is connected and has $|A\cup B|-1$ edges, one per nonroot vertex; hence it is a tree directed towards $x$.
\end{proofsteps}
\end{proof}

\begin{proof}[Proof of Corollary~\ref{cor:resolution-root-parity}]
Lemma~\ref{lem:pivot-polarity} proves the assertions about $x$ and $y$. Let $\Parent$ be any parent map supplied by Theorem~\ref{thm:resolution-tree}. For $v\in A\cup B$, let $d(v)$ be the length of the unique parent path from $v$ to $x$. We prove by induction on $d(v)$ that $v$ and $x$ belong to the same one of $A,B$ exactly when $d(v)$ is even. At distance zero, $v=x$. Suppose the assertion holds at distance $j$. For $d(v)=j+1$, the parent $w=\Parent(v)$ satisfies $d(w)=j$. By Theorem~\ref{thm:resolution-tree}, $v$ and $w$ belong to opposite sets. Thus, $v$ and $x$ belong to the same set exactly when $j$ is odd, equivalently when $j+1$ is even. Under insertion $x\in A$, and under deletion $x\in B$. Since $q\in A$ exactly when $q\in\Wset'$ by~\eqref{eq:q-difference}, the two parity conclusions follow.
\end{proof}

\section{Complexity Proofs}\label{app:complexity}

\begin{proposition}\label{prop:np-membership}
Let $\PiP$ be a safe normal Datalog program treated as constant in data
complexity.  For every pair of distinct $b,c\in\Bthree$, the problem
$\Cause_{b\to c}[\PiP]$ belongs to NP in data complexity; the same holds for
$\DetRadius[\PiP]$.
\end{proposition}

\begin{proof}
The verifier checks that the input encodes an intervention instance as in Definition~\ref{def:instance} and a threshold $k\in\Nat$.
\begin{proofsteps}
\item
For $\Cause_{b\to c}[\PiP]$, reject if $\tau\notin\Univ$. Otherwise guess
$\Gamma\subseteq\Univ\setminus\{\tau\}$ with $|\Gamma|\leq k$ and accept
exactly when
\[
 (\Ans(E)=b)\ \wedge\
 (\Ans(E\toggle\Gamma)=b)\ \wedge\
 (\Ans(E\toggle\Gamma\toggle\{\tau\})=c).
\]
\item
For $\DetRadius[\PiP]$, guess $\Delta\subseteq\Univ$ with
$|\Delta|\leq k$ and accept exactly when
\[
 (\Ans(E)=\Undef)\ \wedge\
 (\Ans(E\toggle\Delta)\in\{\True,\False\}).
\]
\end{proofsteps}
Both guesses use $|\Univ|$ bits. For a fixed program and a grounding domain of size $a$, let $v$ be the maximum number of variables in a rule (zero if there are no variables). Each rule has at most $a^v$ instances. The fixed schema gives polynomially many ground atoms. Each positive reduct reaches its least model after at most one strict increase per ground atom; the same bound holds for strict increases of the alternating lower bound in~\eqref{eq:wfs-bounds}. Thus grounding, the cardinality test, and all three query evaluations take polynomial time. The two acceptance predicates are precisely the respective decision conditions in Section~\ref{subsec:decision-problems}.
\end{proof}

\subsection{\texorpdfstring{Reductions from Vertex Cover}{Reductions from Vertex Cover}}\label{app:selector}

\begin{lemma}\label{thm:selector}
For $C\subseteq\Idx$, the construction with self-loops satisfies
\begin{equation}
 \Ans(E_{\Hyp}^{u}\toggle\Delta_C)=
 \begin{cases}\False,&\bigcup_{j\in C}S_j=X,\\
 \Undef,&\text{otherwise,}\end{cases}
\label{eq:selector-response}
\end{equation}
Thus, $\DetRad(E_{\Hyp}^{u})=\DirRad_{\False}(E_{\Hyp}^{u})=\Cov(\Hyp)$ and $ \DirRad_{\True}(E_{\Hyp}^{u})=\Inf$. For the singleton candidate $\tau$ in $\Hyp^+$, query vertices $q$ and $p$ give, respectively,
\begin{align}
 \Kappa_{E_{\Hyp^+}^{u}}^{\Undef\to\False}(\tau)&=\Cov(\Hyp),
\label{eq:selector-false-cause}\\
 \Kappa_{E_{\Hyp^+}^{u}}^{\Undef\to\True}(\tau)&=\Cov(\Hyp).
\label{eq:selector-true-cause}
\end{align}
\end{lemma}

\begin{proof}
Let $C\subseteq\Idx$.
\begin{proofsteps}
\item \emph{Vertex values.} Let $F=E_{\Hyp}^{u}\toggle\Delta_C$.  For $j\in\Idx$, the only successor of $d_j$ is $d_j$. Thus every play from $d_j$ is infinite, and $d_j\in\Dset_F$ by the game semantics of Eq.~\eqref{eq:win-intro}.

Each $s_j$ has at most one successor. Thus
\begin{itemize}
    \item $j\in C
 \Longleftrightarrow \alpha_j\notin F
 \Longleftrightarrow \Succ_F(s_j)=\varnothing
 \Longleftrightarrow s_j\in\Lset_F,$
 \item $j\notin C\Longleftrightarrow \alpha_j\in F
 \Longleftrightarrow \Succ_F(s_j)=\{d_j\}
 \Longleftrightarrow s_j\in\Dset_F.$
\end{itemize}

Since $\Succ_F(e_x)=\{s_j\mid x\in S_j\}$ is nonempty,
\begin{itemize}
\item $x\in\bigcup_{j\in C}S_j
 \Longleftrightarrow
 \Succ_F(e_x)\cap\Lset_F\neq\varnothing
 \Longleftrightarrow e_x\in\Wset_F$,
 \item $x\notin\bigcup_{j\in C}S_j
 \Longleftrightarrow
 \Succ_F(e_x)\subseteq\Dset_F
 \Longleftrightarrow e_x\in\Dset_F$.
\end{itemize}

Finally, $q$ has all $e_x$ as successors and no losing successor,
and therefore
\begin{equation*}
 \bigcup_{j\in C}S_j=X
 \Longleftrightarrow
 \Succ_F(q)\subseteq\Wset_F
 \Longleftrightarrow q\in\Lset_F.
\end{equation*}
Otherwise, $q$ has a draw successor and is drawn.  These
equivalences prove Eq.~\eqref{eq:selector-response}; the unique representation of
every intervention as $\Delta_C$ gives $\DetRad(E_{\Hyp}^{u})=\DirRad_{\False}(E_{\Hyp}^{u})=\Cov(\Hyp)$ and $ \DirRad_{\True}(E_{\Hyp}^{u})=\Inf$.

\item \emph{Candidate transitions.} For the observation $E_{\Hyp^+}^{u}$ and $C\subseteq\Idx$, let $F=E_{\Hyp^+}^{u}\toggle\Delta_C$ and $F'=F\setminus\{\tau\}$. The sole move from $e_\ast$ is to $s_{j_\ast}$. Before deletion, $s_{j_\ast}$ moves only to the drawn self-loop; after deletion it is terminal. Hence
\[
 (\Val_F(s_{j_\ast}),\Val_F(e_\ast))=(\Undef,\Undef) \quad\text{and}\quad
 (\Val_{F'}(s_{j_\ast}),\Val_{F'}(e_\ast))=(\False,\True).
\]
For $x\in X$, the successors of $e_x$ and their moves are unchanged, so $e_x$ retains its value in the game for $\Hyp$. Writing $U_C=\bigcup_{j\in C}S_j$ gives
\[
 (\Val_F(q),\Val_{F'}(q))=
 \begin{cases}
 (\Undef,\False),&U_C=X,\\
 (\Undef,\Undef),&U_C\ne X .
 \end{cases}
\]
At $p$ the sole successor $q$ complements each value by~\eqref{eq:single-successor-table}. Every contingency candidate is a unique $\Delta_C$ and $|\Delta_C|=|C|$. Taking minima gives~\eqref{eq:selector-false-cause} and~\eqref{eq:selector-true-cause}.
\end{proofsteps}
\end{proof}

\begin{lemma}\label{lem:total-selector}
For the construction with terminals and $\tau=\alpha_{j_\ast}$, the query positions $q$ and $p$ give, respectively,
\begin{align}
    \Kappa_{E_{\Hyp^+}^{t}}^{\True\to\False}(\tau) &=\Cov(\Hyp), \nonumber\\
 \Kappa_{E_{\Hyp^+}^{t}}^{\False\to\True}(\tau) &=\Cov(\Hyp).
\label{eq:total-selector-causes}
\end{align}
Minimum contingencies are in bijection with
minimum covers of $\Hyp$.
\end{lemma}

\begin{proof}
Let $C\subseteq\Idx$.
\begin{proofsteps}
\item \emph{Vertex values.} In the game for $\Hyp$, let $F=E_{\Hyp}^{t}\toggle\Delta_C$. Every $\ell_j$ is terminal and losing,
and
\begin{itemize}
 \item $j\in C
 \Longleftrightarrow \alpha_j\notin F
 \Longleftrightarrow \Succ_F(s_j)=\varnothing
 \Longleftrightarrow s_j\in\Lset_F,$
 \item $j\notin C
 \Longleftrightarrow \alpha_j\in F
 \Longleftrightarrow s_j\in\Wset_F.$
\end{itemize}
For every $x\in X$,
\begin{equation*}
 x\in\bigcup_{j\in C}S_j
 \,\Longleftrightarrow\,
 \Succ_F(e_x)\cap\Lset_F\neq\varnothing
 \,\Longleftrightarrow\, e_x\in\Wset_F.
\end{equation*}
If $x\notin\bigcup_{j\in C}S_j$, every successor of $e_x$ is
winning, and $e_x\in\Lset_F$.  Therefore, all positions are determinate and
\begin{equation*}
 \bigcup_{j\in C}S_j=X
 \,\Longleftrightarrow\,
 \Succ_F(q)\subseteq\Wset_F
 \,\Longleftrightarrow\, q\in\Lset_F;
\end{equation*}
otherwise, $q$ has a losing successor and belongs to $\Wset_F$.

\item \emph{Candidate transitions.} In the game for $\Hyp^+$, let $F=E_{\Hyp^+}^{t}\toggle\Delta_C$ and $F'=F\setminus\{\tau\}$. The original element vertices have the same successors and values as in the game for $\Hyp$. The move to the losing terminal $\ell_{j_\ast}$ is present in $F$ and absent in $F'$. As $e_\ast$ moves only to $s_{j_\ast}$,
\[
 (\Val_F(s_{j_\ast}),\Val_F(e_\ast))=(\True,\False)\quad\text{and}\quad
 (\Val_{F'}(s_{j_\ast}),\Val_{F'}(e_\ast))=(\False,\True).
\]
Thus $q$ wins in $F$. At $F'$, all successors of $q$ win exactly when $C$ covers $X$; if $C$ misses $x$, $e_x$ loses and $q$ wins. Since $p$ moves only to $q$,
\[
 \Val_F(q)=\True,\quad\Val_F(p)=\False,\quad\text{and}\quad
 \bigcup_{j\in C}S_j=X
 \iff \Val_{F'}(q)=\False
 \iff \Val_{F'}(p)=\True .
\]
Other than $\tau$, precisely the $\alpha_j$ are mutable. Hence $C\mapsto\Delta_C$ restricts to a cardinality-preserving bijection from covers to contingencies for either query. Taking the minimum gives~\eqref{eq:total-selector-causes}.
\end{proofsteps}
\end{proof}

\begin{lemma}\label{lem:draw-opening}
For the construction with a conditional move, query vertices $r$ and $p$ give, respectively,
\begin{align}
 \Kappa_{E_{\Hyp}^{o}}^{\False\to\Undef}(\tau) &=\Cov(\Hyp),\nonumber\\
 \Kappa_{E_{\Hyp}^{o}}^{\True\to\Undef}(\tau) &=\Cov(\Hyp).
\label{eq:draw-opening-causes}
\end{align}
Minimum contingencies are in bijection with
minimum covers of $\Hyp$.
\end{lemma}

\begin{proof}
Let $C\subseteq\Idx$ index the inserted moves $\beta_j$.
\begin{proofsteps}
\item \emph{Circuit value.} For $j\in\Idx$, the total input position $p_j$ has a move to a private terminal exactly when $j\in C$. Thus, $\Val(p_j)=\True$ iff $j\in C$. Let $c_x$ represent the disjunction of $p_j$ over $j$ with $x\in S_j$, and let $c$ represent the conjunction of $c_x$ over $x\in X$. These gates form an acyclic Boolean circuit with total inputs. By Lemma~\ref{lem:boolean-game-circuit},
\[
 \Val(c)=\True
 \ \Longleftrightarrow\ 
 \bigl[\forall x\in X\ \exists j\in C\ (x\in S_j)\bigr]
 \ \Longleftrightarrow\ 
 \bigcup_{j\in C} S_j=X.
\]
\item \emph{Conditional move.} Let the only move from $d$ be a fixed self-loop. Then $\Val(d)=\Undef$. For $F=\Delta_C$ and $F'=F\cup\{\tau\}$, let $p_\tau$ be the total input for $\tau$ and $h$ the conjunction of $c$ and $p_\tau$. Then,
\begin{align*}
    \Val_F(h)=&\False,\\
 \Val_{F'}(h)=&\True\iff\bigcup_{j\in C}S_j=X.
\end{align*}
Let $g$ be the sole successor of $r$ in the implementation of the conditional move to $d$ controlled by $h$, and $p$ the one-move predecessor of $r$. For $K\in\{F,F'\}$, equations~\eqref{eq:conditional-move-gate} and~\eqref{eq:single-successor-table} give
\[
 (\Val_K(g),\Val_K(r),\Val_K(p))=
 \begin{cases}
 (\True,\False,\True),&\Val_K(h)=\False,\\
 (\Undef,\Undef,\Undef),&\Val_K(h)=\True .
 \end{cases}
\]
Thus $C$ covers $X$ iff $r$ changes from $\False$ to $\Undef$, equivalently iff $p$ changes from $\True$ to $\Undef$. The observed values are the respective source values also for $C=\varnothing$. Since every candidate contingency consists solely of the moves $\beta_j$ with $j\in\Idx$, $C\mapsto\Delta_C$ gives a bijection preserving cardinality and hence both minima in~\eqref{eq:draw-opening-causes}.
\end{proofsteps}
\end{proof}

\subsubsection{\texorpdfstring{Degree Reduction and Cover Correspondence}{Degree Reduction and Cover Correspondence}}\label{app:selector-degree}

Let $H=(V_H,E_H)$ be a finite simple undirected graph, and let
$\VC(H)$ denote the minimum vertex-cover size of $H$.  If
$E_H=\varnothing$, then $\VC(H)=0$ and the empty set is the unique minimum
vertex cover. Thus, for Vertex Cover with nonnegative threshold, we may replace an edgeless decision instance by a single edge with threshold $1$. Assume henceforth that $E_H\neq\varnothing$.  Delete the isolated vertices of $H$;
every minimum vertex cover omits them; hence, the operation preserves the optimum
and the family of minimum covers. Let $X=E_H$, $\Idx=V_H$, $S_v=\{e\in E_H\mid v\in e\}$ for $v\in\Idx$, and $\Hyp=(X,(S_v)_{v\in\Idx})$.
Distinct indices give distinct mutable moves even when their sets $S_v$ coincide, as occurs at the endpoints of an isolated-edge component.  For every
$C\subseteq\Idx$, the vertex-cover condition is
\begin{equation}
 \bigcup_{v\in C}S_v=X
 \Longleftrightarrow
 (\forall e\in E_H:\ e\cap C\ne\varnothing).
\label{eq:vc-cover}
\end{equation}
Thus,
\begin{equation}
 \Cov(\Hyp)=\VC(H).
\label{eq:vc-opt}
\end{equation}

If $H$ has maximum degree at most three, each element vertex $e_x$
has out-degree two because each graph edge has two endpoints, and each position $s_j$ has in-degree at most three. The extra element $e_\ast$ has a single successor. The self-loop at $d_j$ and the input edge from $s_j$ give $d_j$ in-degree two. The terminal $\ell_j$ in the construction with terminals has in-degree one. Replace the moves from $q$ to the element positions by a binary tree with those positions as leaves. At an internal node with child outputs $u,v$, introduce $a,b$ with moves $a\to u$, $a\to v$, and $b\to a$; the output is $b$. By the win--move rule for $a$ and~\eqref{eq:single-successor-table} for $b$,
\[
\begin{aligned}
 \Val(b)=\True &\iff \Val(a)=\False
       \iff (\Val(u)=\True)\wedge(\Val(v)=\True),\\
 \Val(b)=\False &\iff \Val(a)=\True
       \iff (\Val(u)=\False)\vee(\Val(v)=\False).
\end{aligned}
\]
For a subtree with leaf set $L$ and output $z$, the first condition becomes $\forall v\in L:\ \Val(v)=\True$, and the second becomes $\exists v\in L:\ \Val(v)=\False$. Induct on $|L|$. For $|L|=1$, $z$ is the leaf. At an internal node, the child leaf sets partition $L$. Conjunction of the two universal conditions gives the universal condition on $L$, and disjunction of the two existential conditions gives the existential condition on $L$. If neither holds, $z$ is drawn. A sole move from $q$ to the root thus makes $q$ lose if all element positions win, win if one loses, and draw otherwise, exactly the value in the original game.

Implement the Boolean condition by gates of fan-in at most two.
A copy tree distributes one total input position $u$ to several occurrences. Before connecting these occurrences, regard $u$ as the sole leaf, with no incoming move and at most one outgoing move. To split a leaf $v$, introduce three fresh positions $t,v_1,v_2$ with moves $v_1\to t$, $v_2\to t$, and $t\to v$. Each of $v_1,v_2$ has value $\Val(v)$ by two applications of the one-successor rule. The old leaf acquires one incoming move, $t$ has in-degree two and out-degree one, and both new leaves have in-degree zero and out-degree one. Induction on the number of splits gives $m$ equal-valued leaves after $m-1$ splits. Attaching each occurrence by one incoming move to a distinct leaf preserves the bounds. Each mutable input has a private terminal.

For the conditional move, each graph edge gives a disjunction with two one-successor complements and one two-successor position. Combine the outputs and $p_\tau$ by the binary conjunction tree. When a circuit input occurs more than once, use a separate copy for each occurrence. Every circuit position has in-degree and out-degree at most two. In the conditional-move construction of Eq.~\eqref{eq:edge-expansion}, $g_e$ has two outgoing moves, $n_e$ has one, and the draw vertex $d$ has two incoming moves, including the self-loop. The input position $p_\tau$ for the designated move, the query position $r$, and its predecessor $p$ also have both degrees at most two.  The copy and degree-reduction transformations
use no new mutable edge.  Every resulting vertex has
in-degree at most $3$ and out-degree at most $2$: only a position $s_j$ associated with a degree-three graph vertex can attain in-degree $3$.

\begin{proof}[Proof of Theorem~\ref{thm:np-completeness}]
\leavevmode\par
\begin{proofsteps}
\item \emph{Membership.} The program is safe and treated as constant in every claimed language; hence,
\Cref{prop:np-membership} supplies NP membership.

\item \emph{Hardness.} We give a
polynomial, threshold-preserving reduction from Vertex Cover restricted to graphs of maximum degree at most three with at least one edge, which is NP-complete \cite{garey1974some}.

After deleting isolated vertices, take the indexed set system of Eq.~\eqref{eq:vc-cover}. It has $|E_H|$ elements, $|V_H|$ sets, and $2|E_H|$ incidences. By Eq.~\eqref{eq:vc-opt}, the minimum cover size is $\VC(H)$.

For $\DetRadius$, use the construction with self-loops for $\Hyp$, with no designated candidate. Since $X\ne\varnothing$, $C=\varnothing$ leaves every element uncovered and the observed answer is $\Undef$.
\Cref{thm:selector} gives
\[
 \DetRad(E_{\Hyp}^{u})\leq k
 \,\Longleftrightarrow\,
 \Cov(\Hyp)\leq k\,
 \Longleftrightarrow
 \VC(H)\leq k.
\]
For a prescribed $b\to c$, choose the construction, candidate and query position in Table~\ref{tab:six-transitions}. Let $E$ be its observation. Lemmas~\ref{thm:selector}, \ref{lem:total-selector} and~\ref{lem:draw-opening} give $\Ans(E)=b$ in each row and, for every $C\subseteq V_H$,
\[
\begin{aligned}
 &\Ans(E\toggle\Delta_C)=b,\\
 &\Ans(E\toggle\Delta_C\toggle\{\tau\})=c
 \iff \bigcup_{v\in C}S_v=X
 \iff \forall e\in E_H:\ e\cap C\ne\varnothing .
\end{aligned}
\]
Every subset of the mutable moves other than $\tau$ is a unique $\Delta_C$. Therefore,
\[
 (H,k)\in\mathsf{VertexCover}
 \iff \exists C\subseteq V_H:\ |C|\le k\ \wedge\ \bigcup_{v\in C}S_v=X
 \iff \Kappa_E^{b\to c}(\tau)\le k ,
\]
where the last equivalence uses $|\Delta_C|=|C|$. The construction preserves the threshold.
\item \emph{Size and degree bounds.} There are $2|E_H|$ incidences. The binary trees add a constant number of positions and moves per incidence or leaf, giving $O(N)$ positions and moves for $N=|V_H|+|E_H|$. Numbering the positions consecutively gives an $O(N\log(N+2))$-bit move list that can be written in polynomial time. The degree bounds follow from the binary tree construction in Section~\ref{app:selector-degree}.  The degree
transformation introduces no mutable move, preserves every start value, and
preserves the map $C\mapsto\Delta_C$.  Finally, all mutable objects are move facts of the program in Eq.~\eqref{eq:win-intro}.
\end{proofsteps}
\end{proof}

\begin{proof}[Proof of Theorem~\ref{thm:exact-complexity}]
\leavevmode\par
\begin{proofsteps}
\item \emph{Membership.} On malformed inputs return zero. Otherwise let $n=|\Univ|$. For the contingency problem, $\tau\notin\Univ$ or $\Ans(E)\ne b$ gives $n$ by~\eqref{eq:total-kappa}. For the intervention problem, $\Ans(E)\ne\Undef$ gives $n+1$ by~\eqref{eq:total-radius}. The remaining inputs have the respective upper bounds $m=n-1$ and $m=n$ on a finite optimum. In the first case $\tau\in\Univ$, so $n\ge1$. For either problem let $P(k)$ be acceptance of the corresponding NP decision instance at threshold $k$. Its totalized optimum is
\[
 \min\bigl(\{k\in\{0,\ldots,m\}\mid P(k)\}\cup\{m+1\}\bigr).
\]
A witness of size at most $k$ is a witness at threshold $k+1$. Query $P(m)$ to test feasibility. If it holds, binary search on $\{0,\ldots,m\}$ takes at most $\lceil\log_2(m+1)\rceil$ further queries; if it fails, return $m+1$. In total there are at most $1+\lceil\log_2(n+1)\rceil$ queries. The explicit input lists the $n$ mutable facts, giving membership in $\FPNPlog$.
\item \emph{Optimization hardness.} Maximum clique size on finite simple graphs is $\FPNPlog$-complete under polynomial-time metric reductions~\cite{krentel1986complexity}. Let $G=(V,A)$ be such a graph, $H$ its complement, and $\omega(G)$ its maximum clique size. For $K\subseteq V$,
\[
 K\text{ is a clique of }G
 \iff \forall e\in E(H):\ e\nsubseteq K
 \iff V\setminus K\text{ covers }H.
\]
Hence $\omega(G)=|V|-\VC(H)$. For each claimed optimum $f$ and each $H$ with an edge, discard isolated vertices and let $I_f(H)$ be the corresponding construction of Section~\ref{app:selector}, without a degree restriction. The endpoint and cardinality equivalences in~\eqref{eq:selector-response}, \eqref{eq:selector-false-cause}, \eqref{eq:selector-true-cause}, \eqref{eq:total-selector-causes}, and~\eqref{eq:draw-opening-causes} give $f(I_f(H))=\VC(H)$. The observed value is the required source value, and the candidate, when required, belongs to $\Univ$. Thus the finite optimum is also the totalized output. Writing $K_2$ for the graph consisting of a single edge, define the polynomial-time maps
\[
 h_f(G)=
 \begin{cases} I_f(H),&E(H)\ne\varnothing,\\ I_f(K_2),&E(H)=\varnothing,\end{cases}
 \qquad
 d(G,z)=
 \begin{cases}|V|-z,&E(H)\ne\varnothing,\\ |V|,&E(H)=\varnothing.\end{cases}
\]
If $H$ has no edge, $V$ is a clique of $G$, including $V=\varnothing$. Thus $d(G,f(h_f(G)))=\omega(G)$ in both cases.
\item \emph{Counting hardness.} Counting the vertex covers of a bipartite graph is $\SharpP$-hard under polynomial-time Turing reductions~\cite{provanBall1983}. Given such a graph $G=(V,A)$, form $H$ by adjoining one private leaf $v'$ and the edge $\{v,v'\}$ for each $v\in V$. The $|V|$ disjoint leaf edges give $\VC(H)\ge |V|$; $V$ covers $H$, giving equality. A minimum cover $K$ of $H$ therefore satisfies
\[
 |K\cap\{v,v'\}|=1\quad(v\in V)\quad\text{and}\quad
 \forall e\in A:\ e\cap(K\cap V)\ne\varnothing .
\]
Conversely, for any cover $C$ of $G$, the set $C\cup\{v'\mid v\in V\setminus C\}$ meets all leaf edges and covers $A$, with cardinality $|V|$. Hence the mutually inverse maps are
\[
 C\longmapsto C\cup\{v'\mid v\in V\setminus C\}\quad\text{and}\quad
 K\longmapsto K\cap V.
\]
For $V=\varnothing$, the answer is one. Otherwise $H$ has an edge. For each transition, compose the first bijection with the map $C\mapsto\Delta_C$ from minimum covers of $H$ to minimum contingencies in Section~\ref{app:selector}. The number of minimum contingencies is exactly the number of covers of $G$. The two constructions have polynomial size and running time, proving the counting lower bound.
\end{proofsteps}
\end{proof}

\section{Subtree Min--Max Theorem}\label{app:arboreal}

\subsection{\texorpdfstring{Transversal--Packing Equality}{Transversal--Packing Equality}}
\paragraph*{Two restrictions of Example~\ref{example_intro_2}.}
We write $\tau_a=(s,a)$ and $\tau_\ell=(s,\ell)$. First let $\Mu_{\mathsf f}=E_{\mathsf f}=\{\tau_a\}$, with $(a,s)$ fixed. With $\tau_a$ present, $s$ lies on a forced cycle and is drawn; with $\tau_a$ absent, $s$ is terminal and loses. Thus $f_{\Undef}$ has the single prime implicant $(\{\tau_a\},\varnothing)$. Next let $\Mu_{\mathsf t}=\{\tau_\ell\}$, $E_{\mathsf t}=\varnothing$, and fix the cycle $s\to a\to s$. Without $\tau_\ell$, $s$ is drawn; with $\tau_\ell$, $s$ can move to the losing terminal $\ell$ and wins. The single prime implicant is $(\varnothing,\{\tau_\ell\})$. In each case the two state values verify Clauses~1 and~2 of Definition~\ref{def:subtree-representation}, and the one-vertex tree verifies Clause~3. In each game the singleton containing the mutable move is the minimum intervention. The family consisting of its one support is a maximum packing.

\begin{lemma}\label{lem:requirement-support-toggle}
Let $\Req\in\mathcal O_E^{\min}$, $\Delta\subseteq\Mu$, $F=E\toggle\Delta$, and
$B(\Req)=\Req^+\cup\Req^-$.  Then,
\begin{equation*}
 F\not\models\Req
 \,\,\,\Longleftrightarrow\,\,\,
 \Delta\cap B(\Req)\neq\varnothing.
\end{equation*}
\end{lemma}

\begin{proof}
Since $\Req^+\subseteq E$ and $\Req^-\cap E=\varnothing$, toggling $\Delta$ gives
\[
 F\models\Req\iff(\Req^+\cap\Delta=\varnothing)\wedge(\Req^-\cap\Delta=\varnothing)
 \iff B(\Req)\cap\Delta=\varnothing .
\]
Negating the equivalence proves the claim.
\end{proof}

\begin{lemma}\label{lem:subtree-konig}
Let $T$ be a finite tree and let $\mathcal B=(B_j)_{j\in J}$ be a finite indexed family of nonempty subsets of $V(T)$, each inducing a connected subgraph. The minimum size of a transversal equals the maximum size of a pairwise disjoint subfamily~\cite[Cor.~66.5a]{schrijver2003optimization}. Given $T$ and the indexed sets explicitly, both optima and witnesses are computable in polynomial time.
\end{lemma}

\begin{proof}
For $J=\varnothing$, the empty set and the empty family attain both optima. Assume $J\ne\varnothing$. Root $T$ at $r$, let $d(v)$ be the distance from $r$ to $v$, and write $[u,v]$ for the vertex set of the unique $u$--$v$ path. Choose $t_j\in B_j$ minimizing $d$. For $x\in B_j$, let $w$ be the last common vertex of the paths from $r$ to $t_j$ and to $x$. Since $w\in[t_j,x]\subseteq B_j$, minimality gives $d(t_j)\le d(w)\le d(t_j)$. Thus $w=t_j$. It follows that $t_j\in[r,x]$ for each $x\in B_j$, and a second vertex of the same minimum depth must equal $t_j$.

Let $H\subseteq J$ be the active indices, initially $J$. While $H\ne\varnothing$, choose $j\in H$ maximizing $d(t_j)$. For $i\in H$ and $x\in B_i\cap B_j$, both $t_i,t_j$ lie on $[r,x]$ and $d(t_i)\le d(t_j)$. Thus $t_j\in[t_i,x]\subseteq B_i$. Conversely, $t_j\in B_i$ implies $t_j\in B_i\cap B_j$. The indices removed at this iteration are therefore
\[
 R_j=\{i\in H\mid t_j\in B_i\}
    =\{i\in H\mid B_i\cap B_j\ne\varnothing\}.
\]
Starting from $X=P=\varnothing$, update
\[
 (X,P,H)\longleftarrow
 (X\cup\{t_j\},\ P\cup\{j\},\ H\setminus R_j).
\]
Since $j\in R_j$, after at most $|J|$ iterations $H=\varnothing$. The first expression for $R_j$ shows that $X$ hits each removed set. The second shows that no later chosen set meets an earlier one. Hence $(B_j)_{j\in P}$ is a packing, and the vertices $t_j$ chosen in those disjoint sets are distinct. Every transversal meets each set of any packing, so
\[
 |P|\le\nu(\mathcal B)\le\Trans(\mathcal B)\le|X|=|P|,
\]
where $\nu$ and $\Trans$ refer to the hypergraph with vertices $V(T)$ and indexed edges $\mathcal B$. Computing the depths and scanning the active lists once per iteration takes polynomial time in $|V(T)|+|J|+\sum_{j\in J}|B_j|$.

For a nonempty $H\subseteq J$ with pairwise intersecting sets $(B_i)_{i\in H}$, choose $j\in H$ maximizing $d(t_j)$ over $H$. Then $R_j=H$, giving $t_j\in\bigcap_{i\in H}B_i$.
\end{proof}

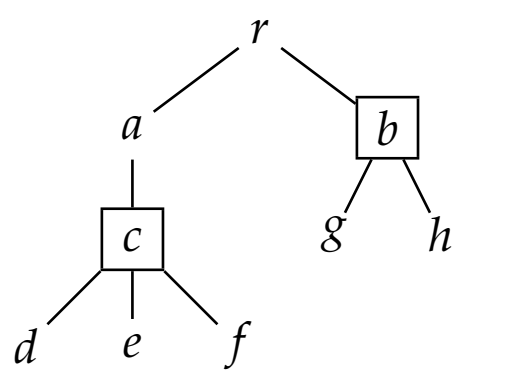
Figure~\ref{fig:greedy-subtrees} applies Lemma~\ref{lem:subtree-konig} to five supports $B_1,\ldots,B_5$. Root the tree at $r$ and let $t_j$ be the vertex of $B_j$ nearest $r$. Let $P$ record the chosen indices and $X$ the chosen vertices. Initially $t_1=c$ is the unique deepest such vertex. Removing the three supports containing $c$ leaves $B_4$ and $B_5$, with $t_4=b$ deeper than $t_5=r$. Thus $X=\{c,b\}$ meets all five supports. The disjoint supports $B_1$ and $B_4$ require two distinct vertices in every transversal, so both witnesses are optimal. Choosing $B_3$ first and deleting the supports containing $t_3=r$ would leave $B_1$ and $B_4$ active even though $B_3$ meets both.

\begin{figure}[htbp]
\centering
{\small
\begin{tabular}{@{}c@{\hspace{10pt}}c@{}}
\begin{tabular}{@{}c@{}}
\setlength{\unitlength}{1pt}%
\begin{picture}(88,72)(0,6)
\linethickness{.5pt}
\put(44,69){\line(-4,-3){16}}
\put(52,69){\line(4,-3){14}}
\put(24,48){\line(0,-1){9}}
\put(18,27){\line(-1,-1){10}}
\put(24,27){\line(0,-1){9}}
\put(30,27){\line(1,-1){10}}
\put(69,48){\line(-1,-2){5}}
\put(75,48){\line(1,-2){5}}
\put(48,72){\makebox(0,0){$r$}}
\put(24,54){\makebox(0,0){$a$}}
\put(66.5,48.5){\framebox(11,11){$b$}}
\put(18.5,27.5){\framebox(11,11){$c$}}
\put(4,13){\makebox(0,0){$d$}}
\put(24,13){\makebox(0,0){$e$}}
\put(44,13){\makebox(0,0){$f$}}
\put(62,34){\makebox(0,0){$g$}}
\put(82,34){\makebox(0,0){$h$}}
\end{picture}
\end{tabular}
&
{\setlength{\tabcolsep}{4pt}\renewcommand{\arraystretch}{1.1}%
\begin{tabular}{|c|l c|}
\hline
$j$ & Support $B_j$ & $t_j$\\ \hline
1 & $\{\textcolor{red}{\boldsymbol{c}},d,e,f\}$ & $c$\\
2 & $\{a,\textcolor{red}{\boldsymbol{c}},d\}$ & $a$\\
3 & $\{r,a,b,\textcolor{red}{\boldsymbol{c}}\}$ & $r$\\ \hline
4 & $\{\textcolor{red}{\boldsymbol{b}},g,h\}$ & $b$\\
5 & $\{r,\textcolor{red}{\boldsymbol{b}},g\}$ & $r$\\ \hline
\end{tabular}}
\end{tabular}
\par\smallskip
{\setlength{\tabcolsep}{5pt}\renewcommand{\arraystretch}{1.1}%
\begin{tabular}{|c|c c|l|}
\hline
Step & Add to $P$ & Add to $X$ & Removed indices\\ \hline
1 & $1$ & $\textcolor{red}{\boldsymbol{c}}$ & $\{1,2,3\}$\\
2 & $4$ & $\textcolor{red}{\boldsymbol{b}}$ & $\{4,5\}$\\ \hline
\end{tabular}}\par}
\caption{Greedy packing and transversal on a branching host tree. Squares mark the selected vertices. The $\textcolor{red}{red}$ vertex in each support is the one selected when that support is removed.}
\label{fig:greedy-subtrees}
\ifdefined\Description
\Description{In the rooted tree, r has children a and b. The only child of a is c, whose children are d, e and f. The children of b are g and h. The first step selects support 1 and vertex c and removes supports 1 through 3. The second selects support 4 and vertex b and removes supports 4 and 5.}
\fi
\end{figure}

\begin{proof}[Proof of Theorem~\ref{thm:arboreal-minmax}]
Let $\Delta\subseteq\Mu$ and $F=E\toggle\Delta$. By Clauses~1 and~2 of Definition~\ref{def:subtree-representation} and Lemma~\ref{lem:requirement-support-toggle},
\[
\begin{aligned}
 \Ans(F)=c
 &\iff \Ans(F)\ne\Undef\\
 &\iff \forall\Req\in\mathcal O_E^{\min}: F\not\models\Req\\
 &\iff \forall\Req\in\mathcal O_E^{\min}: \Delta\cap B(\Req)\ne\varnothing .
\end{aligned}
\]
In particular, $\Delta=\Mu$ resolves the query because all supports are nonempty. The feasible interventions are precisely the transversals, with the same cardinality. Each support induces a connected subgraph of $T$ by Clause~3.

Let $n=|\Mu|$ and $N=|V|+|\Mx|+n$. To test whether $\Delta$ is a transversal, form $E\toggle\Delta$ in $O(n)$ time and evaluate the query there. Each retrograde stage takes $O(N)$ time, and each strict stage adds a vertex. Thus at most $|V|+1$ stages, including the final test, suffice. The transversal test takes $S=O((|V|+1)N)$ time. For a subtree hypergraph on $n$ vertices, the transversal algorithm of van den Heuvel and Johnson~\cite{vandenHeuvelJohnson2008} uses only the vertices and this test and returns a minimum transversal in $O(n^3S)$ time. The returned transversal is therefore a minimum $c$-intervention.

Given $T$ and the terms explicitly, Lemma~\ref{lem:subtree-konig} returns a minimum transversal $X$ and indices of $|X|$ pairwise disjoint supports, and the indices identify the corresponding prime implicants. The equivalence gives $\Ans(E\toggle X)=c$. Retrograde evaluation at $E\toggle X$ computes the winning and losing regions and their first-entry ranks. By \Cref{cor:canonical-partition,lem:canonical-ranks}, these regions and ranks form the required certificate for $q$. The intervention and packing attain Eq.~\eqref{eq:arboreal-minmax}.
\end{proof}

\subsection{\texorpdfstring{Realizability and Strictness}{Realizability and Strictness}}
A hypergraph is \emph{Sperner} if no hyperedge properly contains another.
For a finite indexed Sperner hypergraph
$\mathcal B=(A,(B_i)_{i\in I})$ with $I\neq\varnothing$ and nonempty pairwise distinct hyperedges, let
\[
\begin{aligned}
 D_{\mathcal B}^{\mathsf x}
 &=\{\mathsf{Edge}(i)\mid i\in I\}
  \cup\{\mathsf{In}(i,a)\mid (i\in I)\wedge(a\in B_i)\},\\
 \Univ_{\mathcal B}&=\{\mathsf{On}(a)\mid a\in A\}.
\end{aligned}
\]
The observed state is $E_{\mathcal B}=\Univ_{\mathcal B}$, and
$I_{\mathcal B}=(\PiP_{\mathsf{hyp}},D_{\mathcal B}^{\mathsf x},
\Univ_{\mathcal B},E_{\mathcal B},q)$ is the resulting intervention
instance.  In the following statements, $\Ans$ denotes the response map of
$I_{\mathcal B}$.

\begin{lemma}\label{lem:sperner-realization}
For $X\subseteq A$, let
$F_X=E_{\mathcal B}\toggle\{\mathsf{On}(a)\mid a\in X\}$.  Then,
\begin{equation}
 \Ans(F_X)=
 \begin{cases}
  \False,&\forall i\in I:\ X\cap B_i\neq\varnothing,\\
  \Undef,&\exists i\in I:\ X\cap B_i=\varnothing.
 \end{cases}
\label{eq:sperner-response}
\end{equation}
The prime implicants of $f_{\Undef}$ satisfied by $E_{\mathcal B}$ are precisely the pairs
$\Req_i=(\{\mathsf{On}(a)\mid a\in B_i\},\varnothing)$ for $i\in I$.
Under $a\mapsto\mathsf{On}(a)$, their supports are the indexed hyperedges $B_i$, and we use the same identification for host trees.
\end{lemma}

\begin{proof}
Let $D=D_{\mathcal B}^{\mathsf x}\cup F_X$, let $B=\{\mathsf{Bad}(i)\mid i\in I\}$ and $L=\{\mathsf{Draw}(i)\mid i\in I\}$. We also use
\[
 B_X=\{\mathsf{Bad}(i)\mid i\in I,\ X\cap B_i\ne\varnothing\},
 \qquad
 L_X=\{\mathsf{Draw}(i)\mid i\in I,\ X\cap B_i=\varnothing\}.
\]
Let $Q_X=\{q\}$ for $L_X\ne\varnothing$ and $Q_X=\varnothing$ otherwise. Apply the alternating reduct construction~\eqref{eq:wfs-bounds} to the ground program with facts $D$. Each reduct contains exactly the extensional facts in $D$ in its least model; $\mathsf{Edge}(b)$ is false for $b\notin I$. The bounds are
\[
\begin{aligned}
 K_0&=\varnothing, & U_0&=D\cup B\cup L\cup\{q\},\\
 K_1&=D\cup B_X,    & U_1&=D\cup B_X\cup L_X\cup Q_X,\\
 K_2&=K_1 .
\end{aligned}
\]
At $K_0$ all negative tests pass. Each $B_i$ is nonempty, and $I\ne\varnothing$, giving $U_0$. At $U_0$, $\neg\mathsf{On}(a)$ survives exactly for $a\in X$ among the listed members; every rule for $\mathsf{Draw}(i)$ with $i\in I$ is discarded. For $b\notin I$, any surviving rule for $\mathsf{Draw}(b)$ has the false premise $\mathsf{Edge}(b)$. This gives $K_1$. At $K_1$, no $\mathsf{Draw}(i)$ is present, and its rule survives exactly when $X\cap B_i=\varnothing$; $q$ is derived exactly when $L_X\ne\varnothing$. At $U_1$, for each $i\in I$, either $\mathsf{Bad}(i)\in B_X$ or $\mathsf{Draw}(i)\in L_X$, so no draw rule with a listed index survives and $K_2=K_1$. Thus, $q\notin K_1$ and $q\in U_1\iff L_X\ne\varnothing$, proving~\eqref{eq:sperner-response}. Taking $X=\varnothing$ gives $\Ans(E_{\mathcal B})=\Undef$.

For $P\subseteq A$, let $G_P=\{\mathsf{On}(a)\mid a\in P\}$. Since $E_{\mathcal B}=G_A$, the terms it satisfies are precisely $(G_P,\varnothing)$. By~\eqref{eq:sperner-response},
\[
\begin{aligned}
 (G_P,\varnothing)\text{ implies } f_{\Undef}
 &\iff \forall Y\subseteq A:\ (P\subseteq Y\Rightarrow \exists i\in I:\ B_i\subseteq Y)\\
 &\iff \exists i\in I:\ B_i\subseteq P .
\end{aligned}
\]
For the last equivalence take $Y=P$ in the forward direction and use $B_i\subseteq P\subseteq Y$ in the reverse direction. A minimal such $P$ must equal some $B_i$; no proper subset of any $B_i$ contains a listed $B_j$, by incomparability. Thus all and only the stated terms are prime.
\end{proof}

\begin{proof}[Proof of Proposition~\ref{prop:subtree-strictness}]
\begin{proofsteps}
\item \emph{Subtree representation.} For $n\geq 2$, let $T_n$ be the star with centre $o$ and leaves
$a_1,\ldots,a_n$, and let
\begin{equation*}
 \mathcal B_n^{\star}=
 \{\{o,a_i,a_j\}\mid 1\leq i<j\leq n\}.
\end{equation*}
The family $\mathcal B_n^{\star}$ is Sperner, and every member
induces a connected subtree of $T_n$.  Lemma~\ref{lem:sperner-realization} gives values only in
$\{\Undef,\False\}$, gives $\Ans(F)=\Undef$ iff $F$ contains one of the triples, and identifies the triples as the supports of the prime implicants of $f_{\Undef}$ satisfied by $E$. Hence $E$, $c=\False$, and $T_n$ satisfy Clauses~1--3.

\item \emph{Strict containment.} For $n=3$, the supports are
$\{o,a_1,a_2\}$, $\{o,a_1,a_3\}$ and $ \{o,a_2,a_3\}.$
A three-element subset of a four-vertex path is an interval only when the
omitted vertex is an endpoint.  The three supports omit three distinct
leaves, while a path has two endpoints.  No path host represents all three
supports.  Since every path is a tree, this proves strict containment.

\item \emph{Unbounded incidence treewidth.} Let $G_n$ be the incidence graph of $\mathcal B_n^{\star}$.  Delete the
centre vertex $o$ from $G_n$.  The remaining graph is obtained from
$K_n$ by subdividing every edge once, with subdivision vertex indexed by
$\{o,a_i,a_j\}$.  For each $i<j$, contract the edge from the corresponding subdivision vertex to $a_i$. Each remaining edge then joins $a_i$ to $a_j$, giving a $K_n$ minor. Deletion and contraction cannot increase treewidth: remove a deleted vertex from all bags of a tree decomposition, and replace the endpoints of a contracted edge by the merged vertex. The endpoint subtrees meet in a bag for the edge, thus their union is connected. In a tree decomposition of $K_n$ the subtrees for the $n$ vertices pairwise meet; the subtree intersection property proved in Lemma~\ref{lem:subtree-konig}~\cite{golumbic2004algorithmic} gives a bag containing all $n$. Hence, $\operatorname{tw}(G_n)\geq n-1$, proving that the realized support hypergraphs have unbounded incidence treewidth.
\end{proofsteps}
\end{proof}

\begin{proof}[Proof of Corollary~\ref{thm:packing-failure}]
\begin{proofsteps}
\item
Let $n\geq2$, let $A=[n]=\{1,\ldots,n\}$, let $k=\lfloor n/2\rfloor+1$, and let
\begin{equation*}
 \mathcal B_n=\{B\subseteq A\mid |B|=k\}.
\end{equation*}
Since $1\le k\le n$, the family is nonempty, and distinct members have the same size and are inclusion-incomparable. Lemma~\ref{lem:sperner-realization} realizes it as
$\mathcal H_E$ for program~\eqref{eq:hypergraph-program}.
\item
For distinct
$B,B'\in\mathcal B_n$, $|B\cap B'|=2k-|B\cup B'|\ge 2k-n>0$. Thus, a packing contains at most one hyperedge, and any single member attains this bound:
\begin{equation*}
 \nu(\mathcal H_E)=1.
\end{equation*}
\item
A set $X\subseteq A$ intersects every member of $\mathcal B_n$ if and
only if $|A\setminus X|<k$: if $|A\setminus X|\geq k$, a $k$-subset of $A\setminus X$ is disjoint from $X$; if $|A\setminus X|<k$, no $k$-subset of $A$ is disjoint from $X$. The condition is $|X|\ge n-k+1$, attained by any $X$ of that size. Thus $\Trans(\mathcal H_E)=n-k+1=\lceil n/2\rceil$.
Lemma~\ref{lem:sperner-realization} implies Clauses~1 and~2 for $c=\False$. The ratio equals $\lceil n/2\rceil$ and is unbounded as $n$ grows.
\end{proofsteps}
\end{proof}

\subsection{Testing the coverage condition}\label{app:coverage-test}

\begin{proposition}\label{prop:recognizing-coverage}
Let $E\subseteq\Mu$ be an undefined observation. Clause~\ref{item:observed-coverage} of Definition~\ref{def:subtree-representation} is equivalent to~\eqref{eq:coverage-restoration}. Given a game for the fixed win--move rule, its query, and $E$, testing this clause is $\mathsf{coNP}$-complete under polynomial-time many-one reductions, even if $E=\varnothing$ and $\False$ is the only reachable determinate value.
\end{proposition}

\begin{proof}
\emph{Equivalence.} Suppose Clause~\ref{item:observed-coverage} holds and $\Ans(E\toggle\Delta)=\Undef$. Choose a prime implicant $\Req$ satisfied by $E$ and $E\toggle\Delta$. By Lemma~\ref{lem:requirement-support-toggle}, $B(\Req)\cap\Delta=\varnothing$. For $e\in\Delta$ the support also misses $\Delta\setminus\{e\}$, so $E\toggle(\Delta\setminus\{e\})\models\Req$. This proves~\eqref{eq:coverage-restoration}.

Conversely, suppose~\eqref{eq:coverage-restoration} holds and $\Ans(E\toggle\Delta)=\Undef$. For any $\Gamma\subseteq\Delta$, list $\Delta\setminus\Gamma=\{e_1,\ldots,e_k\}$ and let $\Delta_i=\Delta\setminus\{e_1,\ldots,e_i\}$ for $0\le i\le k$. The answer at $E\toggle\Delta_0$ is undefined by assumption. For $i<k$, apply~\eqref{eq:coverage-restoration} to $\Delta_i$ and $e_{i+1}$; thus, the answer remains undefined through $\Delta_k=\Gamma$. Let $\Req=(E\setminus\Delta,\Mu\setminus(E\cup\Delta))$. For $F\subseteq\Mu$,
\[
 F\models\Req
 \iff F\setminus\Delta=E\setminus\Delta
 \iff E\toggle F\subseteq\Delta
 \implies \Ans(F)=\Undef ,
\]
where the last implication takes $\Gamma=E\toggle F$. Thus $\Req$ is an implicant. Its finitely many subterms include an inclusion-minimal implicant, which remains true at $E$ and $E\toggle\Delta$. This proves coverage; satisfaction of an implicant implies undefinedness by definition, including when $\Delta=\varnothing$.

\emph{Membership.} Failure of~\eqref{eq:coverage-restoration} has a witness $(\Delta,e)$ with $e\in\Delta$, $\Ans(E\toggle\Delta)=\Undef$, and $\Ans(E\toggle(\Delta\setminus\{e\}))\ne\Undef$. The set and the element have polynomial-size encodings, and two evaluations of the fixed program check the witness in polynomial time.

\emph{Hardness.} We reduce unsatisfiability of formulas in conjunctive normal form, the complement of the satisfiability problem in Cook's theorem~\cite[Theorem~2]{karp1972reducibility}. Let $\varphi$ have variable set $Y$ and let $p,s$ be fresh. Use one mutable move per member of $\Mu=Y\cup\{p,s\}$, identifying each member with its move. For $F\subseteq\Mu$, let $F\cap Y\models\varphi$ mean that $\varphi$ holds with precisely the variables of $F\cap Y$ true. Use the Boolean condition
\[
 B(F)=1\iff (F=\varnothing)\ \vee\
       (\{p,s\}\subseteq F\ \wedge\ F\cap Y\models\varphi).
\]
Testing emptiness by a conjunction of negative literals and evaluating the clauses of $\varphi$ require $O(|Y|+|\varphi|+1)$ gates and wires. Lemma~\ref{lem:boolean-game-circuit} gives a total game position for $B$. From a fresh query position $q$, add one move conditioned on $B$ to a fresh position whose only move is a fixed self-loop. By~\eqref{eq:conditional-move-gate}, the successor of $q$ in the compiled game is drawn when $B(F)=1$ and winning when $B(F)=0$. Hence
\[
 \Ans(F)=
 \begin{cases}
  \Undef & B(F)=1,\\
  \False & B(F)=0.
 \end{cases}
\]
The construction takes polynomial time. With $E=\varnothing$, the states $E$ and $\{s\}$ attain $\Undef$ and $\False$, respectively.
\begin{enumerate}
\item If $\varphi$ is unsatisfiable, $B(F)=1$ iff $F=\varnothing$. At the only undefined state there is no changed fact to restore, so~\eqref{eq:coverage-restoration} holds.
\item If $A\subseteq Y$ satisfies $\varphi$, let $F=A\cup\{p,s\}$. Then $B(F)=1$ and $B(F\setminus\{p\})=0$, since the latter state contains $s$ and omits $p$. Restoring $p$ violates~\eqref{eq:coverage-restoration}.
\end{enumerate}
The input formula is unsatisfiable exactly when the game satisfies Clause~\ref{item:observed-coverage}.
\end{proof}

\section{Boolean Responses and Deletion Causality}\label{app:conservativity}

\begin{proposition}\label{prop:conservativity}
Suppose $\forall F\subseteq\Univ:\ \Ans(F)\in\{\True,\False\}$. Let $E\subseteq\Univ$, $\tau\in\Univ$, $b=\Ans(E)$ and $c\in\{\True,\False\}\setminus\{b\}$. A set $\Gamma\subseteq\Univ\setminus\{\tau\}$ is a $b\to c$ contingency exactly when
\[
 \Ans(E\toggle\Gamma)=\Ans(E)\ \wedge\
 \Ans(E\toggle\Gamma\toggle\{\tau\})\ne\Ans(E).
\]
If $\PiP$ is positive, $E=\Univ$ and $b=\True$, let $\mathcal S$ consist of the inclusion-minimal sets $S\subseteq\Univ$ with $\Ans(S)=\True$. Then $\Gamma$ is a deletion contingency precisely when it misses a member of $\mathcal S$ and $\Gamma\cup\{\tau\}$ meets each member.
\end{proposition}

\begin{proof}
Let $F=E\toggle\Gamma$ and $F'=F\toggle\{\tau\}$. Boolean totality and $c\ne b$ give $\Ans(F')=c\iff\Ans(F')\ne b$. Definition~\ref{def:typed-cause} now gives the first equivalence.

Assume $\PiP$ is positive, $E=\Univ$ and $b=\True$. For any $F\subseteq F'\subseteq\Univ$, the instance has the same ground rules and fixed facts, and $D_F\subseteq D_{F'}$. Let $L_F,L_{F'}$ be the least models of the respective positive ground programs. Then
\[
 L_{F'}\models\PiP\cup D_F
 \,\,\Longrightarrow\,\, L_F\subseteq L_{F'}
 \,\,\Longrightarrow\,\, (q\in L_F\Rightarrow q\in L_{F'}).
\]
The well-founded and least models agree on positive programs~\cite[Theorem~3.7]{van1991well}. Thus, $\Ans(F)=\True$ implies $\Ans(F')=\True$.
For each $F\subseteq\Univ$ with $\Ans(F)=\True$, the finite nonempty family $\{S\subseteq F\mid \Ans(S)=\True\}$ contains an inclusion-minimal member. Every proper subset of that member lies in $F$, so the member belongs to $\mathcal S$. Conversely, if $S\in\mathcal S$ and $S\subseteq F$, monotonicity gives $\Ans(F)=\True$. Hence,
\begin{equation}\label{eq:support-cover}
 \forall F\subseteq\Univ:\quad
 \Ans(F)=\True\ \Longleftrightarrow\ \exists S\in\mathcal S:\ S\subseteq F .
\end{equation}
Since $E=\Univ$ and $\tau\notin\Gamma$, the two states are $\Univ\setminus\Gamma$ and $\Univ\setminus(\Gamma\cup\{\tau\})$. Substitution in~\eqref{eq:support-cover} expresses the $\True\to\False$ contingency condition as
\begin{align*}
 &(\Ans(\Univ\setminus\Gamma)=\True)\ \wedge\ (\Ans(\Univ\setminus(\Gamma\cup\{\tau\}))=\False)\\
 &\quad\Longleftrightarrow
 (\exists S\in\mathcal S:\ S\cap\Gamma=\varnothing)\ \wedge\ (\forall S\in\mathcal S:\ S\cap(\Gamma\cup\{\tau\})\ne\varnothing).
\end{align*}
If $\varnothing\in\mathcal S$, then $\varnothing\cap(\Gamma\cup\{\tau\})=\varnothing$ for every $\Gamma$, and the second conjunct fails. The equivalence gives the stated deletion contingencies, hence the same minima and responsibilities.
\end{proof}

\end{document}